\documentclass{siamart1116}

\usepackage{lipsum}
\usepackage{amsfonts}
\usepackage{graphicx}
\usepackage{epstopdf}
\usepackage{algorithmic}
\usepackage{amssymb}
\usepackage{amsmath}
\usepackage{amscd}
\usepackage{graphicx}
\usepackage{graphics}
\usepackage{xcolor,graphicx}
\usepackage{csquotes}
\usepackage{mathtools}
\usepackage{caption}
\usepackage{float}
\usepackage{subcaption}
\usepackage{cite}
\usepackage{booktabs}
\usepackage{multirow}
\usepackage{rotating}
\usepackage[algo2e]{algorithm2e} 

\ifpdf
  \DeclareGraphicsExtensions{.eps,.pdf,.png,.jpg}
\else
  \DeclareGraphicsExtensions{.eps}
\fi

\newcommand{\creflastconjunction}{, and~}

\newsiamremark{remark}{Remark}
\newsiamremark{hypothesis}{Hypothesis}
\crefname{hypothesis}{Hypothesis}{Hypotheses}
\newsiamthm{claim}{Claim}

\headers{Complex-valued Imaging}{H. S. Aghamiry, A. Gholami, and S. Operto}

\title{Complex-valued Imaging with Total Variation Regularization: An Application to Full-Waveform Inversion in Visco-acoustic Media\thanks{.
\funding{This study was partially funded by the WIND consortium (\textit{https://www.geoazur.fr/WIND}), sponsored by Chevron, Shell, and Total.}}}

\author{Hossein S. Aghamiry \thanks{Universit\'e C\^ote d'Azur, CNRS, Observatoire de la C\^ote d'Azur, IRD , G\'eoazur, Valbonne, France 
  (\email{aghamiry@geoazur.unice.fr}).}
\and Ali Gholami \thanks{Institute of Geophysics, University of Tehran, Tehran, Iran
  (\email{agholami@ut.ac.ir}).}
\and St\'ephane Operto \thanks{Universit\'e C\^ote d'Azur, CNRS, Observatoire de la C\^ote d'Azur, IRD , G\'eoazur, Valbonne, France
  (\email{operto@geoazur.unice.fr}).}}

\usepackage{amsopn}

\ifpdf
\hypersetup{
  pdftitle={Complex-valued Seismic Imaging in Visco-acoustic Media by TV-Regularized Wavefield Inversion},
  pdfauthor={H. S. Aghamiry, A. Gholami, and S. Operto}
}
\fi

\begin{document}

\maketitle

\begin{abstract}
Full waveform inversion (FWI) is a nonlinear PDE constrained optimization problem, which seeks to estimate constitutive parameters of a medium such as phase velocity, density, and anisotropy, by fitting waveforms. Attenuation is an additional parameter that needs to be taken into account in viscous media to exploit the full potential of FWI.
Attenuation is more easily implemented in the frequency domain by using complex-valued velocities in the time-harmonic wave equation. These complex velocities are frequency-dependent to guarantee causality and account for dispersion.
Since estimating a complex frequency-dependent velocity at each grid point in space is not realistic, the optimization is generally performed in the real domain by processing the phase velocity (or slowness) at a reference frequency and attenuation (or quality factor) as separate real parameters. This real parametrization requires an a priori empirical relation (such as the nonlinear Kolsky-Futterman (KF) or standard linear solid  (SLS) attenuation models) between the complex velocity and the two real quantities, which is prone to generate modeling errors if it does not represent accurately the attenuation behavior of the subsurface. Moreover, it leads to a multivariate inverse problem, which is twice larger than the actual size of the medium and ill-posed due to the cross-talk between the two classes of real parameters. 
To alleviate these issues, we present a mono-variate algorithm that solves directly the optimization problem in the complex domain by processing in sequence narrow bands of frequencies under the assumption of band-wise frequency dependence of the sought complex velocities. 
The algorithm relies on the iteratively-refined wavefield reconstruction inversion method (IR-WRI). IR-WRI extends the linear regime of FWI by processing the wave equation as a weak constraint with the alternating-direction method of multipliers (ADMM) to mitigate the risk of spurious local minima generated by the ill-famed cycle skipping pathology.
To mitigate the ill-posedness of the inversion, three total variation (TV) regularization schemes based upon ADMM and proximity algorithms are presented. In the first, regularization is applied directly on the complex velocities. In the two others, separate TV regularizations are tailored to different attributes of the complex velocities (real and imaginary parts, magnitude, and phase).
The real phase velocity and attenuation factor are then reconstructed a posteriori at each spatial position from the estimated complex velocity using arbitrary empirical relation.
In the numerical experiments, the recorded data are generated with the SLS attenuation model and the real physical parameters are extracted from the recovered complex velocities with both the SLS and KF models. The numerical results first show that the regularization of the amplitude and phase provides the most reliable results. Moreover, they show that the band-by-band design of the inversion limits the sensitivity of the recovered phase velocity and attenuation factor to the attenuation model used for their a posteriori extraction. 
\end{abstract}



\section{Introduction}
Complex-valued image reconstruction arises frequently in different fields of imaging sciences, 
such as seismic imaging \cite{Tarantola_1988_TBI,Ribodetti_1998_ATI,Mulder_2009_AAS}, magnetotellurics \cite{Constable_1998_MMF,Habashy_2004_AGF}, synthetic aperture radar (SAR) imaging \cite{Baraniuk_2007_CRI,Moreira_2013_ATO}, and  magnetic resonance imaging (MRI) \cite{Liang_2000_POM,Lustig_2007_SMT},
in which a complex-valued image is determined from a set of observations by solving an optimization problem which is inherently ill-conditioned.
The focus of this paper is on seismic imaging by Full Waveform Inversion (FWI). FWI is nonlinear PDE constrained optimization problem, which estimates the constitutive parameters of the subsurface (wavespeeds, density, attenuation and anisotropy) by minimizing a distance between the recorded and the simulated waveforms \cite{Tarantola_1984_ISR,Pratt_1999_SWIb,Virieux_2009_OFW}.\\
In virtue of correspondence principle, viscous effects are easily implemented in frequency-domain FWI by using complex-valued velocities in the time-harmonic wave equation \cite{Aki_1980_QST,Toksoz_1981_GRS}. 
Although such complex-valued optimization problems are necessarily nonanalytic, computing the gradient of real functions with respect to complex variables is possible by means of Wirtinger derivatives \cite{Brandwood_1983_CGO,Sorber_2012_UOO}. 
However, traditional approaches  solve the optimization problem in the real domain by treating two real-valued physical attributes of the complex velocity (like phase velocity at a reference frequency and attenuation)
as separate parameters \cite{Hak_2011_SAI,Kamei_2013_ISV} (hereafter we refer to these methods as C2R).
The main reason is that dispersion makes the complex velocities frequency dependent, whose estimation at each spatial position would make the number of unknowns unpractical \cite{Innanen_2016_SAD,Keating_2019_PCA}. In return, inversion in the real domain requires to define  an a priori attenuation mechanism, such as the nonlinear Kolsky-Futterman (KF) or standard linear solid  (SLS) models (see \cite{Ursin_2002_CDA} for a review of dispersion and attenuation models and Appendix~\ref{Appa}), which can generate significant modeling errors when it does not represent accurately the attenuation behavior of the subsurface \cite{Innanen_2016_SAD,Keating_2019_PCA}. Moreover, the real parametrization leads to a multivariate optimization problem, which is twice larger than the actual size of the medium and ill-posed. Ill-posedness mainly results from the cross-talks between the two real parameters, which co-determine the dispersion term of the attenuation mechanism \cite{Keating_2019_PCA}. These cross-talks can be efficiently mitigated by inverting a broad band of frequencies during the late stage of a multiscale inversion \cite{Bunks_1995_MSW} to increase the sensitivity of the FWI to dispersion and hence better decouple the two classes of parameter. However, \cite{Keating_2019_PCA} show the detrimental effects of the modeling errors generated by inaccurate an a priori attenuation model when a broad band of frequencies is involved in the inversion. As a compromise to deal with both the cross-talk and the modeling error issues, \cite{Keating_2019_PCA} design a frequency band-by-band FWI such that the piecewise frequency dependence of the subsurface attenuation can be more easily matched with the a priori attenuation model during each narrow-band inversion. In this framework, the width of the frequency bands controls the trade-off between the needs to reduce parameter cross-talk and sensitivity to modeling errors. \\
In this paper, we follow another road and solve FWI for complex velocities in the framework of the iteratively-refined wavefield reconstruction inversion (IR-WRI) method \cite{Aghamiry_2019_IWR}. IR-WRI extends the search space of FWI with the alternating-direction method of multipliers (ADMM) \cite{Boyd_2011_DOS} to mitigate the risk of spurious local minima generated by the so called cycle skipping pathology. In IR-WRI, the method of multiplier (or augmented Lagrangian method) allows for wave equation errors during wavefield reconstruction to match the data (i.e., satisfy the observation equation) from the first iteration before updating the model parameters from the wavefields by minimizing the wave equation violations (i.e., source residuals). Performing these two tasks in alternating way recasts IR-WRI as a sequence of two linear subproblems, relaying on the bilinearity of the wave equation in wavefield and subsurface parameters. Then, the dual variables or Lagrange multipliers are updated by the constraint violations according to a gradient ascent step. This workflow is iterated until both the observation equation and the wave equation are satisfied with prescribed accuracies. \cite{Aghamiry_2019_VWR} implements IR-WRI for visco-acoustic imaging in the framework of C2R methods where squared slowness and attenuation in the KF model are processed as real optimization parameters. To achieve this goal, the original nonlinear multi-parameter problem was replaced by three recursive linear mono-parameter subproblems for wavefield, squared slowness and attenuation factor.
In this paper, we instead implement IR-WRI as a complex-valued biconvex optimization problem for wavefield and complex velocities.  \\
Our motivations to implement the inversion in the complex domain is two fold. First, our approach does not require an a priori attenuation model since we directly estimate complex velocities. Accordingly, our inversion scheme is not sensitive to modeling errors that would be generated by an inaccurate an a priori attenuation model. Second, we solve a monovariate optimization problem in the complex domain. This avoids the tedious scaling of a multivariate Hessian and the cross-talk issue at the optimization stage. 
To address the issue of the frequency dependency of the complex velocities during multi-frequency inversion, our approach proceeds over contiguous slightly-overlapping frequency batches, in a similar manner to \cite{Keating_2019_PCA}. During each frequency band inversion, we assume a piecewise constant behavior of the sought complex velocities with respect to frequency and we use the final velocity of one batch inversion as the starting velocity for the next batch. This means that, although our band-by-band inversion design is similar to that of \cite{Keating_2019_PCA}, our motivations differ: \cite{Keating_2019_PCA} aims to mitigate the footprint of modeling errors, while we bypass the frequency dependency of the complex velocity with a piecewise constant approximation. Once the final complex velocity model has been recovered from the last frequency batch, the real attributes can be extracted with arbitrary attenuation model. Since the last velocity model was recovered from a narrow band of high frequencies, we do not expect a significant sensitivity of the extraction to the chosen attenuation model according to the conclusions of \cite{Keating_2019_PCA}. Also, although FWI is performed over narrow frequency batches, we expect a much lower imprint of parameter cross-talk than in \cite{Keating_2019_PCA} since the complex-domain inversion is monovariate. Instead, cross-talks are potentially generated during the a posteriori nonlinear extraction of the two real attributes from the complex velocities, the strengths of these cross-talks depending on the accuracy of the inversion results.

\begin{figure}
\centering
\includegraphics[width=0.8\textwidth]{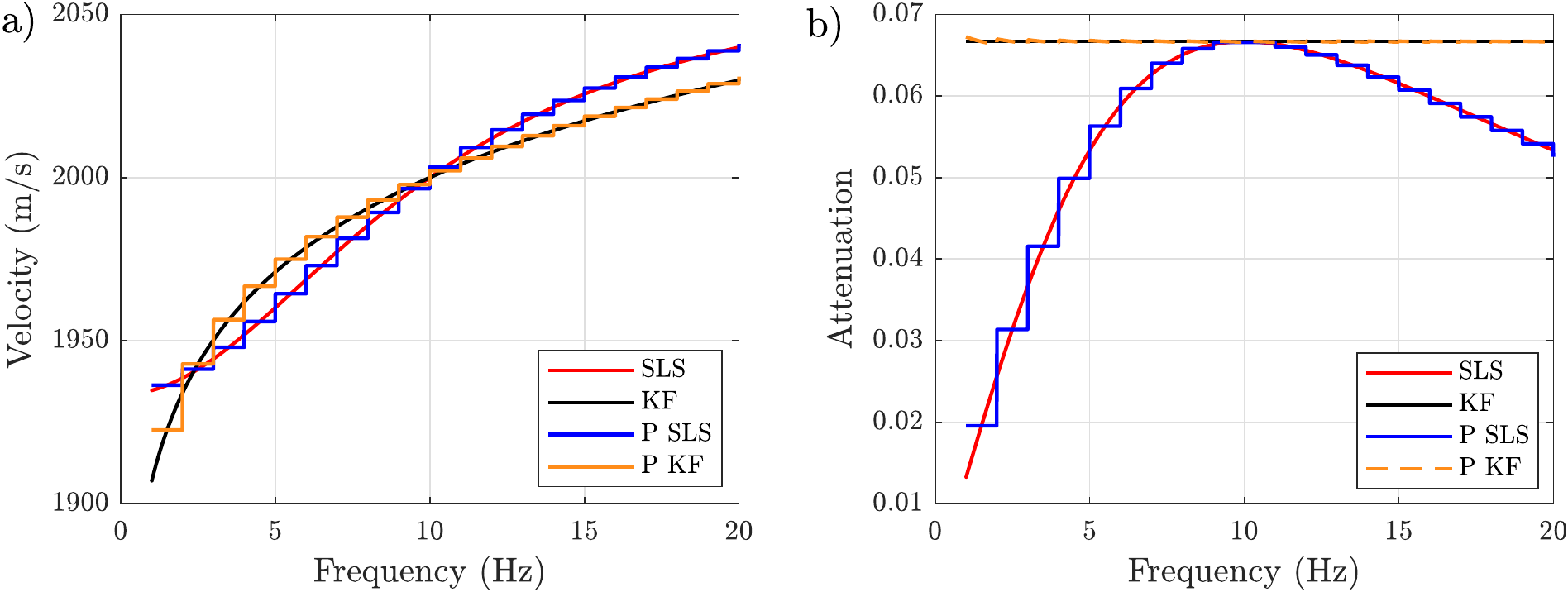}
\caption{Comparison of SLS (red), KF (black), piecewise SLS or P SLS (blue) and piecewise KF or P KF (orange) for frequency range [1 20]Hz. (a) velocity and
(b) attenuation factor. The model is homogeneous and chosen to have a velocity of 2000 m/s and attenuation factor 0.066 at the reference frequency 10 Hz. }
\label{fig:SLS_KF}
\end{figure}

\begin{figure}
\centering
\includegraphics[width=0.8\textwidth]{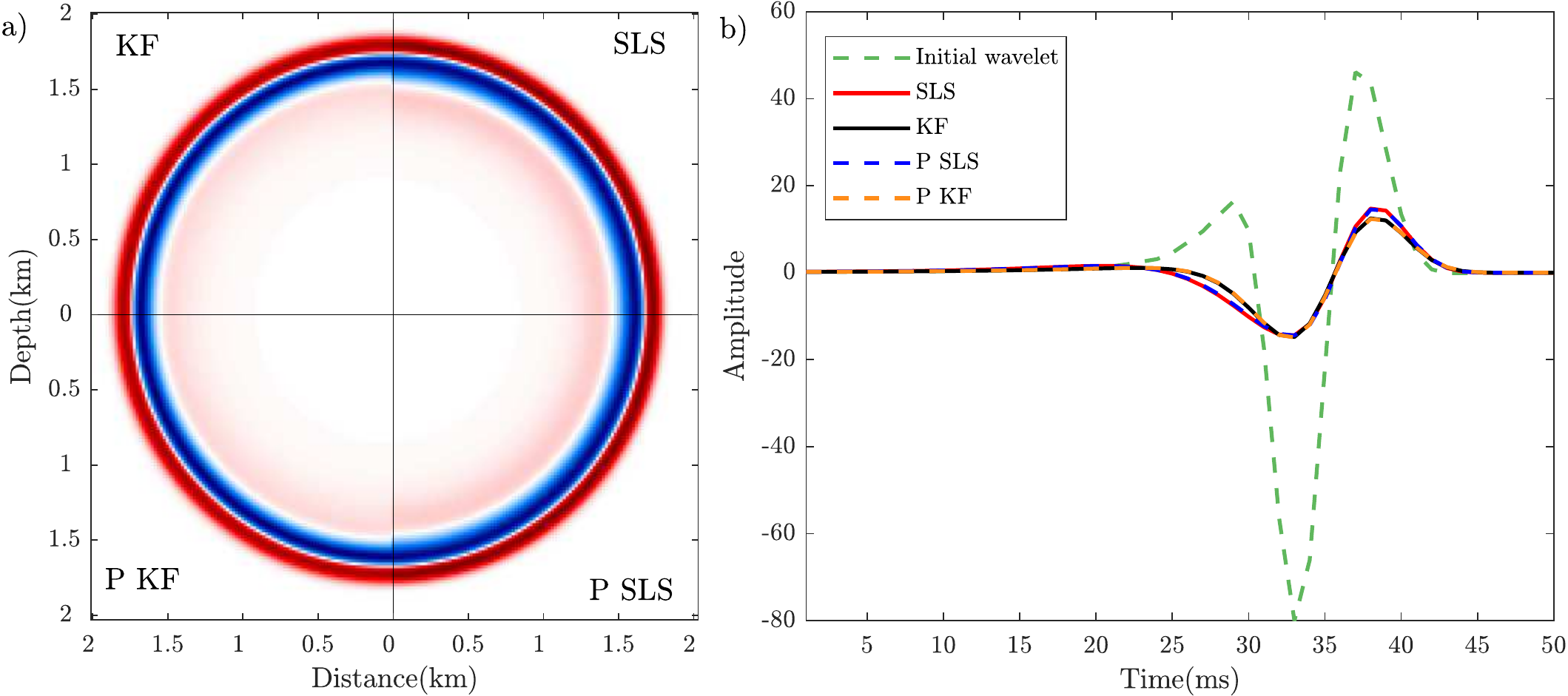}
\caption{Comparison between propagated wavefield with SLS and KF model and their piecewise approximation in 2km $\times$ 2km homogeneous model with velocity 2000 m/s and attenuation factor 0.066 at the reference frequency 10 Hz. (a) Propagated wavefield after 1 s  using KF (top left), SLS (top right), piecewise KF or P KF (bottom left), and piecewise SLS or P SLS (bottom right). (b) A direct comparison between initial source wavelet (dashed green) and the above mentioned attenuated wavefields after 1 s. }
\label{fig:SLS_KF_time}
\end{figure}

To illustrate the validity of the band-wise frequency dependency approximation used in this study,  we compare in Fig. ~\ref{fig:SLS_KF} the phase velocity and attenuation factor of the SLS \cite{Zhu_2013_ACS} and KF \cite{Kolsky_1956_PSP} models (see Appendix~\ref{Appa} for their expression) with those generated with the band-wise frequency dependency approximation for a [1 20]~Hz frequency band. To perform this comparison, we use a velocity of 2000~m/s, an attenuation factor of 0.006 and a reference frequency of 10~Hz. The width of the frequency bands is 1~Hz. It can be seen that the piecewise constant velocities and attenuation factors match more closely the true ones at high frequencies relative to low frequencies. Therefore, we expect a limited imprint of the piecewise approximation at high frequencies, when the extraction of the real attributes from the final complex velocity model is performed. In Fig. ~\ref{fig:SLS_KF}, we centered the horizontal segments of the piecewise constant approximation of the phase velocity and attenuation on the continuous curves at the central frequency of the batch for sake of illustration. During inversion, we indeed do not control what would be the vertical coordinate of the segment. To show the effects of this approximation on the propagated wavefield, we put a point source in the middle of a 2km $\times$ 2km homogeneous model with the above velocity, attenuation and reference frequency, and record the propagated wavefield using the SLS and KF attenuation mechanism after 1 s. These wavefields are shown in the top right and left panels  of Fig. \ref{fig:SLS_KF_time}a for SLS and KF models, respectively. Also, the propagated wavefields with the piecewise SLS (P SLS) (blue lines in Fig. \ref{fig:SLS_KF}) and piecewise KF (P KF) (orange lines in Fig. \ref{fig:SLS_KF}) are shown in the bottom right and left panels of Fig. \ref{fig:SLS_KF_time}a, respectively.  Finally, a direct comparison between these attenuated wavefields and the initial source wavelet are shown in Fig. \ref{fig:SLS_KF_time}b. It can be seen that the errors in the approximated wavefields generated by the band-wise frequency dependency approximation are negligible. \\
The second main objective of this study is to design the most suitable regularization scheme for complex-domain IR-WRI. \cite{Aghamiry_2019_IBC,Aghamiry_2019_CRO} implemented bound constraints and total variation (TV) regularizations in IR-WRI with the split Bregman recipe \cite{Goldstein_2009_SBM}, a particular instance of ADMM for $\ell{1}$ regularized convex problems, to image large-contrast media. Later, \cite{Aghamiry_2019_VWR} applies TV regularization and bound constraints on the squared slowness and attenuation factor for stabilizing the C2R implementation of visco-acoustic IR-WRI.
When considering complex-valued parameters as in this study, the regularization can be applied directly on the complex parameter in a manner similar to that for real parameters or separately on pair of real attributes (real and imaginary parts, magnitude and phase). The second approach is more flexible than the first one in the sense that it allows one to tailor the regularization to the distinctive feature of each attribute.
In this paper, we apply TV regularization first on the real and imaginary parts of the complex velocity and then on the magnitude and phase.
The magnitude and phase separation may be suitable if we assume that the phase mostly depends on attenuation as suggested by the empirical KF and SLS models \cite{Kolsky_1956_PSP,Zhu_2013_ACS}. 
Separate regularization of magnitude and phase images is however more difficult than regularization of real and imaginary parts. 
The former has already been considered in \cite{Zhao_2012_SMA} by using non-linear conjugate gradient method.
In this paper, we propose a novel algorithm based on a generalized proximal point algorithm \cite{Fukushima_1981_AGP}, which can  be easily implemented in IR-WRI. \\
We assess our method against a set of synthetic examples and a more realistic one representative of the North Sea. The simulations are generated using exact SLS attenuation mechanism and the extraction of the real physical parameters (phase velocity and attenuation factor at the reference frequency) is done using both SLS and KF mechanisms to show the robustness of the proposed algorithm against the chosen attenuation mechanism used for real-attribute extraction.  We also compare the relevance of the three different implementations of the TV regularization. 
\subsection{Organization of the Paper}
The paper is organized as follows. 
Section \ref{notation} presents the mathematical symbols adopted in this paper.
Section \ref{FWI0} introduces the regularized FWI problem in visco-acoustic media.
Section \ref{CVOPT} reviews the basics of complex-valued optimization including some preliminaries about convex optimization in subsection \ref{prem} and different algorithms for TV regularization of complex-valued models such as separate real and imaginary regularization (subsection \ref{SRIR}) and separate magnitude and phase regularization (subsection \ref{SMPR}). 
Section \ref{FWI} recasts complex-valued regularized FWI in the framework of the ADMM-based Wavefield Reconstruction Inversion (IR-WRI) method. Complex-valued TV regularized IR-WRI is assessed against numerical examples in 
section \ref{numerical}. Section \ref{conclusions} ends the paper with conclusions. Also, in Appendix~\ref{Appa}, we review the SLS and KF attenuation mechanisms. These models will be used to simulate the observables (seismic data) during the synthetic tests presented in section \ref{numerical} and convert back the reconstructed complex velocities into real velocity and attenuation factor for assessment of the FWI results.

\section{Notation} \label{notation}
The mathematical symbols adopted in this paper are as follows:
We use italics for scalar quantities, boldface lowercase letters for vectors, and boldface capital letters for matrices and tensors.
We use the superscripts $\cdot^T$ to denote the Hermitian conjugate of an operator. 
The $i$th entry of the column vector $\boldsymbol{z}$ is shown by $\boldsymbol{z}_i$. Also, for a complex number $\boldsymbol{z}=\Re(\boldsymbol{z})+i\Im(\boldsymbol{z})=|\boldsymbol{z}|e^{i\angle\boldsymbol{z}}$, $\Re(\boldsymbol{z})$ and $\Im(\boldsymbol{z})$ refer
to the real and imaginary parts, $|\boldsymbol{z}|=\sqrt{\Re(\boldsymbol{z})^2+\Im(\boldsymbol{z})^2}$ shows the magnitude of $\boldsymbol{z}$ and $\angle \boldsymbol{z}=\text{atan2}(\Im(\boldsymbol{z}),\Re(\boldsymbol{z}))$ shows the phase of $\boldsymbol{z}$ and $i=\sqrt{-1}$. For the $n$-length column vectors $\boldsymbol{x}$ and $\boldsymbol{y}$ the dot product is defined by $\langle \boldsymbol{x},\boldsymbol{y}\rangle=\boldsymbol{x}^T\boldsymbol{y}=\sum_{i=1}^n\boldsymbol{x}_i\boldsymbol{y}_i$ and
their Hadamard product, denoted by $\boldsymbol{x}\circ \boldsymbol{y}$, is another vector made up of their component-wise products, i.e. $(\boldsymbol{x}\circ \boldsymbol{y})_i=\boldsymbol{x}_i\boldsymbol{y}_i$. 
The $\ell_2$- and $\ell_1$-norms of $\boldsymbol{x}$ are, respectively, defined by $\|\boldsymbol{x}\|_2=\sqrt{\langle \boldsymbol{x},\boldsymbol{x}\rangle}=\sqrt{\sum_{i=1}^n|\boldsymbol{x}_i|^2}$ and
$\|\boldsymbol{x}\|_1=\sum_{i=1}^n|\boldsymbol{x}_i|$.

\section{Full-waveform Inversion in Visco-acoustic Media} \label{FWI0}
Frequency-domain FWI in visco-acoustic media can be formulated as the following nonlinear PDE-constrained optimization problem 
\begin{align} \label{main0}
&\min_{\boldsymbol{u},\boldsymbol{v},\boldsymbol{\alpha}}~~~~~~~~~~~\mathcal{R}(\boldsymbol{v},\boldsymbol{\alpha}) \\
&\text{subject to}~~~~\boldsymbol{A}(\omega,\boldsymbol{v},\boldsymbol{\alpha})\boldsymbol{u}=\boldsymbol{b}, \nonumber\\
&     \hspace{2cm}                \boldsymbol{Pu}=\boldsymbol{d},   \nonumber
\end{align} 
where $\boldsymbol{u} \in \mathbb{C}^{n\times 1}$ is the state variable (wavefield), 
$\boldsymbol{d} \in \mathbb{C}^{m\times 1}$ is the recorded wavefield at receiver locations (data) via the linear observation operator $\boldsymbol{P} \in \mathbb{R}^{m\times n}$  that maps the state to the observation space,
$\boldsymbol{b} \in \mathbb{C}^{n\times 1}$ is the source term, and  
\begin{equation}  
\label{helmholtz}
 \boldsymbol{A} = \bold{\Delta} + \omega^2 \bold{C} \text{diag}\left(\mathcal{M}(\omega,\boldsymbol{v}, \boldsymbol{\alpha})\right)\bold{B},
\end{equation}
where $\boldsymbol{A}\equiv \boldsymbol{A}(\omega,\boldsymbol{v},\boldsymbol{\alpha})  \in \mathbb{C}^{n\times n}$
is the matrix representation of the discretized PDE for Helmholtz equation, $\omega$ is the angular frequency and $\bold{\Delta}$ is the discretized Laplace operator. We review the method for one source and one frequency. However, the algorithm easily generalizes to multi-source and multi-frequency configuration as discussed in the next subsection \cite{Aghamiry_2019_AMW}.
The operator $\bold{C}$ encloses boundary conditions, e.g., perfectly-matched layers \cite{Berenger_1994_PML}, and the linear operator $\bold{B}$ is the ``mass" matrix  \cite{Marfurt_1984_AFF}.   
More importantly, $\boldsymbol{v}$ and $\boldsymbol{\alpha} \in \mathbb{R}^{n \times 1}$ are the real-valued phase velocity and attenuation factor (inverse of quality factor) at the reference frequency, and $\mathcal{M}$ is a nonlinear mapping, whose expression is provided in equations \ref{KFmodel0} and \ref{SLSmodel0} for the KF model and SLS model, respectively. We refer the reader to \cite{Ursin_2002_CDA} and Appendix \ref{Appa} for a more detailed review of $\mathcal{M}$ for KF and SLS attenuation mechanisms. \\
The optimization problem defined by \eqref{main0} is highly nonconvex and traditionally it is solved in the framework of C2R method with gradient-based Newton type algorithms either for the original parameters $\boldsymbol{v}$ and $\boldsymbol{Q}$ \cite{Ribodetti_2000_AVD,Gao_2016_SIF,Operto_2018_MFF} or using re-parametrization by a variable transformation as reviewed by \cite{Hak_2011_SAI,Kamei_2013_ISV}. 
The fact that the optimization problem is solved using C2R method rather than for a complex parameter results from the dependency of $\mathcal{M}$ on the frequency. A potential issue in C2R method, beside increasing the size, is that this approach requires to introduce a prior in the inversion through the particular form of $\mathcal{M}$. A second issue is potential cross-talks between $\boldsymbol{v}$ and $\boldsymbol{\alpha}$, which makes the multi-variate inversion ill-posed. This cross-talk issue is well illustrated by the fact that $\boldsymbol{v}$ and $\boldsymbol{\alpha}$ co-determine the dispersion term in the KF model in \eqref{KFmodel0} \cite{Keating_2019_PCA}.  \\
In this paper, we seek to avoid using a prior $\mathcal{M}$ and solve problem \ref{main0} directly in the complex domain, leading to
\begin{align} \label{main}
&\min_{\boldsymbol{u},\boldsymbol{m}}~~~~~~~~~~~\mathcal{R}(\boldsymbol{m}) \\
&\text{subject to}~~~~\boldsymbol{A}\boldsymbol{u}=\boldsymbol{b}, \nonumber\\
&     \hspace{2cm}                \boldsymbol{Pu}=\boldsymbol{d},   \nonumber
\end{align} 
which is a bi-convex optimization problem with complex-valued variables $\boldsymbol{u}$ and $\boldsymbol{m}=\mathcal{M}(\omega,\boldsymbol{v}, \boldsymbol{\alpha})$, where the bi-convexity results from the bilinearity of the wave equation in wavefield and subsurface parameters \cite{Aghamiry_2019_IWR}.
The optimization gives $\hat{\boldsymbol{m}}$ (an estimate of ${\boldsymbol{m}}$) and subsequently the velocity and attenuation factor parameters can be obtained by the inverse mapping of arbitrary $\mathcal{M}$. This inverse mapping is given by \eqref{IKFv}-\eqref{IKFalpha} and \eqref{ISLSalpha}-\eqref{ISLSv} for the KF model and the SLS model, respectively. 
In this new approach, we therefore do not consider any attenuation mechanism for the inversion.
After inversion, one can test any attenuation relation \cite{Aki_2002_QST,Toverud_2005_CSA} to check which one provides the most geologically plausible parameters and this is a significant advantage of the proposed method when dealing with real data.

The main issues with \eqref{main} are two fold: 
First, during a multi-frequency inversion, we assume that $\boldsymbol{m}$ is constant with respect to frequencies in the band considered for inversion, and second the regularization must be applied on the complex-valued parameter. The first issue is tackled by proceeding hierarchically over small batches of frequencies, from low frequencies to higher ones following a classical multiscale approach. This means that we introduce implicitly the dependency of frequency in $\boldsymbol{m}$, when we move from one frequency batch to the next using the final model of the previous batch as the initial model for the next one. Put simply, this means that we solve the forward problem for each frequency batch with an average dispersion term. This approximation is discussed in the introduction section (Fig.~\ref{fig:SLS_KF}) and further in Section \ref{numerical}. The second issue related to regularization, which is addressed in the next section.

\section{Complex-Valued Optimization} \label{CVOPT}
\subsection{Preliminaries} \label{prem}
Through this paper, we frequently need to solve constrained optimization problem  of the form
\begin{align} \label{COOP}
\min_{\boldsymbol{x}}~~\mathcal{R}(\boldsymbol{x}) ~~ \text{subject to}~~~~ \boldsymbol{y}=\boldsymbol{Gx},
\end{align}
or the equivalent unconstrained problem
\begin{align} \label{UCOP}
\min_{\boldsymbol{x}}~~\mathcal{R}(\boldsymbol{x})+\frac{\lambda}{2} \|\boldsymbol{y}-\boldsymbol{Gx}\|_2^2,
\end{align}
where $\mathcal{R}(\boldsymbol{x})$ is a (convex) regularizer or regularization function,  $\boldsymbol{x}\in \mathbb{C}^n$,  
$\boldsymbol{y}\in \mathbb{C}^m$, $\boldsymbol{G}\in \mathbb{C}^{m\times n}$, and $\lambda>0$. 
Thus some basic concepts for solving them are reviewed here. 
Equation \eqref{UCOP} is a quadratic penalty formulation of \eqref{COOP} and both of these equations are equivalent in the sense that the solution of \eqref{UCOP} converges to that of \eqref{COOP} as $\lambda\to \infty$. Otherwise, for a finite $\lambda$, \eqref{UCOP} generates a solution that is only an approximate solution of \eqref{COOP} \cite{Nocedal_2006_NO}.
However, \eqref{UCOP} is simpler to solve.

The method of multipliers \cite{Bertsekas_1982_COA} solve \eqref{COOP} by solving problems of the form \eqref{UCOP} several times.
It is based on the corresponding augmented Lagrangian (AL) function 
\begin{align} 
\mathcal{L}(\boldsymbol{x},\boldsymbol{v})=\mathcal{R}(\boldsymbol{x})+ \langle \boldsymbol{\rho},\boldsymbol{y}-\boldsymbol{Gx}\rangle+ 
\frac{\lambda}{2}\|\boldsymbol{y}-\boldsymbol{Gx}\|_2^2,
\end{align}
where $\boldsymbol{\rho}$ is the vector of Lagrange multipliers or dual variable. The optimum solution now corresponds a saddle point of the AL function, given by 
\begin{equation} \label{AL}
\min_{\boldsymbol{x}} \max_{\boldsymbol{\rho}}~ \mathcal{L}(\boldsymbol{x},\boldsymbol{\rho}),
\end{equation}
which is solved in an alternating mode: fix $\boldsymbol{\rho}$ and solve \eqref{AL} for $\boldsymbol{x}$ and then fix $\boldsymbol{x}$ and  solve \eqref{AL} for $\boldsymbol{\rho}$. This strategy with a dual variable ascent (with step size $\lambda/2$) leads to
\begin{subequations} \label{MM}
\begin{align}
\boldsymbol{x}^{k+1} &= \arg\min_{\boldsymbol{x}}~ \mathcal{L}(\boldsymbol{x},\boldsymbol{\rho}^k),\\
\boldsymbol{\rho}^{k+1} &= \boldsymbol{\rho}^k + \lambda/2(\boldsymbol{y} - \boldsymbol{Gx}^{k+1}),
\end{align}  
\end{subequations}
beginning with $\boldsymbol{\rho}^0=\boldsymbol{0}$. 
We can simplify \eqref{MM} by a change of variable $\boldsymbol{y}^k = 2\boldsymbol{\rho}^k/\lambda$ to arrive at a more familiar form 
\begin{subequations} \label{MM_s}
\begin{align}
\boldsymbol{x}^{k+1} &= \arg\min_{\boldsymbol{x}}~ \mathcal{R}(\boldsymbol{x})+\frac{\lambda}{2} \|\boldsymbol{y}^k+\boldsymbol{y}-\boldsymbol{Gx}\|_2^2, \label{MM_s_a} \\
\boldsymbol{y}^{k+1} &= \boldsymbol{y}^k + \boldsymbol{y} - \boldsymbol{Gx}^{k+1}, \label{MM_s_b}
\end{align}  
\end{subequations}
where the scaled dual variable $\boldsymbol{y}$ in (\ref{MM_s_b})  records the running sum of the constraint violations in iterations.
The same algorithm can also be derived by using the concepts of Bregman iteration \cite{Goldstein_2009_SBM}. 
As can be seen, the constraint optimization in \eqref{COOP} is tackled by iteratively solving the approximate (and simpler) problem \eqref{UCOP} whose solution is refined at each iteration from the residual error by adding back the running sum of the constraint violation to the right-hand side.  
This is similar to the classical iterative refinement (IR) method for solving linear systems from an approximate inverse operator \cite{Wilkinson_1963_REA,Aghamiry_2018_IFW,Gholami_2019_3DD,Gholami_2017_CNA}.
Therefore, through the paper, we call the IR solve of a problem when the corresponding right-hand side is iteratively updated as in \eqref{MM_s_b}. 
\subsection{Complex-Valued TV Regularization} \label{C-Reg}
In this section, we consider solving the optimization problem \eqref{UCOP} when $\mathcal{R}(\boldsymbol{x})$ is the  TV function \cite{Rudin_1992_NTV}.
It should be noted that such a TV regularization has been extensively studied for real-valued model parameters \cite{Goldstein_2009_SBM,Afonso_2010_AAL,Wahlberg_2012_AAA,Chan_2013_CTV}.
When the model parameters are complex valued, the sensitivity of the inversion to the real coordinates of the complex quantity, i.e. real and imaginary parts, magnitude and phase, can be quite different. Thus, we need to have the necessary flexibility to tailor the TV regularization to each of the real coordinates. 
In the next subsections, we address this issue in detail.
\subsubsection{Treating $\boldsymbol{x}$ as a Real-Valued Variable} \label{RVR}
The traditional approach is to treat the model parameters as a real-valued variable even though they are complex valued.
In this case, using the (isotropic) TV regularization, \eqref{UCOP} reads as
\begin{align} \label{TV_main}
\min_{\boldsymbol{x}}~~\|\boldsymbol{x}\|_{\text{TV}}+\frac{\lambda}{2} \|\boldsymbol{y}-\boldsymbol{Gx}\|_2^2,
\end{align}
where
\begin{equation} \label{TV}
\|\boldsymbol{x}\|_{\text{TV}}=\sum  \sqrt{|{\boldsymbol{\nabla}}_{\!\! x} \boldsymbol{x}|^2 + |{\boldsymbol{\nabla}}_{\!\! z}\boldsymbol{x}|^2},
\end{equation}
in which $\boldsymbol{\nabla}_{\!\! x,z}$ are finite-difference operators in directions $x$ and $z$. The subscripts $x,z$ denotes either $x$ or $z$.
In this paper, we consider 2D models otherwise its is stated. For 2D models $\boldsymbol{x}$ is an $n_z\times n_x$ matrix which is resorted into an $n=n_z\times n_x$ column vector. 
 Furthermore, with notation abuse, the absolute sign, square power, and the square root operations are done component-wise, and the sum runs over all elements.
The TV model as defined in \eqref{TV} was essentially developed for regularization of real-valued model parameters \cite{Rudin_1992_NTV}, whereby the sum of the gradient field of the model is minimized to enforce its sparsity, leading to regularized solutions with piecewise-constant features. 
%
The unconstrained problem \eqref{TV_main} can be solved by the following easy tricks as presented in \cite{Goldstein_2009_SBM,Boyd_2011_DOS,Gholami_2019_3DD}. 
(I) Using \textit{variable splitting} scheme to split the TV term from the misfit term by introducing auxiliary variables  $\boldsymbol{p}_x={\boldsymbol{\nabla}}_{\!\! x} \boldsymbol{x}$ (horizontal gradient) and $\boldsymbol{p}_z={\boldsymbol{\nabla}}_{\!\! z} \boldsymbol{x}$ (vertical gradient), appearing as constraints. 
(II) Using a \textit{penalty formulation} to replace the resulting constrained problem by an equivalent (approximate) unconstrained problem. This is achieved simply by adding the constraints to the objective via a quadratic penalty method (using positive penalty parameters $\gamma_x, \gamma_z$).
(III) Using the \textit{IR} scheme to iteratively refine the solution of the approximate problem (with the help of refinement or dual variables $\boldsymbol{q}_x$ and $\boldsymbol{q}_z$). 
Recipes (II) and (III) are the same as those used in AL methods to solve constrained problem as highlighted by \eqref{MM_s}.
\begin{subequations}
 \label{TV_main_AL}
\begin{align}
(\boldsymbol{x}^{k+1},\boldsymbol{p}_x^{k+1},\boldsymbol{p}_z^{k+1}) &=\underset{\boldsymbol{x},\boldsymbol{p}_x,\boldsymbol{p}_z}{\arg\min}~\mathcal{L}(\boldsymbol{x},\boldsymbol{p}_x,\boldsymbol{p}_z,\boldsymbol{q}^k_x,\boldsymbol{q}^k_z),\\
\boldsymbol{q}^{k+1}_x &= \boldsymbol{q}^k_x + \boldsymbol{p}_x^{k+1} - {\boldsymbol{\nabla}}_{\!\! x}\boldsymbol{x}^{k+1},\\
\boldsymbol{q}^{k+1}_z &= \boldsymbol{q}^k_z + \boldsymbol{p}_z^{k+1} - {\boldsymbol{\nabla}}_{\!\! z}\boldsymbol{x}^{k+1},
\end{align}
\end{subequations}
where
\begin{align} \label{ALL}
\mathcal{L}(\boldsymbol{x},\boldsymbol{p}_x,\boldsymbol{p}_z,\boldsymbol{q}^k_x,\boldsymbol{q}^k_z)&=
\sum  \sqrt{|\boldsymbol{p}_z|^2 + |\boldsymbol{p}_x|^2} 
+\frac{\lambda}{2} \|\boldsymbol{Gx}-\boldsymbol{y}\|_2^2 \nonumber \\
&+ \frac{\gamma_x}{2} \|\boldsymbol{p}_x+\boldsymbol{q}^k_x-{\boldsymbol{\nabla}}_{\!\! x}\boldsymbol{x}\|_2^2
+ \frac{\gamma_z}{2} \|\boldsymbol{p}_z+\boldsymbol{q}^k_z-{\boldsymbol{\nabla}}_{\!\! z}\boldsymbol{x}\|_2^2. \nonumber
\end{align}
The resulting Lagrangian function is now multivariate and should be minimized over three primal variable blocks $\boldsymbol{x}, \boldsymbol{p}_x, \boldsymbol{p}_z$. Since simultaneous minimization over these blocks is difficult,   
(IV) a \textit{block relaxation} strategy \cite{Bezdek_1987_LCA,De_1994_BAI} is used to simplify the minimization over each block separately through three steps, leading to the alternating direction method of multipliers \cite{Gabay_1975_ADA,Glowinski_1975_SAP}. 
In every step, only one block is optimized, while the other blocks are fixed (similar to the Gauss-Seidel method for solving linear equations).
Following these tricks, we get that the variable block $\boldsymbol{x}^{k+1}$ is the solution of the following (inconsistent) overdetermined system:
\begin{equation} \label{model}
\begin{pmatrix}
\sqrt{\lambda} \boldsymbol{G}\\
\sqrt{\gamma_x} {\boldsymbol{\nabla}}_{\!\! x} \\
\sqrt{\gamma_z} {\boldsymbol{\nabla}}_{\!\! z}
\end{pmatrix}
\boldsymbol{x}^{k+1}
\approx
\begin{pmatrix}
\sqrt{\lambda} \boldsymbol{y}\\
\sqrt{\gamma_x}(\boldsymbol{p}_x^k+\boldsymbol{q}_x^k) \\
\sqrt{\gamma_z}(\boldsymbol{p}_z^k+\boldsymbol{q}_z^k)
\end{pmatrix},
\end{equation}
which is solved in the least-squares  sense.
The gradient components $\boldsymbol{p}_{x,z}$ are initialized to $\boldsymbol{0}$ and are updated via the following  proximal operator 
\begin{equation} \label{prox}
   \boldsymbol{p}_{x,z}^{k+1} = \frac{\boldsymbol{z}_{x,z}^k}{|\boldsymbol{z}^k|} \max(|\boldsymbol{z}^k| - \frac{1}{\gamma_{x,z}},0),
\end{equation}
where $\boldsymbol{z}_{x,z}^k=\nabla_{\!\!x,z}\boldsymbol{x}^{k+1}-\boldsymbol{q}_{x,z}^k$, and $|\boldsymbol{z}^k|=\sqrt{|\boldsymbol{z}^k_x|^2 + |\boldsymbol{z}^k_z|^2}$. 

It is seen that the effect of TV regularization only appeared in the update of $\boldsymbol{p}_{x,z}^{k}$, equation \eqref{prox}. 
From \eqref{prox}  and noting that $\boldsymbol{z}_{x,z}=\boldsymbol{\nabla}_{\!\!x,z}\boldsymbol{x}^{k+1}-\boldsymbol{q}_{x,z}^k$ are complex-valued, we get that
\begin{align} \label{c_prox}
   \boldsymbol{p}_{x,z}^{k+1} &= \varphi_{\frac{1}{\gamma_{x,z}}}(|\boldsymbol{z}^k|) \circ \boldsymbol{z}_{x,z}^k \\
   &= \varphi_{\frac{1}{\gamma_{x,z}}}(|\boldsymbol{z}^k|) \circ |\boldsymbol{z}_{x,z}^k|\circ e^{i\angle{\boldsymbol{z}_{x,z}^k}} \nonumber \\ 
   &= \varphi_{\frac{1}{\gamma_{x,z}}}(|\boldsymbol{z}^k|)\circ\Re(\boldsymbol{z}_{x,z}^k) + i \varphi_{\frac{1}{\gamma_{x,z}}}(|\boldsymbol{z}^k|) \circ\Im(\boldsymbol{z}_{x,z}^k), \nonumber
\end{align} 
where $\Re$ and $\Im$ denote real and imaginary parts, respectively, and the weighting function $\varphi_{\gamma}(x)$ is defined as
\begin{equation} \label{wf}
\varphi_{\gamma}(x)= \max(1 - \frac{\gamma}{x},0),~~~ x>0.
\end{equation}
The steps of performing this procedure to solve \eqref{TV_main} are summarized in Algorithm \ref{Alg1} and the details can be found in \cite{Goldstein_2009_SBM}.
However, it should be noted that in \eqref{c_prox} 
\begin{align} \label{abs4}
|\boldsymbol{z}^k|  &=\sqrt{|\boldsymbol{z}_x^k|^2 + |\boldsymbol{z}_z^k|^2}
=\sqrt{\Re(\boldsymbol{z}_x^k)^2 + \Re(\boldsymbol{z}_z^k)^2+\Im(\boldsymbol{z}_x^k)^2 + \Im(\boldsymbol{z}_z^k)^2}. 
\end{align}  
Thus, applying the TV regularization to complex-valued parameters only affects the magnitude of the model, which turns out to be similar to the approach developed by \cite{Guven_2016_AAL}. It gives similar weights to the real and imaginary parts of both horizontal and vertical model gradients. The weighting coefficient at each pixel is build by the Euclidean norm of a four component vector whose components are the real and imaginary parts of both horizontal and vertical gradients at that pixel, as in \eqref{abs4}.
This enforces the real and imaginary parts of both horizontal and vertical gradients to follow the same (sparsity) patterns and hence generates a solution with similar structures in its real and imaginary parts. 
Consequently, this method is effective for complex-valued recovery if phase variations are small.
In the next subsection, we perform separate TV regularization on the real and imaginary parts.
%
\begin{algorithm}[!h]
\vspace{.2cm}
 \caption{Complex-valued image reconstruction with isotropic TV regularization in \eqref{TV_main}.}  \label{Alg1}
 initialize: set $\boldsymbol{p}_{x,z}^0= \boldsymbol{q}_{x,z}^0=\boldsymbol{0}$   \\
 \For{$k\leftarrow 0,1,2,\cdots$ }
 {
 Solve \eqref{model} for  $\boldsymbol{x}^{k+1}$ \\
 $\boldsymbol{z}_{x,z}^k=\boldsymbol{\nabla}_{\!\!x,z}\boldsymbol{x}^{k+1}-\boldsymbol{q}_{x,z}^k$ \\
 $|\boldsymbol{z}^k|  = \sqrt{\Re(\boldsymbol{z}^k_x)^2 + \Re(\boldsymbol{z}^k_z)^2+\Im(\boldsymbol{z}^k_x)^2 + \Im(\boldsymbol{z}^k_z)^2}$\\
 $\boldsymbol{p}_{x,z}^{k+1} = \varphi_{\frac{1}{\gamma_{x,z}}}(|\boldsymbol{z}^k|) \circ \boldsymbol{z}_{x,z}^k$ \\
$ \boldsymbol{q}_{x,z}^{k+1} = \boldsymbol{q}_{x,z}^{k} + \boldsymbol{p}_{x,z}^{k+1} - \boldsymbol{\nabla}_{\!\!x,z}\boldsymbol{x}^{k+1}$
 }
\end{algorithm}

\subsubsection{Separate regularization of real and imaginary parts}  \label{SRIR}
In this section, we show how Algorithm \ref{Alg1} can be modified to apply TV regularization separately on the real and imaginary parts of the complex-valued parameters. The reader is referred to \cite{Olafsson_2006_SRA,Zibetti_2017_ICS} for similar regularization strategies developed in the framework of MRI applications.
Mathematically, we seek to solve the following minimization problem:
\begin{align} \label{RITV_main}
\min_{\boldsymbol{x}}~~\tau\|\Re(\boldsymbol{x})\|_{\text{TV}} + (1-\tau)\|\Im(\boldsymbol{x})\|_{\text{TV}}+\frac{\lambda}{2} \|\boldsymbol{y}-\boldsymbol{Gx}\|_2^2,
\end{align}
in which  $\tau$, $0\leq \tau\leq1$, controls the relative amount of weight assigned to the two regularization terms, while $\lambda$ controls the relative weight between TV regularization and the misfit function.
Therefore, the hyperparameter $\tau$ provides the necessary flexibility to tailor the TV regularization to the real and imaginary part of the complex parameter.
We can follow the same procedure as that reviewed in the previous section and outlined in Algorithm 1 to solve problem \eqref{RITV_main}.
The only difference is in the update of $\boldsymbol{p}^k_{x,z}$, which takes the following form:
\begin{align}
   \boldsymbol{p}_{x,z}^{k+1} 
   &= \varphi_{\frac{\tau}{\gamma_{x,z}}}(|\Re(\boldsymbol{z}^k)|) \circ\Re(\boldsymbol{z}_{x,z}^k) + i \varphi_{\frac{(1-\tau)}{\gamma_{x,z}}}(|\Im(\boldsymbol{z}^k)|) \circ\Im(\boldsymbol{z}_{x,z}^k), 
\end{align} 
with 
\begin{equation}
|\Re(\boldsymbol{z}^k)|  =\sqrt{\Re(\boldsymbol{z}^k_x)^2 + \Re(\boldsymbol{z}^k_z)^2}, \quad 
|\Im(\boldsymbol{z}^k)|  =\sqrt{\Im(\boldsymbol{z}^k_x)^2 + \Im(\boldsymbol{z}^k_z)^2}.\nonumber
\end{equation}  
As can be seen, in this case, the real and imaginary parts are weighted independently (Algorithm~\ref{Alg2}). Accordingly, each part is enforced to be piecewise-constant independently of the other.
In the next section, we perform separate TV regularization on the magnitude and phase of the model.
%
\begin{algorithm}[!h]
\vspace{.2cm}
 \caption{Complex-valued image reconstruction with separate TV regularization of real and imaginary parts defined in \eqref{RITV_main}.}  \label{Alg2}
 initialize: set $\boldsymbol{p}_{x,z}^0= \boldsymbol{q}_{x,z}^0=\boldsymbol{0}$   \\
 \For{$k\leftarrow 0,1,2,\cdots$ }
 {
 Solve \eqref{model} for  $\boldsymbol{x}^{k+1}$ \\
 $\boldsymbol{z}_{x,z}^k=\boldsymbol{\nabla}_{\!\!x,z}\boldsymbol{x}^{k+1}-\boldsymbol{q}_{x,z}^k$ \\
$|\Re(\boldsymbol{z}^k)|  = \sqrt{\Re(\boldsymbol{z}^k_x)^2 + \Re(\boldsymbol{z}_z^k)^2}$\\
$|\Im(\boldsymbol{z}^k)|  = \sqrt{\Im(\boldsymbol{z}_x^k)^2 + \Im(\boldsymbol{z}_z^k)^2}$\\
$\boldsymbol{p}_{x,z}^{k+1} = \varphi_{\frac{\tau}{\gamma_{x,z}}}(|\Re(\boldsymbol{z}^k)|) \circ \Re(\boldsymbol{z}_{x,z}^k) + i \varphi_{\frac{(1-\tau)}{\gamma_{x,z}}}(|\Im(\boldsymbol{z}^k)|) \circ \Im(\boldsymbol{z}_{x,z}^k)$ \\
$ \boldsymbol{q}_{x,z}^{k+1} = \boldsymbol{q}_{x,z}^{k} + \boldsymbol{p}_{x,z}^{k+1} - \boldsymbol{\nabla}_{\!\!x,z}\boldsymbol{x}^{k+1}$
 }
\end{algorithm}

\subsubsection{Separate Magnitude and Phase Regularization}  \label{SMPR}
Using the polar form of the complex-valued model parameters, $\boldsymbol{x} = \boldsymbol{a}e^{i\boldsymbol{\theta}}$,  allows us to apply regularization separately on the magnitude and phase 
\begin{align} \label{MTV2_main}
\min_{\boldsymbol{a},\boldsymbol{\theta}}~~\tau\|\boldsymbol{a}\|_{\text{TV}} +(1-\tau)\phi(\boldsymbol{\theta})+
\frac{\lambda}{2} \|\boldsymbol{y}-\boldsymbol{G}\text{diag}(e^{i\boldsymbol{\theta}}) \boldsymbol{a}\|_2^2,
\end{align}
in which $\text{diag}(e^{i\boldsymbol{\theta}})$ denotes a diagonal matrix with vector $e^{i\boldsymbol{\theta}}$ on its main diagonal and $\tau$, $0\leq \tau\leq1$, controls how much weight is assigned to the magnitude relative to the phase.
Here the magnitude is regularized by the TV functional, while the phase is regularized by $\phi$, a general closed proper convex function not necessarily differentiable. This makes the proposed objective function general enough to introduce other regularizers that are designed for specific applications, e.g., one which enforces smoothness of the phase field \cite{Zibetti_2017_ICS}.

A main difficulty with the bi-variate objective function \eqref{MTV2_main} is that the data $\boldsymbol{y}$ are nonlinear in the phase function $\boldsymbol{\theta}$, although they are linear in $\boldsymbol{a}$.
A simple approach to address this issue relies on a block relaxation strategy, which updates each of the two blocks $\boldsymbol{a}$ and $\boldsymbol{\theta}$ in alternating mode, leading to
\begin{subequations}
\begin{align} 
\boldsymbol{a}^{k+1}&=\arg\min_{\boldsymbol{a}}~\tau\|\boldsymbol{a}\|_{\text{TV}}+\frac{\lambda}{2} \|\boldsymbol{y}-\boldsymbol{G}\text{diag}(e^{i\boldsymbol{\theta}^k})\boldsymbol{a}\|_2^2, \label{xsub}\\
\boldsymbol{\theta}^{k+1} &=\arg\min_{\boldsymbol{\theta}}~ (1-\tau)\phi(\boldsymbol{\theta})
+\frac{\lambda}{2} \|\boldsymbol{y}-\boldsymbol{G}\text{diag}(e^{i\boldsymbol{\theta}})\boldsymbol{a}^{k+1}\|_2^2. \label{phisub}
\end{align}
\end{subequations}

In this case, the magnitude subproblem \eqref{xsub} is linear and can be solved efficiently by using traditional TV regularization algorithms like the one given in Algorithm \ref{Alg1} with $\boldsymbol{G}\text{diag}(e^{i\boldsymbol{\theta}^k})$ as the forward operator. However, solving the phase subproblem \eqref{phisub} is more challenging. 
 \cite{Zhao_2012_SMA} solved this subproblem by non-linear conjugate gradient method with back-tracking line search and
 \cite{Zibetti_2017_ICS} regularized the exponential of the phase, $e^{i\boldsymbol{\theta}}$, instead of the phase itself.

The objective function in \eqref{phisub} is in general neither convex nor differentiable.
Although even resolving the question of whether a descent direction exists from a point is NP-hard for a general nonsmooth nonconvex problem \cite{Nesterov_2013_GMF}, the generalized proximal point algorithm \cite{Fukushima_1981_AGP}  or composite gradient mapping \cite{Nesterov_2013_GMF} can be used to find a local minimizer of the structured objective function \eqref{phisub}, which is a sum of two terms (convex+smooth), even if no descent direction exists.
The method is based on defining a locally simple quadratic approximation of the misfit function, $f(\boldsymbol{\theta})=\|\boldsymbol{y}-\boldsymbol{G}\text{diag}(e^{i\boldsymbol{\theta}})\boldsymbol{a}^{k+1}\|_2^2$,  near a reference phase $\boldsymbol{\theta}^k$ as (assuming $\nabla^2 f(\boldsymbol{\theta}^k)\approx c^k\boldsymbol{I}$)
\begin{equation}
\tilde{f}(\boldsymbol{\theta},\boldsymbol{\theta}^k) = f(\boldsymbol{\theta}^k)+ \langle \nabla f(\boldsymbol{\theta}^k), \boldsymbol{\theta} - \boldsymbol{\theta}^k\rangle +
\frac{c^k}{2} \|\boldsymbol{\theta} - \boldsymbol{\theta}^k\|_2^2,
\end{equation}
where $c^k$ are positive parameters and $\nabla f(\boldsymbol{\theta}^k)$ is the gradient of $f$ at $\boldsymbol{\theta}^k$, defined as
\begin{equation}
\nabla f(\boldsymbol{\theta}^k) = \Im(\text{diag}(\boldsymbol{a}^{k+1}\circ e^{-i\boldsymbol{\theta}^k})\boldsymbol{G}^H(\boldsymbol{G}\text{diag}(e^{i\boldsymbol{\theta}^k})\boldsymbol{a}^{k+1}-\boldsymbol{y})          ).
\end{equation}
Accordingly, the generalized proximal point algorithm solves \eqref{phisub}  by a line search method  as 
\begin{subequations}
\label{GPP}
\begin{align}
\Delta \boldsymbol{\theta}^k &=\boldsymbol{\theta}^k- \text{prox}_{\frac{(1-\tau)}{c^k\lambda}\phi}(\boldsymbol{\theta}^k - \frac{1}{c^k} \nabla f(\boldsymbol{\theta}^k)), \label{compgrad}\\
\boldsymbol{\theta}^{k+1} &= \boldsymbol{\theta}^k - \beta^k \Delta \boldsymbol{\theta}^k, \label{linsearch}
\end{align}
\end{subequations}
beginning from an initial point $\boldsymbol{\theta}^0$, in which $\Delta \boldsymbol{\theta}^k$ is a search direction obtained by the composite gradient step \eqref{compgrad} with 
\begin{align}
\text{prox}_{\frac{(1-\tau)}{c^k\lambda}\phi}(\boldsymbol{\theta}^k - \frac{1}{c^k} \nabla f(\boldsymbol{\theta}^k))
=\arg\min_{\boldsymbol{\theta}}~ (1-\tau)\phi(\boldsymbol{\theta})
+\frac{\lambda}{2}\tilde{f}(\boldsymbol{\theta},\boldsymbol{\theta}^k).
\end{align}
It is easily seen that this composite gradient is the sum of an explicit gradient (of $f$) and an implicit subgradient (of $\phi$) which is embedded in the prox function. The step size $\beta^k$ in \eqref{linsearch} is determined according to an Armijo type rule \cite{Armijo_1966_MOF} such that
\begin{equation} \label{Arrnijo}
\phi(\boldsymbol{\theta}^k - \beta^k \Delta \boldsymbol{\theta}^k) \geq 
\phi(\boldsymbol{\theta}^k) - \alpha \beta^k \|\Delta \boldsymbol{\theta}^k\|_2^2,
\end{equation}
for $\alpha >0$, where $\phi(\boldsymbol{\theta})= (1-\tau)g(\boldsymbol{\theta})+\frac{\lambda}{2} f(\boldsymbol{\theta})$. Convergence of this algorithm to a local minimum from which there is no descent direction is proved in \cite{Fukushima_1981_AGP} and \cite{Nesterov_2013_GMF}.

Interestingly, we do not need to reach the convergence point of \eqref{xsub} or \eqref{phisub} at each iteration of the full problem \eqref{MTV2_main}, i.e., only a single iteration of each subproblem leads to the convergence of \eqref{MTV2_main}.
Based on this, the proposed algorithm to solve \eqref{MTV2_main} is summarized in Algorithm \ref{Alg3}, in which the model parameters at each iteration is obtained by solving
\begin{equation} \label{model_abs}
\begin{pmatrix}
\sqrt{\lambda} \boldsymbol{G}\text{diag}(e^{i\boldsymbol{\theta}^k})\\
\sqrt{\gamma_x} {\boldsymbol{\nabla}}_{\!\! x} \\
\sqrt{\gamma_z} {\boldsymbol{\nabla}}_{\!\! z}
\end{pmatrix}
\boldsymbol{a}^{k+1}
\approx
\begin{pmatrix}
\sqrt{\lambda} \boldsymbol{y}\\
\sqrt{\gamma_x}(\boldsymbol{p}_x^k+\boldsymbol{q}_x^k) \\
\sqrt{\gamma_z}(\boldsymbol{p}_z^k+\boldsymbol{q}_z^k)
\end{pmatrix}.
\end{equation}

\begin{algorithm}[!h]
\vspace{.2cm}
 \caption{Complex-valued image reconstruction with separate magnitude and phase regularization in \eqref{MTV2_main}.}  \label{Alg3}
 initialize: set $\boldsymbol{\theta}^0=\boldsymbol{0}$  and  $\boldsymbol{p}_{x,z}^0= \boldsymbol{q}_{x,z}^0=\boldsymbol{0}$   \\
 \For{$k\leftarrow 0,1,2,\cdots$ }
 {
 Solve \eqref{model_abs} for  $\boldsymbol{a}^{k+1}$ \\
 $\boldsymbol{z}_{x,z}^k=\boldsymbol{\nabla}_{\!\!x,z}\boldsymbol{a}^{k+1}-\boldsymbol{q}_{x,z}^k$ \\
 $|\boldsymbol{z}^k|  = \sqrt{|\boldsymbol{z}^k_x|^2 + |\boldsymbol{z}^k_z|^2}$\\
 $\boldsymbol{p}_{x,z}^{k+1} = \theta_{\frac{\tau}{\gamma_{x,z}}}(|\boldsymbol{z}^k|) \circ \boldsymbol{z}_{x,z}^k$ \\
$ \boldsymbol{q}_{x,z}^{k+1} = \boldsymbol{q}_{x,z}^{k} + \boldsymbol{p}_{x,z}^{k+1} - \boldsymbol{\nabla}_{\!\!x,z}\boldsymbol{a}^{k+1}$\\
$\Delta \boldsymbol{\theta}^k =\boldsymbol{\theta}^k- \text{prox}_{\frac{(1-\tau)}{c^k\lambda}\phi}(\boldsymbol{\theta}^k - \frac{1}{c^k} \nabla f(\boldsymbol{\theta}^k))$\\
determine $\beta^k$ according to \eqref{Arrnijo} \\
$\boldsymbol{\theta}^{k+1} = \boldsymbol{\theta}^k - \beta^k \Delta \boldsymbol{\theta}^k$
 }
\end{algorithm}

%
%
\section{Complex-valued Full-waveform Inversion: Solving \eqref{main}} \label{FWI}
The PDE constraint in \eqref{main}, $\boldsymbol{A}\boldsymbol{u}=\boldsymbol{b}$, is nonlinear in $\boldsymbol{m}$, where $\boldsymbol{A}$ is a nonsingular, large and sparse matrix.
The data constraint, $\boldsymbol{Pu}=\boldsymbol{d}$, is linear but the operator $\boldsymbol{P}$ is rank-deficient with a huge null space because the data are recorded only at or near the surface, hence $m\ll n$. Therefore, determination of the optimum solution pair ($\boldsymbol{u}^*,\boldsymbol{m}^*$) satisfying the two constraints simultaneously is an ill-posed problem, which requires sophisticated regularization techniques.

The constrained optimization problem \eqref{main} is traditionally written as an unconstrained, nonlinear regularized least squares problem \cite{Pratt_1998_GNF,Plessix_2006_RAS} after projection of the variable $\boldsymbol{u}$ in the data misfit function.
\begin{align} \label{main_red}
\min_{\boldsymbol{m}}~~\mathcal{R}(\boldsymbol{m}) + \frac{\lambda}{2} \|\boldsymbol{d}-\boldsymbol{P}\boldsymbol{A}^{-1}\boldsymbol{b}\|_2^2.
\end{align} 
This variable projection shrinks the full search space to the parameter space and makes the resulting optimization problem highly nonlinear \cite{Symes_2008_MVA}. This nonlinear problem is usually solved with Gauss-Newton or quasi-Newton methods with the risk to remain stuck in spurious minimum if the starting $\boldsymbol{m}$ is not accurate enough. In order to extend the search space of FWI and mitigate its nonlinearity accordingly,\cite{VanLeeuwen_2013_MLM,vanLeeuwen_2016_PMP} implement the wave equation constraint as a soft constraint with a quadratic penalty method to foster data fitting.

For a convex regularization functional $\mathcal{R}$, the problem described by \eqref{main} is a bi-convex constrained optimization in the complex-valued variables $(\boldsymbol{u},\boldsymbol{m})$.
In this framework, ADMM provides a more efficient optimization algorithm than the penalty method of  \cite{VanLeeuwen_2013_MLM,vanLeeuwen_2016_PMP} to update the two primal variables $(\boldsymbol{u},\boldsymbol{m})$ in alternating mode, thanks to the defect correction action of the dual variables reviewed in section \ref{prem}.
Through this alternating-direction optimization, minimization of the bi-convex objective function is broken down into two linear subproblems that are activated in cycles (see \cite{Aghamiry_2019_IWR,Aghamiry_2019_IBC} for more details):
\begin{subequations}
\label{main30}
\begin{align}
\boldsymbol{u}^{k+1} &=    \underset{\boldsymbol{u}}{\arg\min} ~
\frac{\gamma}{2}\|\boldsymbol{d}^k+\boldsymbol{d}-\boldsymbol{Pu}\|_2^2 + \frac{\lambda}{2}\|\boldsymbol{b}^k+\boldsymbol{b} - \boldsymbol{A}^{k}\boldsymbol{u}\|_2^2 ,\label{main3a}   \\
\boldsymbol{m}^{k+1} &=  \underset{\boldsymbol{m}}{\arg\min} ~~   \mathcal{R}(\boldsymbol{m})+ \frac{\lambda}{2}\|\boldsymbol{b}^k+\boldsymbol{b} - \boldsymbol{A}\boldsymbol{u}^{k+1}\|_2^2, \label{main3b}
\end{align} 
\end{subequations}
where $\bold{A}^k\equiv \bold{A}(\omega^k,\boldsymbol{v}^k,\boldsymbol{\alpha}^k)$, $\boldsymbol{b}^k$ and $\boldsymbol{d}^k$ are Lagrange multiplier (dual) vectors corresponding to each constraint in \eqref{main} and $\lambda, \gamma>0$ are the associated penalty parameters. 
These dual vectors are initialized to zero and are updated at each iteration following a dual ascent method as in \eqref{MM_s_b}  
\begin{subequations} 
\label{main_duals}
\begin{align}
\boldsymbol{b}^{k+1} &= \boldsymbol{b}^k  +\boldsymbol{b}- \boldsymbol{A}^{k+1}\boldsymbol{u}^{k+1}, \label{dual_b}\\ 
\boldsymbol{d}^{k+1} &= \boldsymbol{d}^k  +\boldsymbol{d}- \boldsymbol{P}\boldsymbol{u}^{k+1}.  \label{dual_d}
\end{align} 
\end{subequations}
This ADMM-based FWI workflow, as described in \eqref{main30}-\eqref{main_duals}, is known as IR-WRI and is fully analyzed for real velocity estimation with TV regularization in \cite{Aghamiry_2019_IBC}. 
Using \eqref{helmholtz}, \eqref{main3b} can be written as \cite{Aghamiry_2019_IWR,Aghamiry_2019_IBC} 
\begin{equation} \label{msub}
\boldsymbol{m}^{k+1} =  \underset{\boldsymbol{m}}{\arg\min} ~~   \mathcal{R}(\boldsymbol{m})+ \frac{\lambda}{2}\|\boldsymbol{y}^k-\boldsymbol{L}^{k}\boldsymbol{m}\|_2^2,
\end{equation}  
where 
\begin{equation}
\boldsymbol{L}^{k}=\frac{\partial\boldsymbol{A}}{\partial \boldsymbol{m}}\boldsymbol{u}^{k+1},
\end{equation}
is a diagonal matrix whose columns contain the so-called virtual sources \cite{Pratt_1998_GNF} and $\boldsymbol{y}^k=\boldsymbol{b}+\boldsymbol{b}^k-\Delta \boldsymbol{u}^{k+1}$ is the right-hand side of the corresponding linear system. 
The complex-valued optimization problem \eqref{msub} is now similar to \eqref{UCOP}. Thus Algorithms \ref{Alg1}-\ref{Alg3} can be used (based on the form of $\mathcal{R}$) to solve it as an inner loop.
It should be noted that we do not need to fully solve this subproblem during one IR-WRI iteration, (\ref{main30}-\ref{main_duals}), and in practice only a single iteration of this inner loop is sufficient for convergence of the full algorithm \cite{Goldstein_2009_SBM,Aghamiry_2019_IWR,Aghamiry_2019_IBC}.\\ 
\subsection{Multi-frequency Inversion} \label{MFI}
To mitigate nonlinearity, FWI is usually performed with a multi-scale frequency continuation strategy by proceeding from the low frequencies to the higher ones \cite{Sirgue_2004_EWI}. Moreover,  computationally-efficient frequency-domain algorithms can be designed by processing one frequency at a time between $\omega_{min}$ and $\omega_{max}$ with an interval $d\omega$, leading to $\frac{\omega_{max}-\omega_{min}}{d\omega}$ successive mono-frequency inversions. This parsimonious strategy contrasts with time-domain methods where the frequency band is augmented as the inversion progresses toward higher frequencies such that, at the final inversion step, the full frequency bandwidth is processed in one go \cite{Bunks_1995_MSW}. 
Moreover, the parsimonious strategy is suboptimal for multiparameter inversion in the sense that the redundancy with which frequencies and scattering angles sample the wavenumber spectrum is full removed during one mono-frequency inversion step, which means that one wavenumber component is sampled by one scattering angle. Considering that most of parameter classes are uncoupled by FWI according to their specific scattering pattern, the one-to-one mapping between scattering angle and wavenumber makes the multiparameter FWI fully underdetermined. Note however that this comment does not apply to attenuation, which is instead decoupled from the phase velocity according to the wave dispersion behavior when the FWI is performed in the real domain as discussed in the introduction.

To mitigate the drawback of mono-frequency inversions, it is common to divided the desired frequency range $[\omega_{min},\omega_{max}]$ into  a number of overlapping batches \cite{Brossier_2009_SIC,Operto_2018_MFF,Aghamiry_2019_IWR,Aghamiry_2019_IBC}. For example, if we have 9 frequencies $\omega_1-\omega_9$ in this range and each batch includes three frequencies with one frequency overlap between two consecutive batches then we have the following batches for the inversion:


\begin{equation} \label{batches}
\lefteqn{\underbrace{\phantom{\omega_1\omega_2\omega_3}}_{\text{batch}~ 1}}
\omega_1\omega_2\overbrace{\omega_3\omega_4\lefteqn{\underbrace{\phantom{\omega_5\omega_6\omega_7}}_{\text{batch}~ 3}}
\omega_5}^{\text{batch}~ 2}\omega_6\overbrace{\omega_7\omega_8\omega_9}^{\text{batch}~ 4},
\end{equation}
In this case, the inversion is performed for each batch independently moving from the low frequencies (batch 1) to the higher ones (batch 4) according to a classic frequency continuation strategy but using the final model of the previous batch as the initial model of the current batch.
The frequencies in each batch, however, are inverted simultaneously. When the velocity model is real valued, i.e., is independent of frequency, all the frequencies are inverted for a single model. However, when attenuation effects are taken into account, the complex-valued velocities are frequency dependent, which imposes extra difficulties for the joint inversion of multiple frequencies.
In order to solve this issue, we consider a  band-wise frequency dependence \cite{Keating_2019_PCA} which means that an average velocity model and attenuation factor (independent of frequency) is estimated for each batch. This approximation is reasonable as long as the frequency range of each batch is not large and use the model inferred from current batch as the initial model for the next batch.
Assuming $L$ frequencies $\{\omega_l\},~l=1,2,...,L$ in a batch, the coefficient matrix and the corresponding right-hand-side of \eqref{msub} at iteration $k$th is
\begin{equation} \label{Ly}
\boldsymbol{L}^{k}=
\begin{bmatrix}
 \omega_1^2 \boldsymbol{C} \text{diag}(\boldsymbol{B}\boldsymbol{u}^{k+1}_{\omega_1})\\
  \omega_2^2 \boldsymbol{C} \text{diag}(\boldsymbol{B}\boldsymbol{u}^{k+1}_{\omega_2})\\
  \vdots\\
   \omega_L^2 \boldsymbol{C} \text{diag}(\boldsymbol{B}\boldsymbol{u}^{k+1}_{\omega_L})
\end{bmatrix},
\quad \quad
\boldsymbol{y}^{k}=
\begin{bmatrix}
\boldsymbol{b}^k_{\omega_1}+ \boldsymbol{b}_{\omega_1}-\Delta\boldsymbol{u}^{k+1}_{\omega_1}\\
\boldsymbol{b}^k_{\omega_2}+  \boldsymbol{b}_{\omega_2}-\Delta\boldsymbol{u}^{k+1}_{\omega_2}\\
  \vdots\\
\boldsymbol{b}^k_{\omega_L}+   \boldsymbol{b}_{\omega_L}-\Delta\boldsymbol{u}^{k+1}_{\omega_L}
\end{bmatrix}, 
\end{equation}
and the wavefields $\boldsymbol{u}^{k+1}_{\omega_l}$ are obtained by solving \eqref{main3a} for each frequency $\omega_l$ with the same model $\boldsymbol{m}^{k}$. 

\section{Numerical examples} \label{numerical}
In this section, we consider a set of test problems. One of them is a general complex-valued problem in compress sensing \cite{Donoho_2006_CS} to show the effectiveness of the algorithms and rest of them are complex-valued seismic imaging problems. 
In order to perform seismic forward modeling in our tests, we use a nine-point stencil finite-difference method implemented with anti-lumped mass and PML absorbing boundary conditions, where the stencil coefficients are optimized to the frequency \cite{Chen_2013_OFD}. Also, to show the robustness of the proposed approach against attenuation mechanism, we use SLS model to generate the data, while we use both SLS and KF models to extract the physical parameters from the final estimated complex velocity. We note that choosing SLS or KF model is not related to their accuracy in simulating real seismic data. We just select them to show that the proposed inversion method works without any priors about the attenuation mechanism and the numerical tests do not rely on an inverse crime.  


\subsection{1D compress sensing test}
\begin{figure}
\includegraphics[width=1\textwidth]{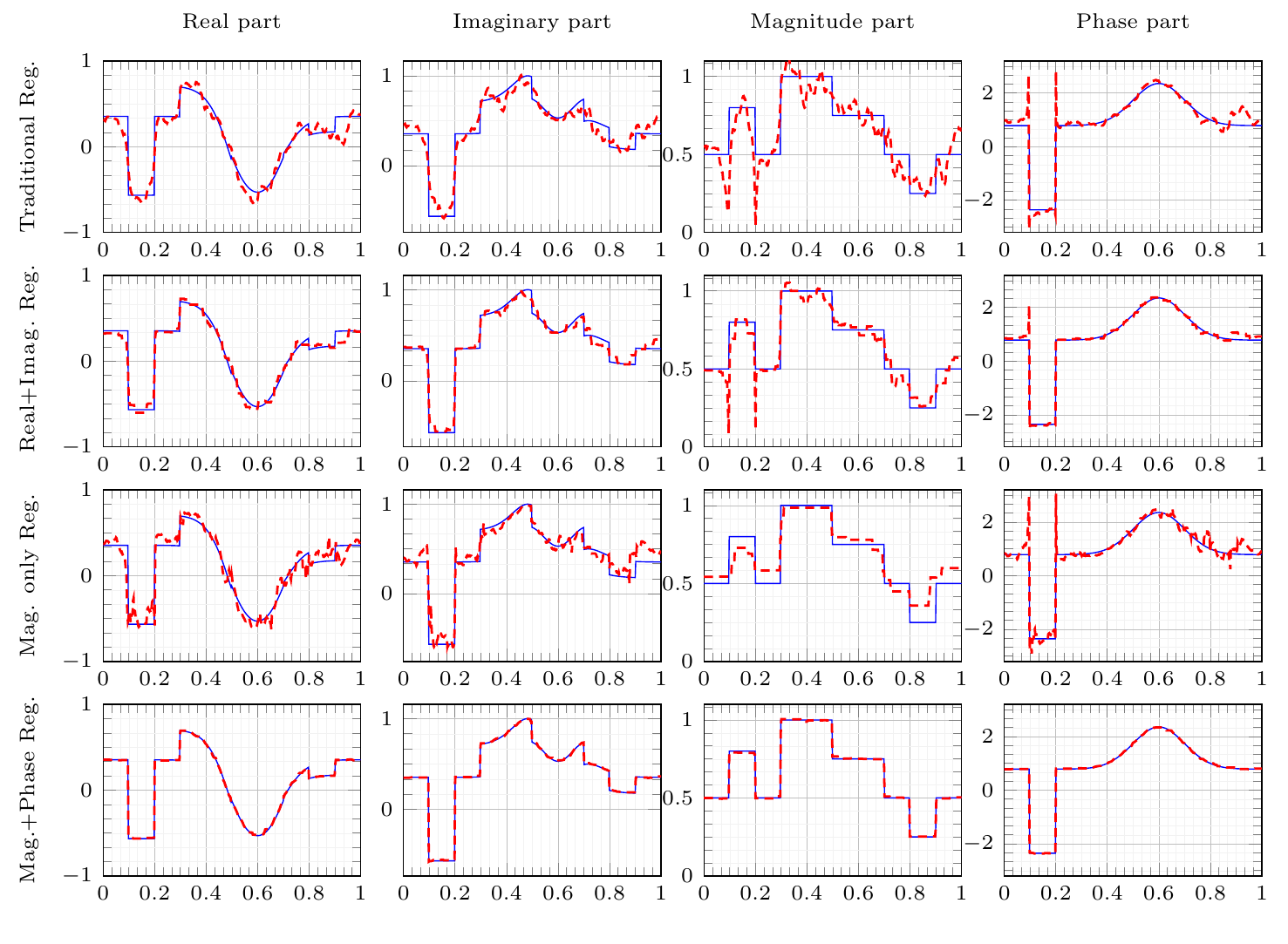}
\caption{Specified characteristics of a signal (in blue) and its reconstruction from random projections (in red) using traditional TV regularization (Algorithm \ref{Alg1}), separate TV regularization applied on real and imaginary parts (Algorithm \ref{Alg2}, $\tau=0.5$), TV regularization applied only on magnitude part (Algorithm \ref{Alg3}, $\tau=1$),  and separate TV regularization applied on magnitude and phase parts (Algorithm \ref{Alg3}, $\tau=0.5$).}
\label{fig:simple_test}
\end{figure}

In this subsection, using a simple 1D signal, we show the performance of the proposed algorithms for complex-valued regularization in Algorithms \ref{Alg1}-\ref{Alg3}.
The simulated discrete signal of length 500 samples with the specified characteristics is shown in Fig. \ref{fig:simple_test}. 
The forward operator $\boldsymbol{G}$ of size $50\times 500$ is a complex Gaussian random matrix. 
We generated the data by random projection of the signal, $\boldsymbol{y=Gm}$, and then inverted them via the proposed algorithm with different regularization models: traditional TV regularization performed by Algorithm \ref{Alg1},  separate TV regularization of real and imaginary parts performed by Algorithm \ref{Alg2} with $\tau=0.5$, TV regularization of magnitude part performed by Algorithm \ref{Alg3} with $\tau=1$,  and separate TV regularization of magnitude and phase performed by Algorithm \ref{Alg3} with $\tau=0.5$.
In all cases, we solved the constrained optimization in \eqref{COOP} by refining the data at each iteration of the algorithms (as in \eqref{MM_s_b}) and performed 500 iterations with penalty parameter $\gamma=1000$. 
Fig. \ref{fig:simple_test} shows the reconstructed signals by different algorithms.
It is seen in Fig. \ref{fig:simple_test} that near perfect reconstruction of the signal with discontinuities in the magnitude
and smoothness in the phase are obtained when TV regularization is applied on the magnitude and phase parts separately. 
\subsection{Simple inclusions test}
\begin{figure}
\centering
\includegraphics[width=0.49\textwidth]{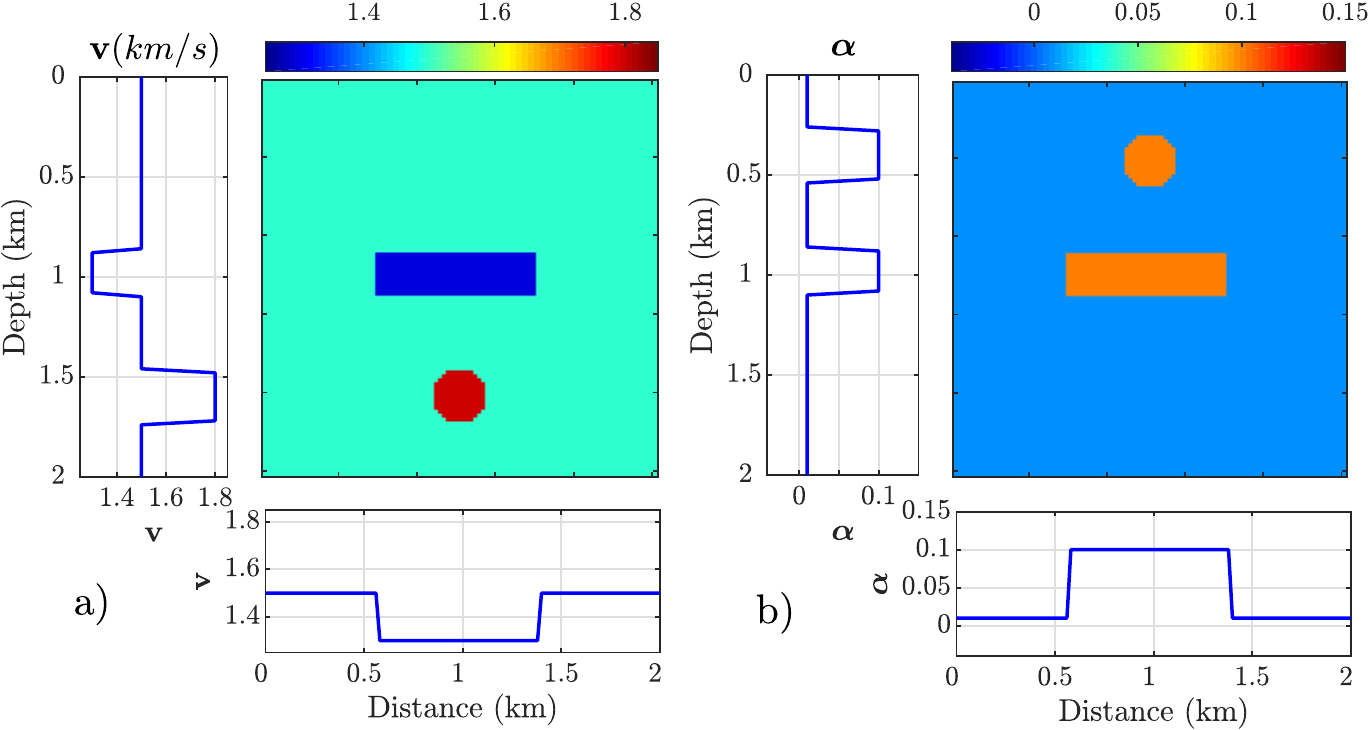}
\caption{Inclusion example. (a) True velocity model. (b) True $\boldsymbol{\alpha}$ model. Profiles of the true models running across the center of the models are also shown.}
\label{fig:Box_true}
\end{figure}

\begin{figure}
\centering
\includegraphics[width=0.81\textwidth]{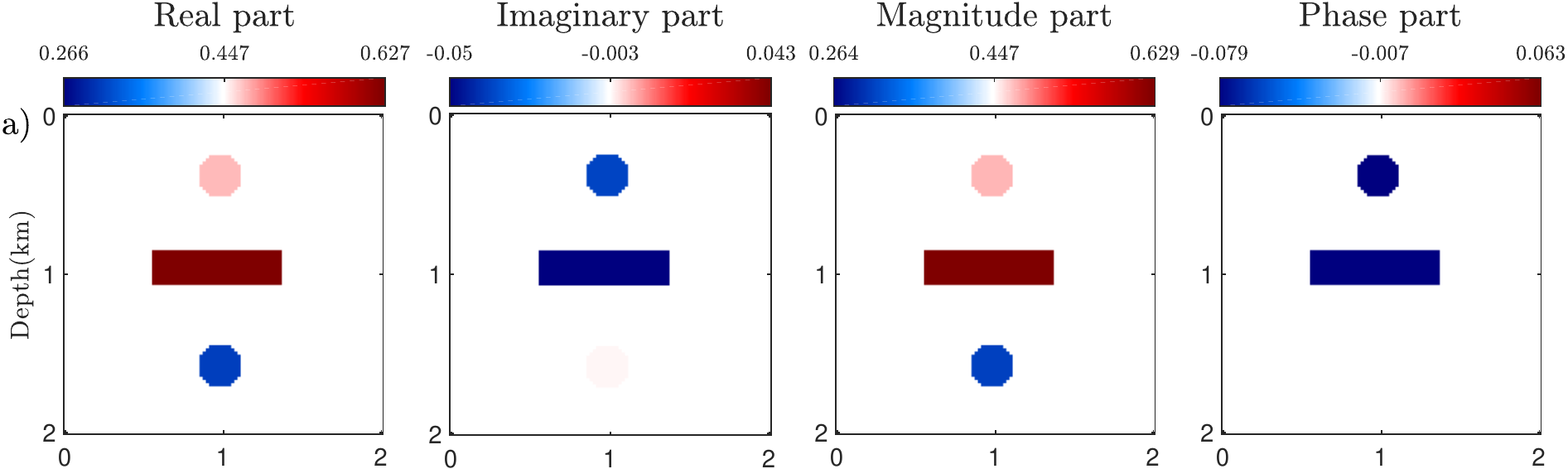}\\
\includegraphics[width=0.8\textwidth]{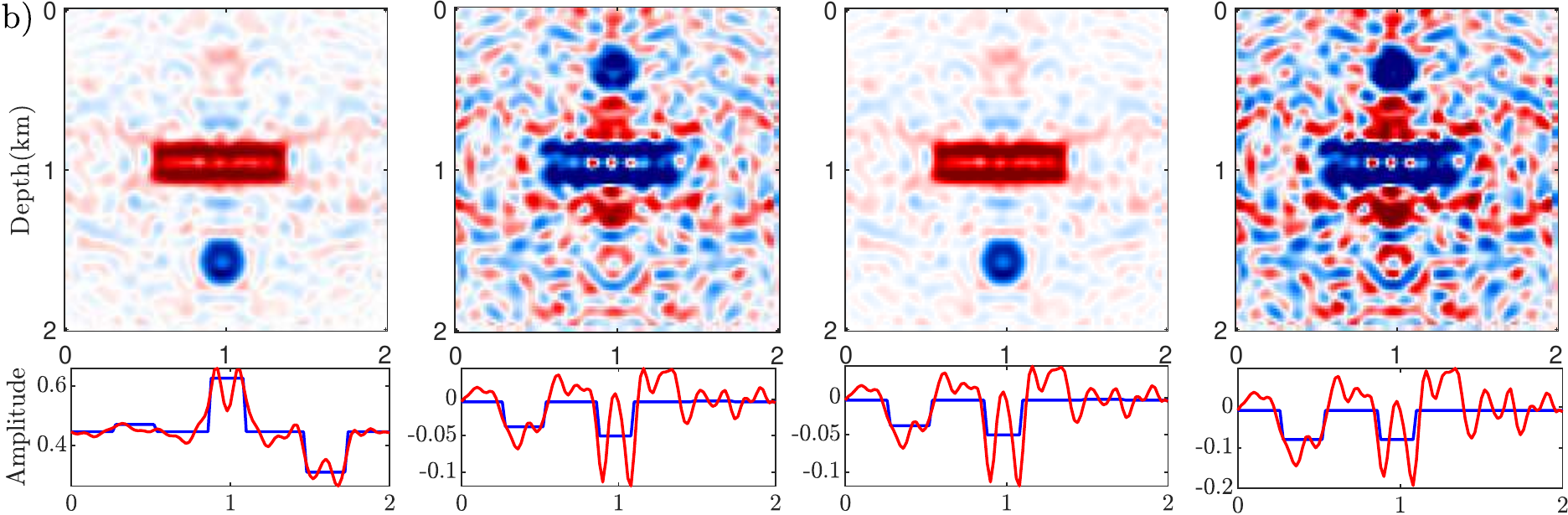}\\
\includegraphics[width=0.8\textwidth]{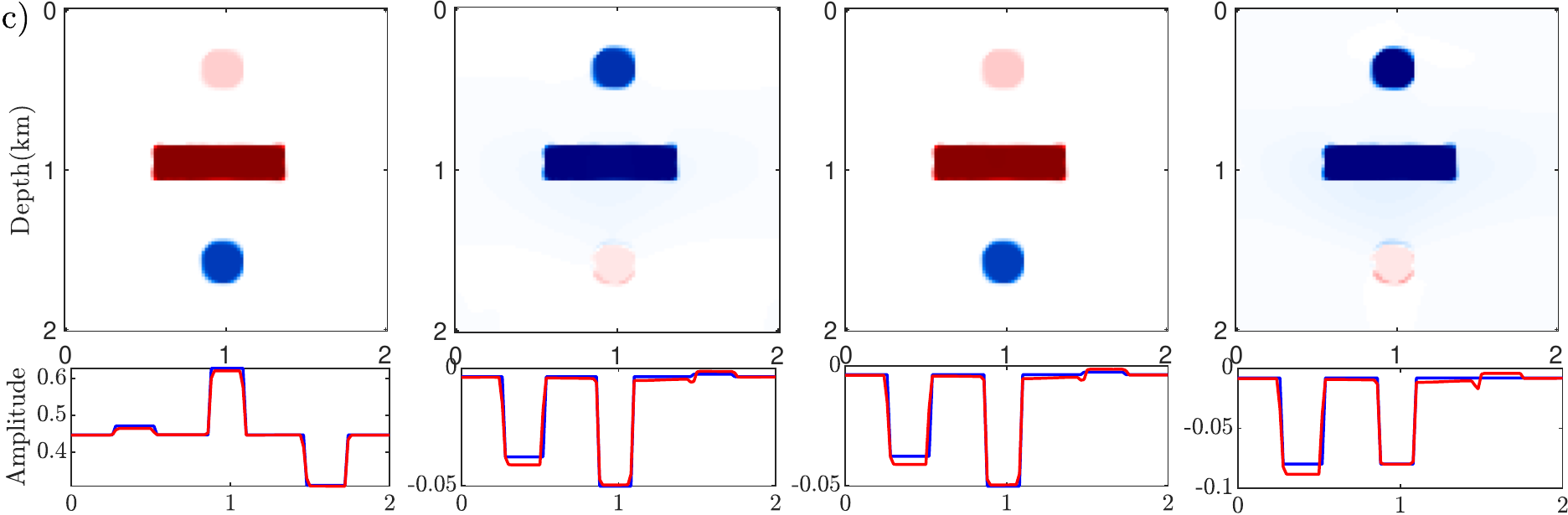}\\
\includegraphics[width=0.8\textwidth]{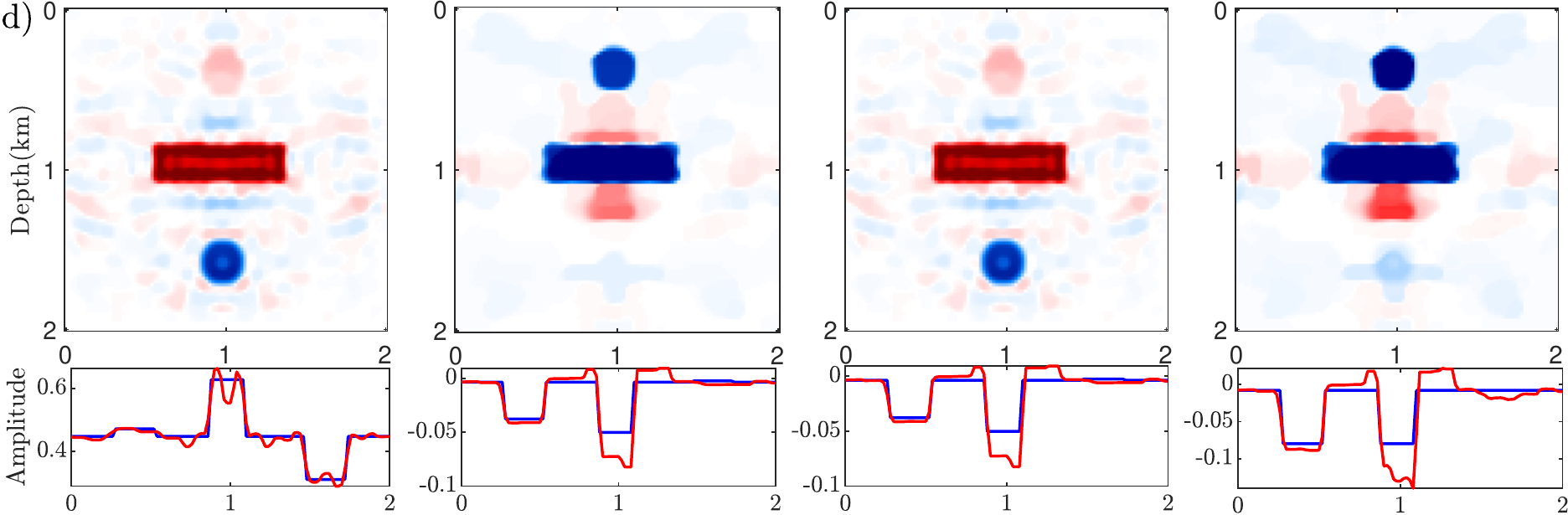}\\
\includegraphics[width=0.8\textwidth]{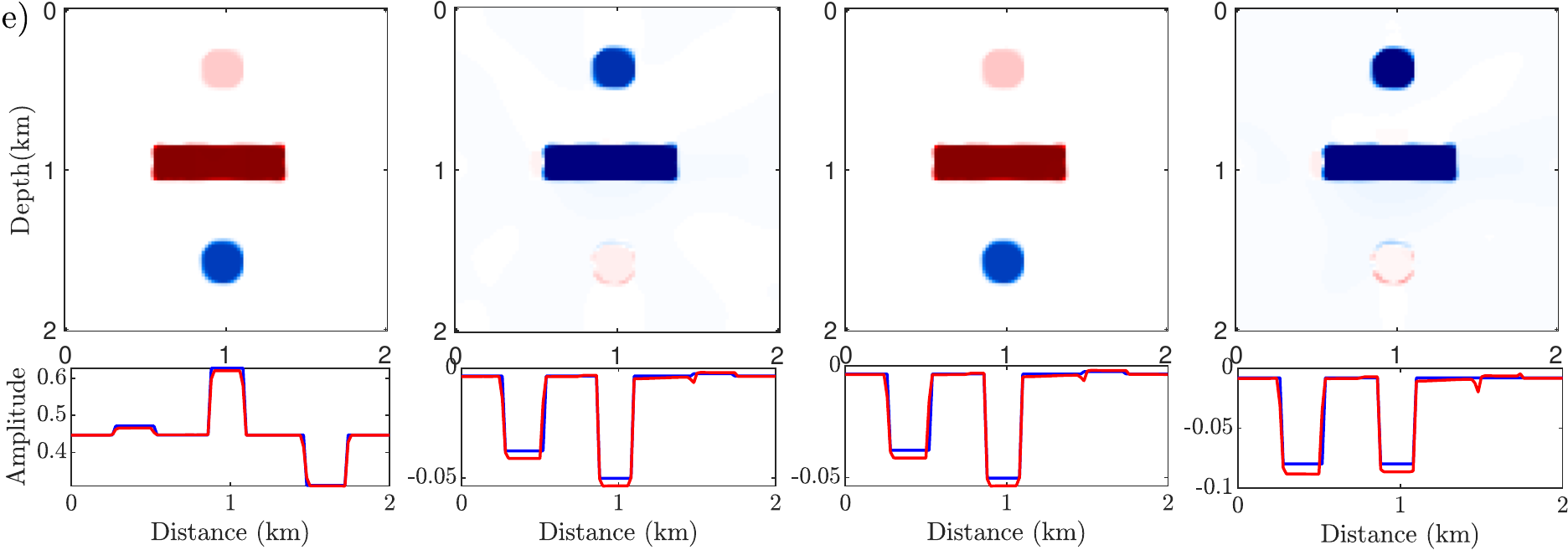}
\caption{Inclusion example. Visco-acoustic IR-WRI results. (a-e) Real part (first column), imaginary part (second column), magnitude part (third column) and phase part (fourth column). (a) True complex-velocity model. (b-e) Reconstructed complex velocity model. (b) Without regularization. (c-e) With TV regularization using (c) Alg. \ref{Alg1}, (d) Alg. \ref{Alg2} and (e) Alg. \ref{Alg3}. Vertical profiles of the true (blue) and reconstructed (red) models running across the center of the models are shown in the bottom of the reconstructed models. }
\label{fig:Box_test_real_imag_abs_phase}
\end{figure}

\begin{figure}
\includegraphics[width=0.49\textwidth]{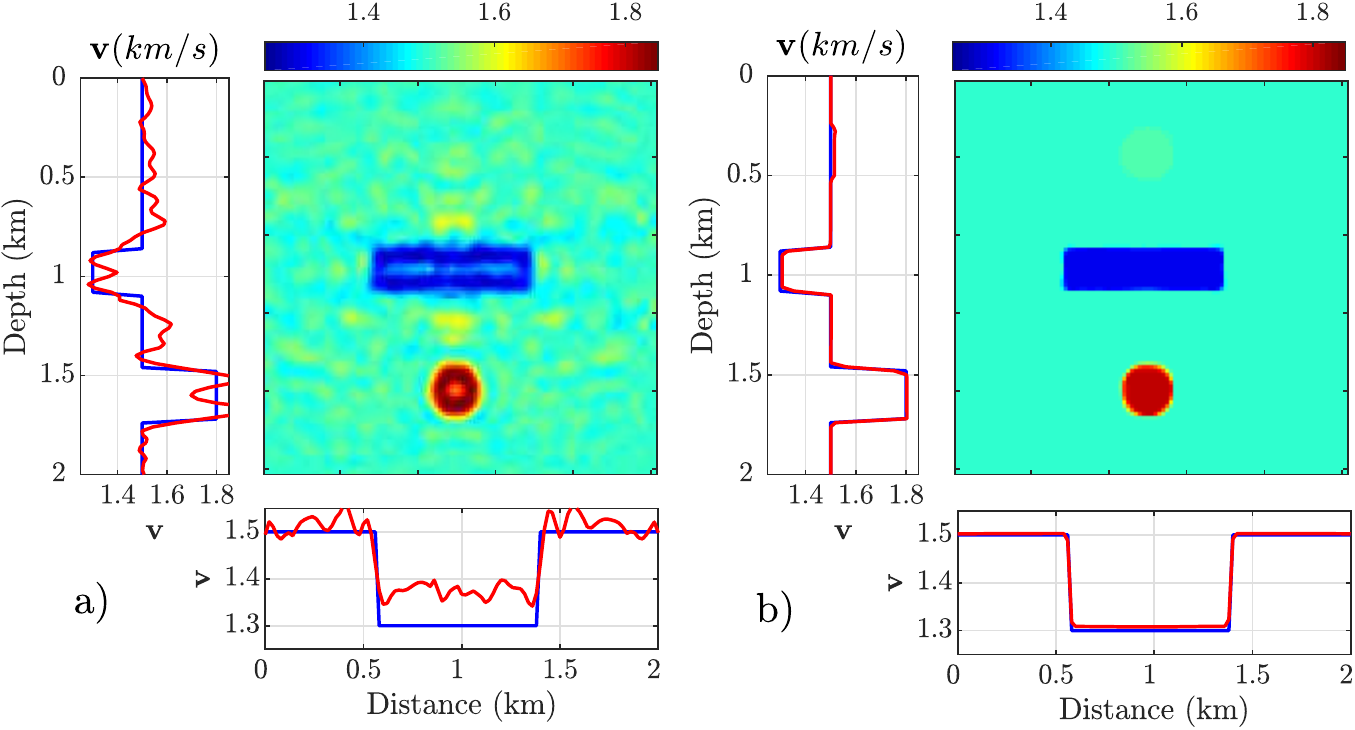}
\includegraphics[width=0.49\textwidth]{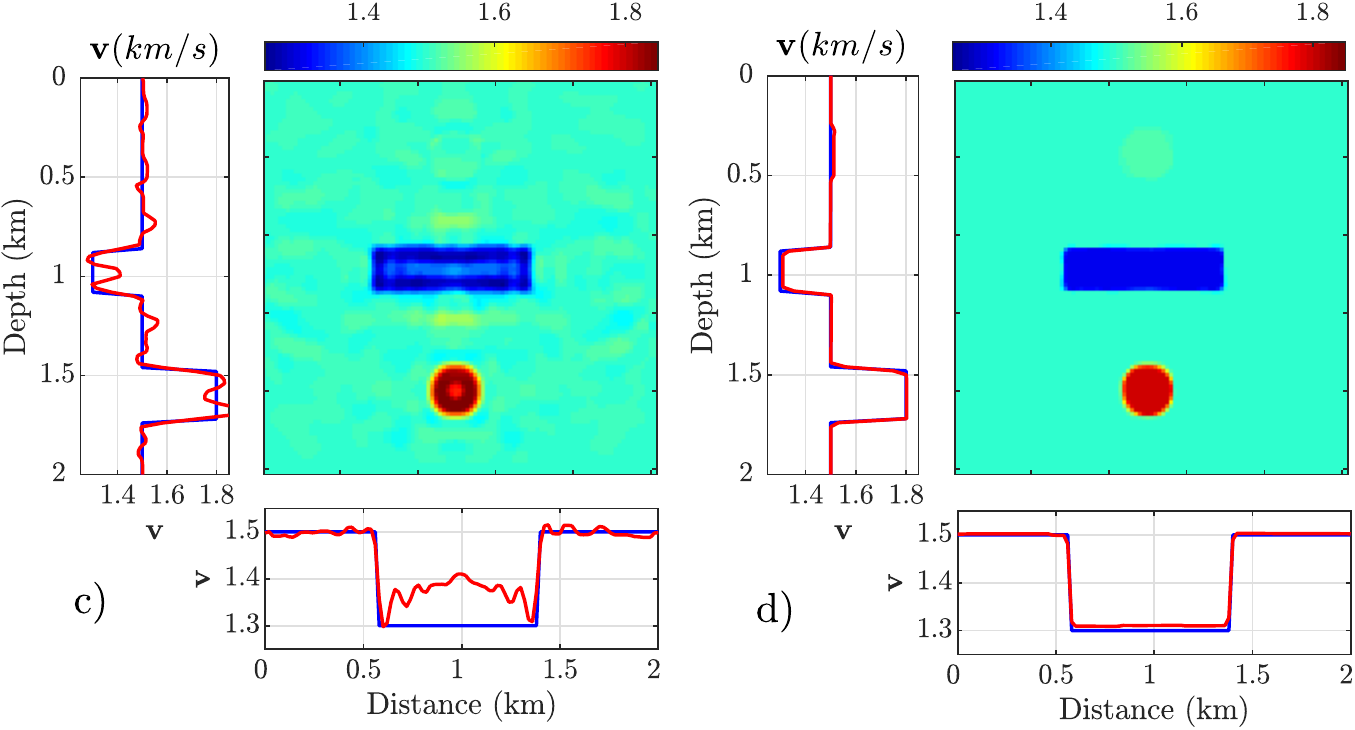}\\
\includegraphics[width=0.49\textwidth]{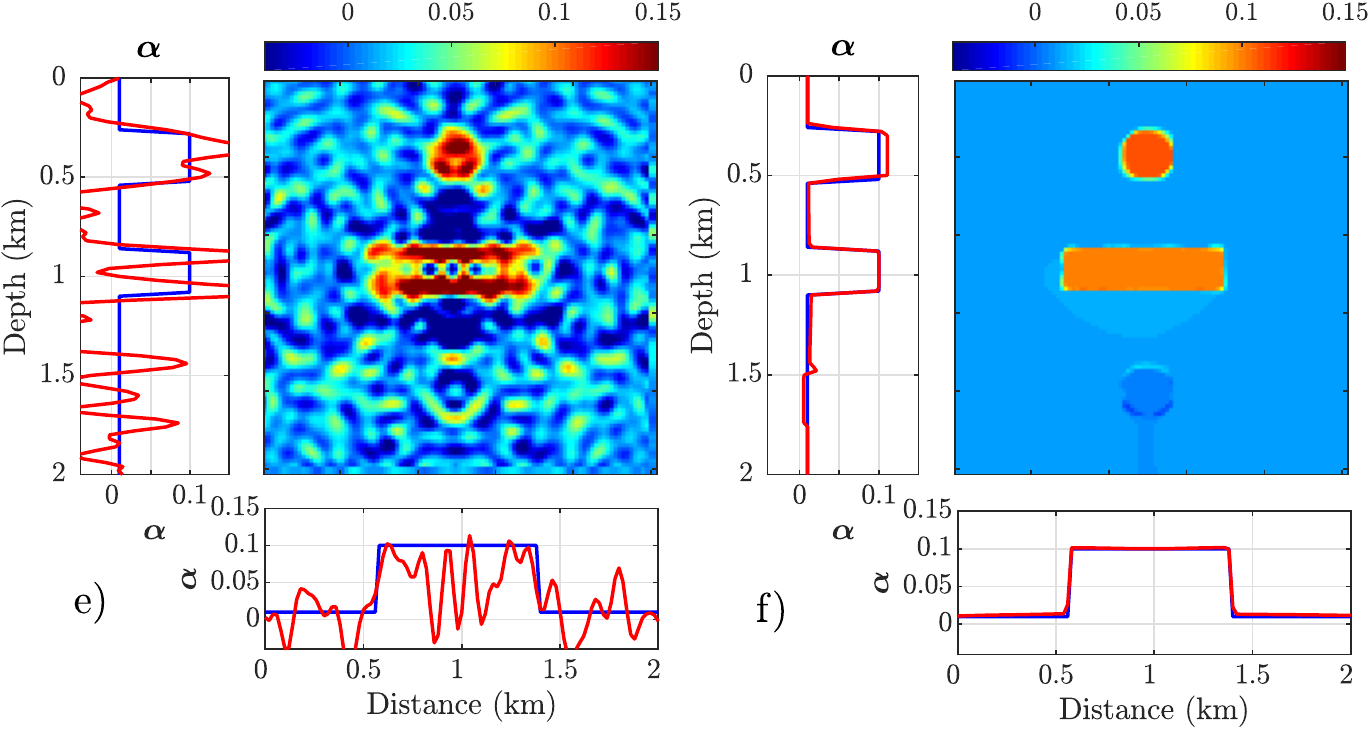}
\includegraphics[width=0.49\textwidth]{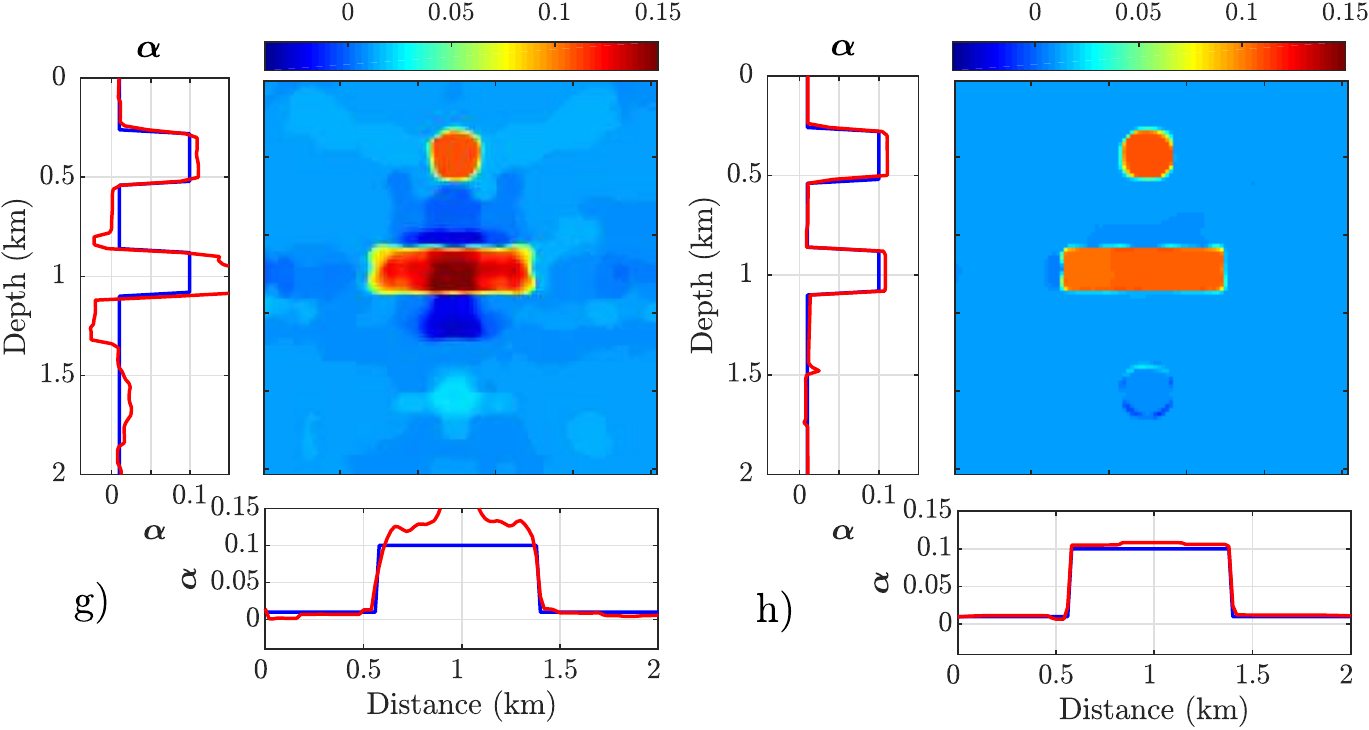}
\caption{Inclusion example. Extracted velocity and attenuation model using SLS mechanism (\eqref{ISLSv} and \eqref{ISLSalpha}) from results of Fig. \ref{fig:Box_test_real_imag_abs_phase}. (a-d) Extracted velocity model   (a) without regularization and (b-d) with TV regularization. (b) Algorithm \ref{Alg1}. (c) Algorithm \ref{Alg2}. (d) Algorithm \ref{Alg3}. (e-h) Same as (a-d) but for extracted $\boldsymbol{\alpha}$ models. Profiles of the true (blue) and reconstructed (red) models running across the center of the models are shown in the left and bottom of the reconstructed models. }
\label{fig:Box_test_vQ-SLS}
\end{figure}

\begin{figure}
\includegraphics[width=0.49\textwidth]{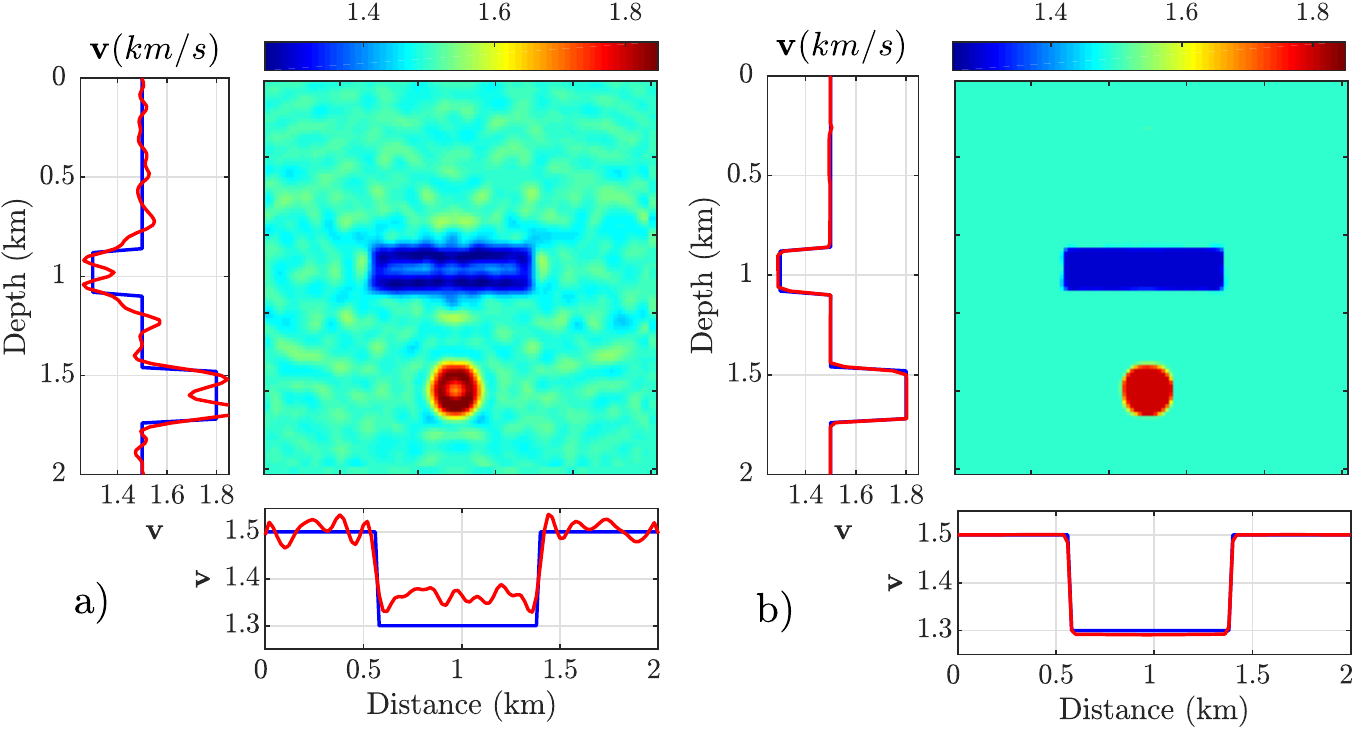}
\includegraphics[width=0.49\textwidth]{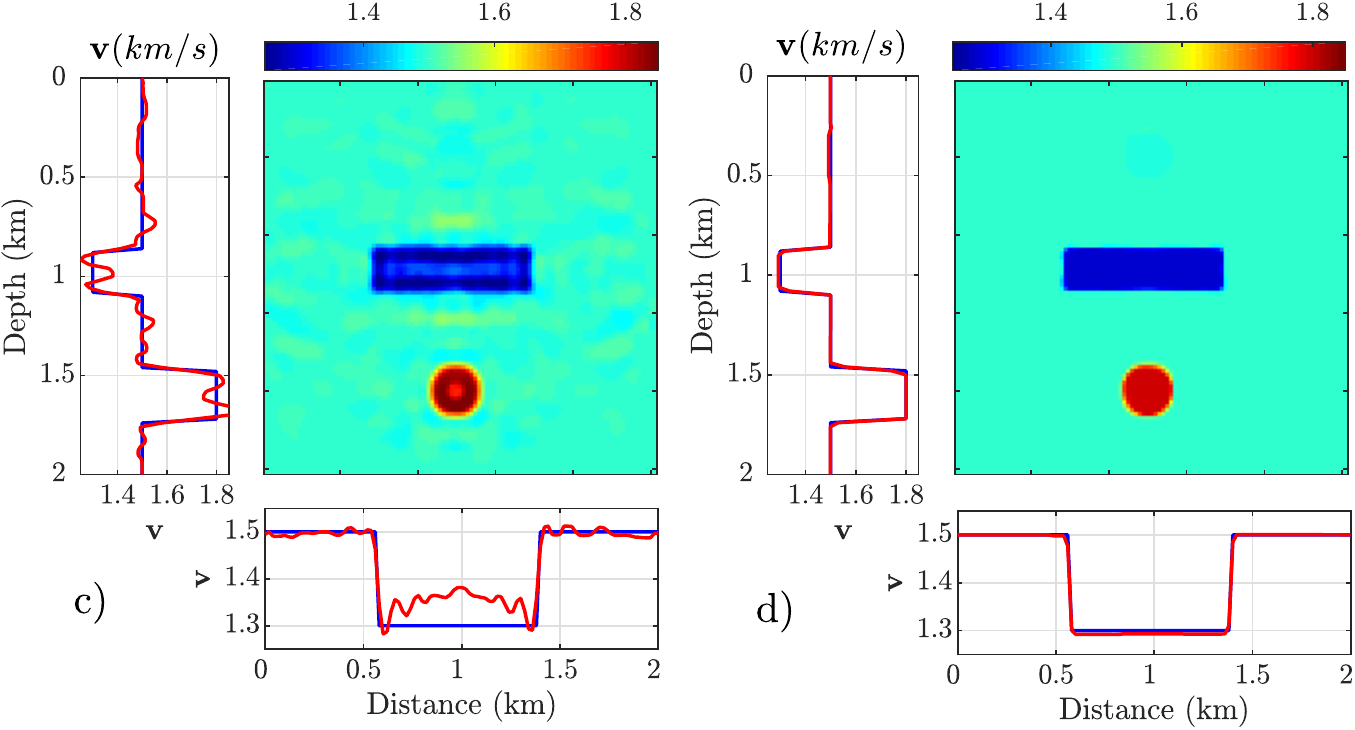}\\
\includegraphics[width=0.49\textwidth]{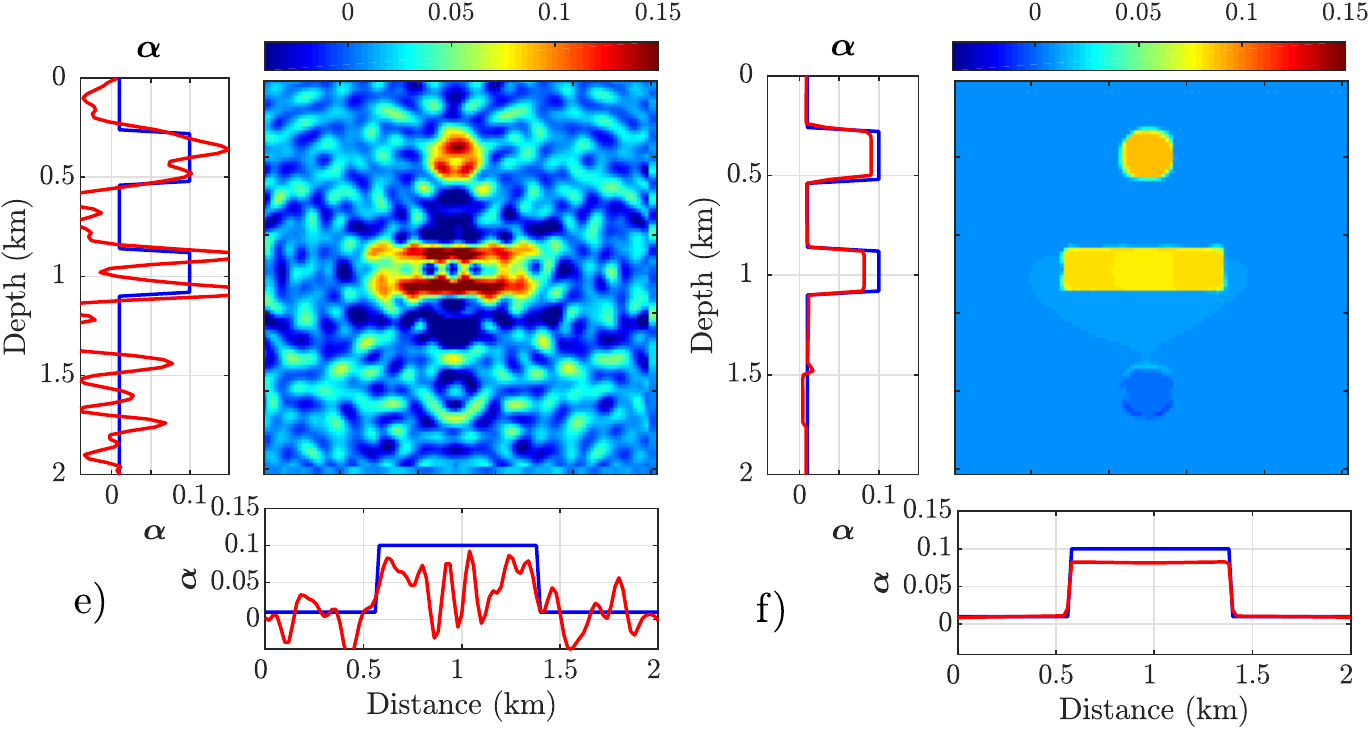}
\includegraphics[width=0.49\textwidth]{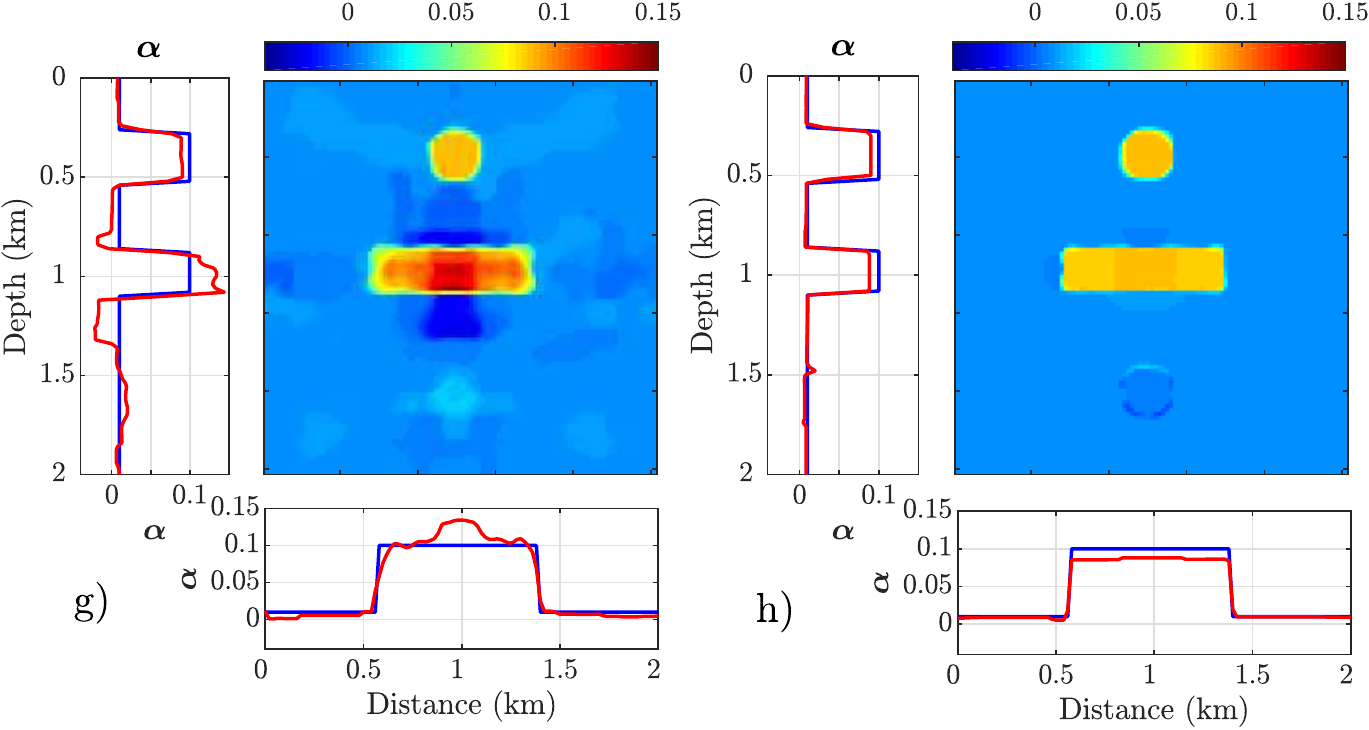}
\caption{Same as Fig. \ref{fig:Box_test_vQ-SLS}, but for extracted velocity and attenuation model using KF mechanism (eqs. \eqref{IKFv} and \eqref{IKFalpha})} 
\label{fig:Box_test_vQ-KF}
\end{figure}

We continue with a simple 2D example to validate visco-acoustic IR-WRI for complex-valued TV regularization with Algorithms \ref{Alg1}-\ref{Alg3}. The true velocity model $\boldsymbol{v}$ is formed by a homogeneous background model of velocity equal to 1.5 km/s to which two inclusions are added: a 250-m-diameter circular inclusion with a velocity of 1.8 km/s at position (1.6~km,1~km), and a 0.2 $\times$ 0.8 km$^2$ rectangular inclusion with a  velocity of 1.3 km/s at the center of the model (Fig. \ref{fig:Box_true}a). Also, the true attenuation model ($\boldsymbol{\alpha}$) is a homogeneous background model with $\boldsymbol{\alpha}=0.01$ to which two inclusions with $\boldsymbol{\alpha}=0.1$ are added: a 250-m-diameter circular inclusion at position (0.4~km,1~km) and a 0.2 $\times$ 0.8 km$^2$ rectangle inclusion at the center of the model (Fig. \ref{fig:Box_true}b).
The acquisition is designed with 8 sources and 200 receivers along the four edges of the model and three frequency components (5, 6, 7~Hz) are jointly inverted with the stopping criterion of iteration being set to 30 iterations. The recorded data are simulated with the SLS attenuation model  \eqref{SLSmodel0} with a reference frequency of 10 Hz. 
%
%
We performed visco-acoustic IR-WRI without and with TV regularization starting from a homogeneous $\boldsymbol{m}=1/\boldsymbol{v}^2$ generated with the KF  \eqref{KFmodel0} or SLS \eqref{SLSmodel0} relation using $\boldsymbol{v}=\bold{1.5}$  km/s, $\boldsymbol{\alpha}=\bold{0}$. 
First, we show different components (Real, Imaginary, Magnitude and Phase) of
the estimated complex-velocity using IR-WRI without regularization (Fig. \ref{fig:Box_test_real_imag_abs_phase}b) and with TV regularization using Algorithms \ref{Alg1}-\ref{Alg3} (Figs. \ref{fig:Box_test_real_imag_abs_phase}c-\ref{fig:Box_test_real_imag_abs_phase}e). 
A vertical log, which crosses the center of each panel of Fig.  \ref{fig:Box_test_real_imag_abs_phase}, is plotted in the bottom side of them. 
The physical parameters $\boldsymbol{v}$ and $\boldsymbol{\alpha}$ at the reference frequency extracted from the final complex-valued model $\boldsymbol{m}$ (Fig. \ref{fig:Box_test_real_imag_abs_phase}) with the inverse of SLS mapping (\eqref{ISLSalpha} and \eqref{ISLSv}) for frequency $\omega=12\pi$ Rad and $\omega_r=20 \pi$ Rad are shown in Figs. \ref{fig:Box_test_vQ-SLS}a-\ref{fig:Box_test_vQ-SLS}d and \ref{fig:Box_test_vQ-SLS}e-\ref{fig:Box_test_vQ-SLS}h, respectively. Also, the extracted $\boldsymbol{v}$ and $\boldsymbol{\alpha}$
with the inverse of KF mapping (eqs. \eqref{IKFv} and \eqref{IKFalpha}) with same $\omega$ and $\omega_r$ are shown in Fig. \ref{fig:Box_test_vQ-KF}. \\
Those  obtained without regularization are quite noisy  (Figs. \ref{fig:Box_test_real_imag_abs_phase}b, \ref{fig:Box_test_vQ-SLS}a, \ref{fig:Box_test_vQ-SLS}e, \ref{fig:Box_test_vQ-KF}a and \ref{fig:Box_test_vQ-KF}e). This can result from the sparsity of the acquisition design leading to wraparound artefacts \cite{Aghamiry_2019_RAR}, parameter cross-talk and the fact that the dependency of $\boldsymbol{m}$ to frequency was not taken into account in this test (since all the frequencies were inverted in one go).
The TV regularizations performed by Algorithms \ref{Alg1} and \ref{Alg3} removes most of this noise, although moderate cross-talk artefacts are still visible, while Algorithm \ref{Alg2} clearly provides the worst result among the regularized inversions. Cross-talk artefacts manifest by small overestimation of the velocity contrasts which are balanced by the underestimation of the attenuation contrasts. Relatively to Algorithm \ref{Alg1}, Algorithm \ref{Alg3} based upon the separate regularization of amplitude and phase better focuses the shape of different parts of models (Figs. \ref{fig:Box_test_real_imag_abs_phase}e) as well as the $\boldsymbol{v}$ and $\boldsymbol{\alpha}$ models (Figs. \ref{fig:Box_test_vQ-SLS}d,\ref{fig:Box_test_vQ-SLS}h and \ref{fig:Box_test_vQ-KF}d,\ref{fig:Box_test_vQ-KF}h). 
This probably results because the effects of $\boldsymbol{\alpha}$ are mostly contained in the phase of $\boldsymbol{m}$ when the data are generated with the SLS relation \eqref{SLSmodel0} (We can have a same conclusion for the data generated by KF model in \eqref{KFmodel0}). Therefore, tailoring the regularization of the phase should implicitly amount to tailor the regularization of $\boldsymbol{\alpha}$, while the regularization of the magnitude mixes the regularization of $\boldsymbol{v}$ and $\boldsymbol{\alpha}$. Therefore, we can expect that the beneficial impact of the tailored regularization of  the phase during the inversion is preserved during the posteriori extraction of $\boldsymbol{\alpha}$ via the nonlinear inverse mapping of SLS (\eqref{ISLSalpha} and \eqref{ISLSv}) or KF mechanism (\eqref{IKFv} and \eqref{IKFalpha}). \\
Comparing the logs of the models obtained with Algorithms \ref{Alg1} and \ref{Alg3} shows that the later manages better the cross-talk artefacts in the phase reconstruction (Figs. \ref{fig:Box_test_real_imag_abs_phase}c,\ref{fig:Box_test_real_imag_abs_phase}e). See also the  $(\boldsymbol{v},\boldsymbol{\alpha})$ rectangular anomaly and the circular $\boldsymbol{\alpha}$ anomaly in Figs. \ref{fig:Box_test_vQ-SLS}b,\ref{fig:Box_test_vQ-SLS}d,\ref{fig:Box_test_vQ-SLS}f,\ref{fig:Box_test_vQ-SLS}h and \ref{fig:Box_test_vQ-KF}b,\ref{fig:Box_test_vQ-KF}d,\ref{fig:Box_test_vQ-KF}f,\ref{fig:Box_test_vQ-KF}h). But, the $\boldsymbol{\alpha}$ model inferred from Algorithm \ref{Alg3} still shows a small-amplitude ghost of the contour of the circular anomaly in the bottom of the reconstructed phase (Fig. \ref{fig:Box_test_real_imag_abs_phase}e) and extracted $\boldsymbol{\alpha}$ (Figs. \ref{fig:Box_test_vQ-SLS}h and \ref{fig:Box_test_vQ-KF}h). One reason might be that the regularization of the phase is inactive in this part of the model where there is no attenuation anomaly, making the reconstruction of  $\boldsymbol{\alpha}$ quite ill posed and hence subject to cross-talk with the dominant parameter $\boldsymbol{v}$. \\
Comparing  Figs.~\ref{fig:Box_test_vQ-SLS} and \ref{fig:Box_test_vQ-KF} allows us to assess the sensitivity of the inversion to the attenuation model used to perform the extraction of $\boldsymbol{v}$ and $\boldsymbol{\alpha}$ from the complex velocity. The results show that, for this simple example, the choice of the attenuation mechanism has a small imprint on the recovered $\boldsymbol{v}$. Only $\boldsymbol{\alpha}$ shows slightly underestimated amplitudes when it is extracted with the KF model, while the data were generated with the SLS model (Fig.~\ref{fig:Box_test_vQ-KF}h).
%
%

\subsection{North Sea case study}
\begin{figure}
\centering
\includegraphics[width=1\textwidth]{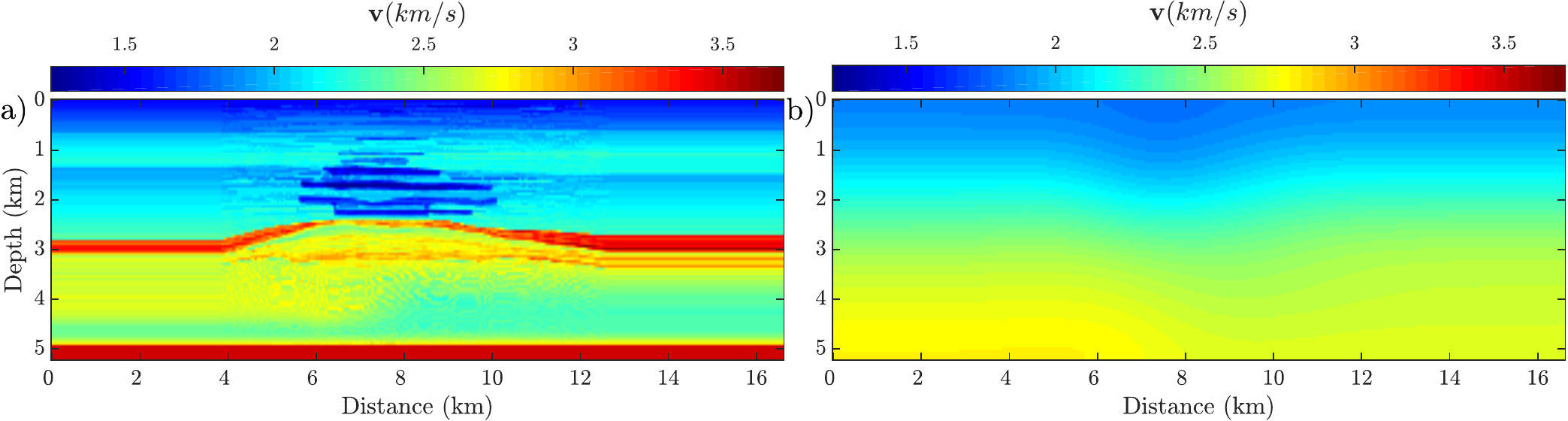}
\caption{North Sea case study. (a) True $\boldsymbol{v}$ model. (b) Initial $\boldsymbol{v}$ model.}
\label{fig:north_test_true}
\end{figure}

\begin{figure}[ht!] 
\centering
   \includegraphics[width=1\textwidth]{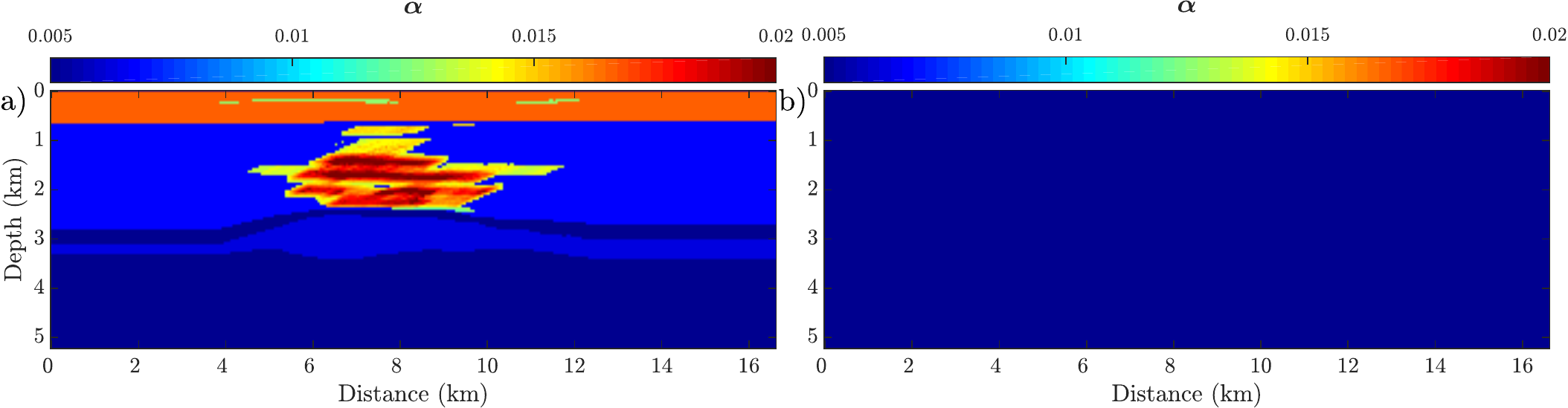} 
\caption{North Sea case study. (a)  True $\boldsymbol{\alpha}$ model. (b)  Initial $\boldsymbol{\alpha}$ model.}
\label{fig:val_initial}
\end{figure}

We continue by considering a more realistic 16 km $\times$ 5.2 km shallow-water model representative of the North Sea \cite{Munns_1985_VFG}. The true models for $\boldsymbol{v}$ and $\boldsymbol{\alpha}$ at the reference frequency are shown in Figs. \ref{fig:north_test_true}a and \ref{fig:val_initial}a, respectively. 
The velocity model is formed by soft sediments in the upper part, a pile of low-velocity gas layers above a chalk reservoir, the top of which is indicated by a sharp positive velocity contrast at around 2.5~km depth, and a flat reflector at 5~km depth (Fig. \ref{fig:north_test_true}a). 
The $\boldsymbol{\alpha}$ model has two highly attenuative zones, in the upper soft sediments and gas layers, and the $\boldsymbol{\alpha}$ value is relatively high elsewhere (Fig. \ref{fig:val_initial}a).
The initial model for $\boldsymbol{v}$ is a highly Gaussian filtered version of the true model (Fig. \ref{fig:north_test_true}b), while the starting attenuation model is homogeneous with $\boldsymbol{\alpha}=0$ (Fig. \ref{fig:val_initial}b).
%
%
%

\begin{figure}
\includegraphics[width=1\textwidth]{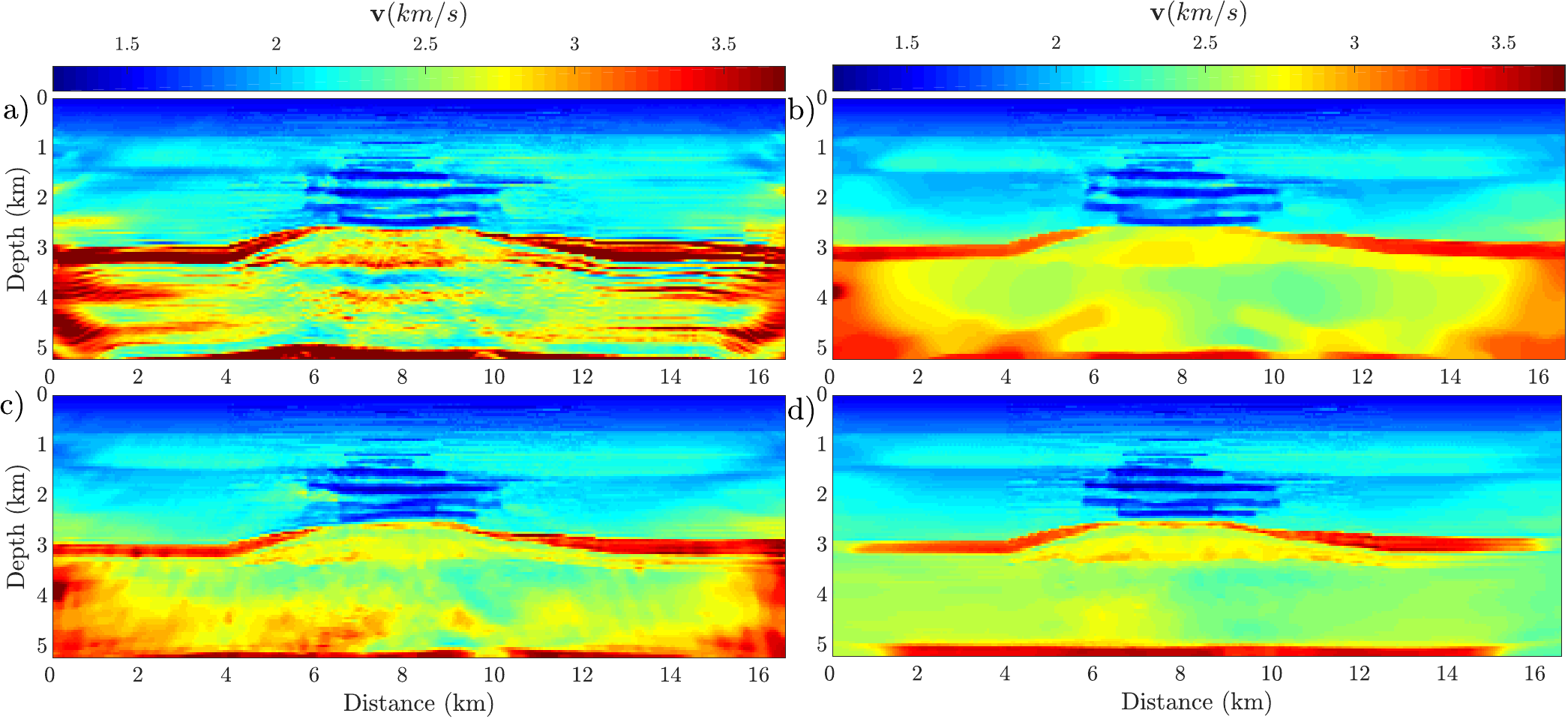}
\caption{North Sea case study. Extracted velocity using KF mechanism from visco-acoustic IR-WRI results, (a) without regularization, (b-d) with TV regularization using (b) Algorithm \ref{Alg1}, (c) Algorithm \ref{Alg2}, and (d)  Algorithm \ref{Alg3}.}
\label{fig:north_test}
\end{figure}

\begin{figure}
\includegraphics[width=1\textwidth]{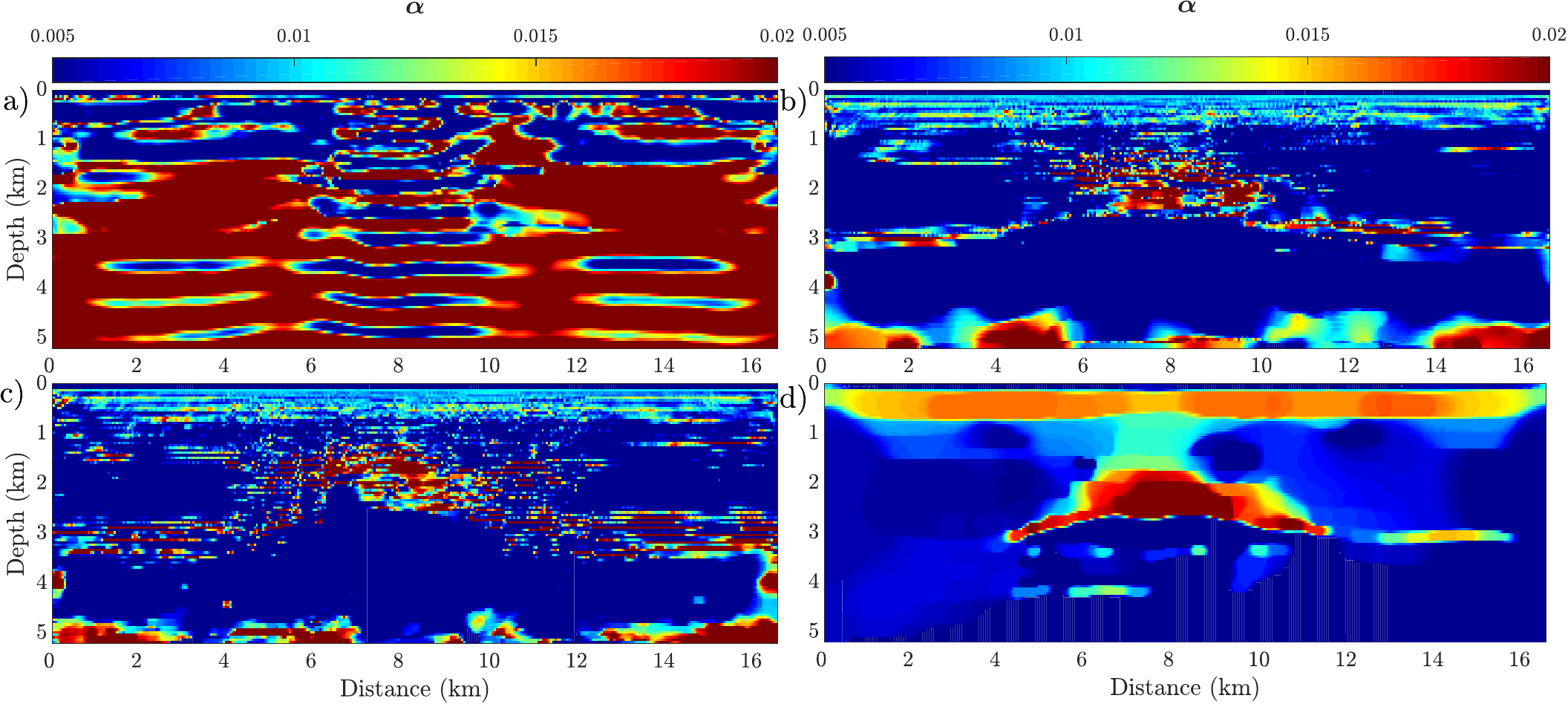}
\caption{Same as Fig. \ref{fig:north_test}, but for extracted $\boldsymbol{\alpha}$.}
\label{fig:north_test_Q}
\end{figure}

\begin{figure}
\includegraphics[width=0.49\textwidth]{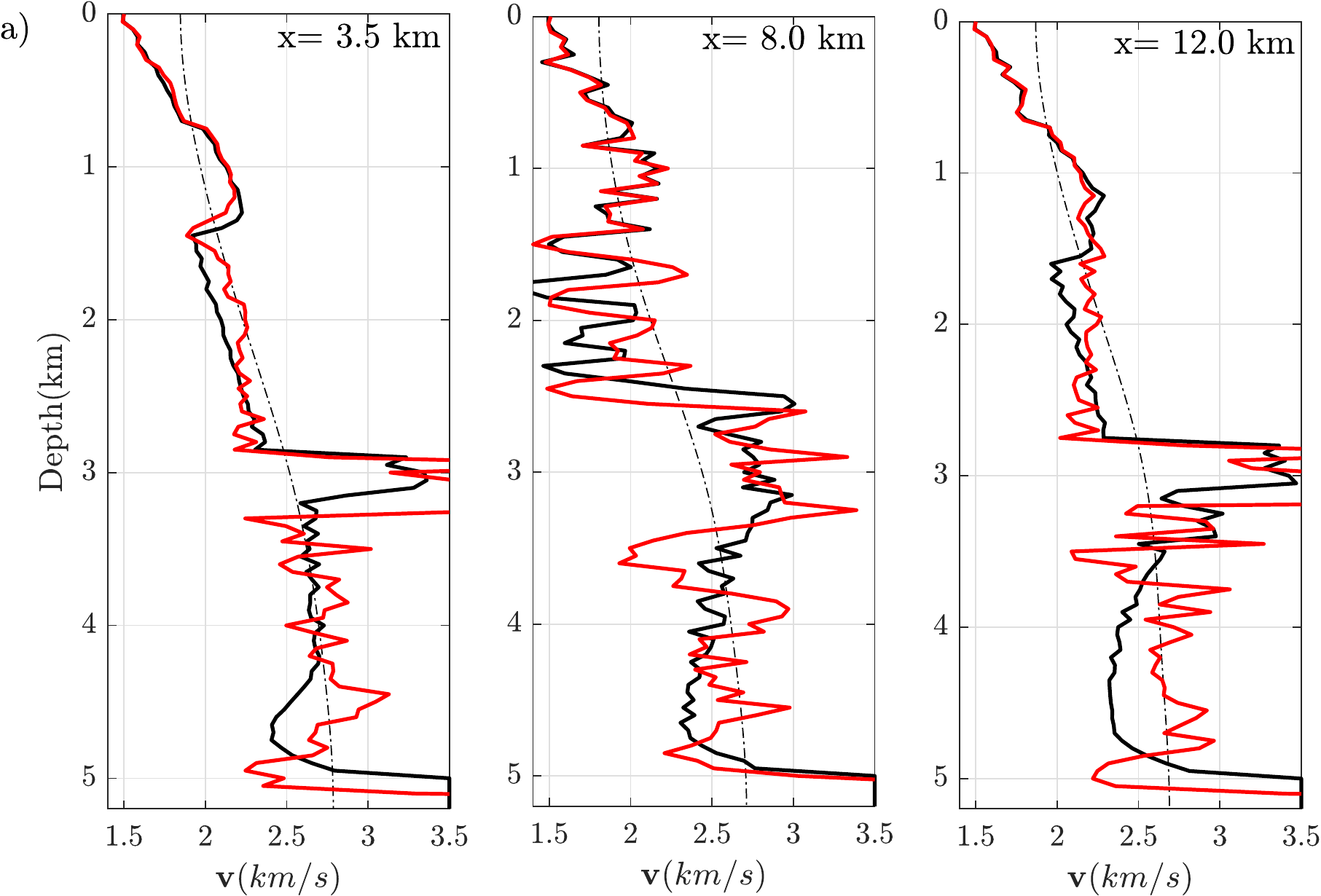}
\includegraphics[width=0.47\textwidth]{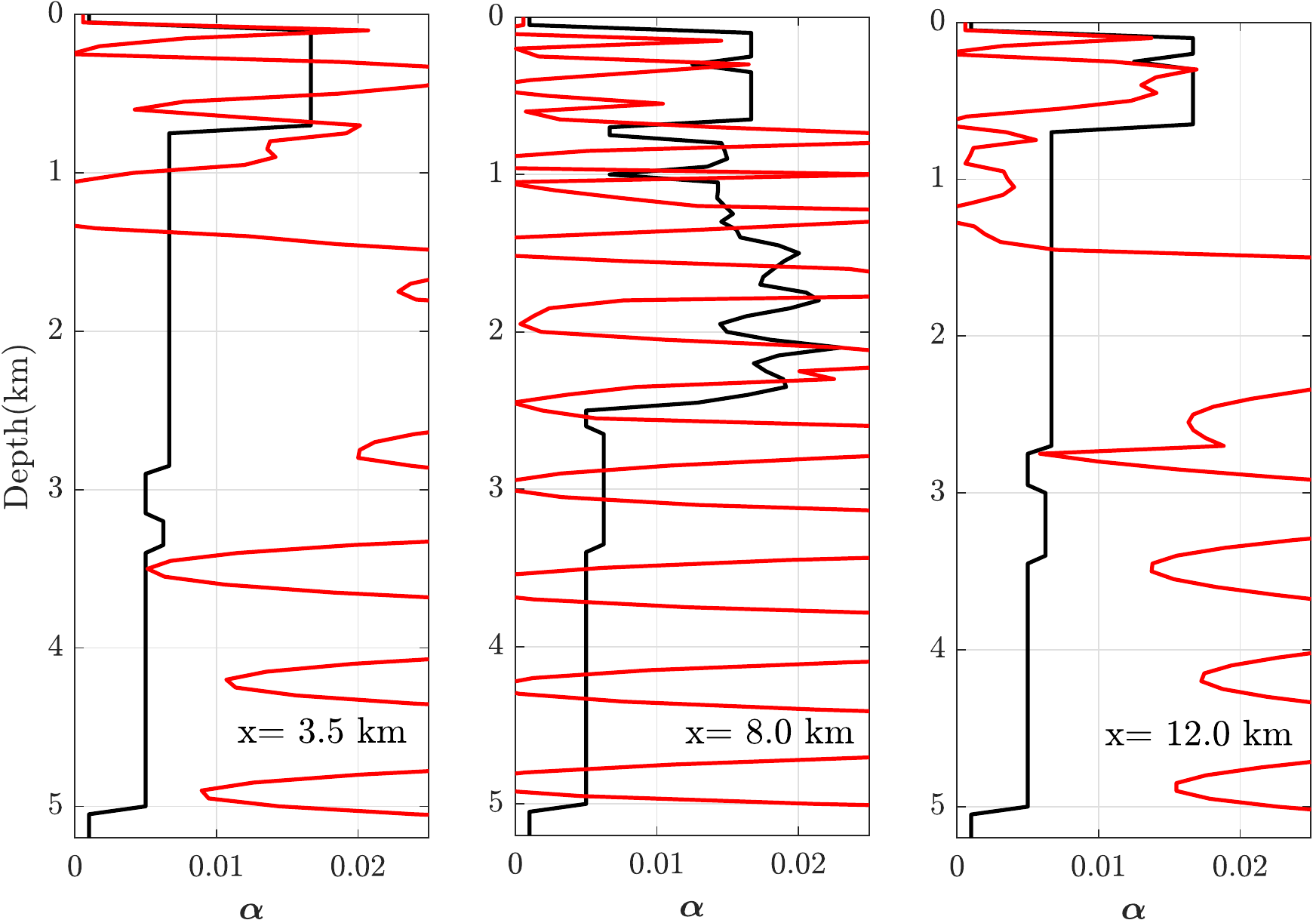}\\
\includegraphics[width=0.49\textwidth]{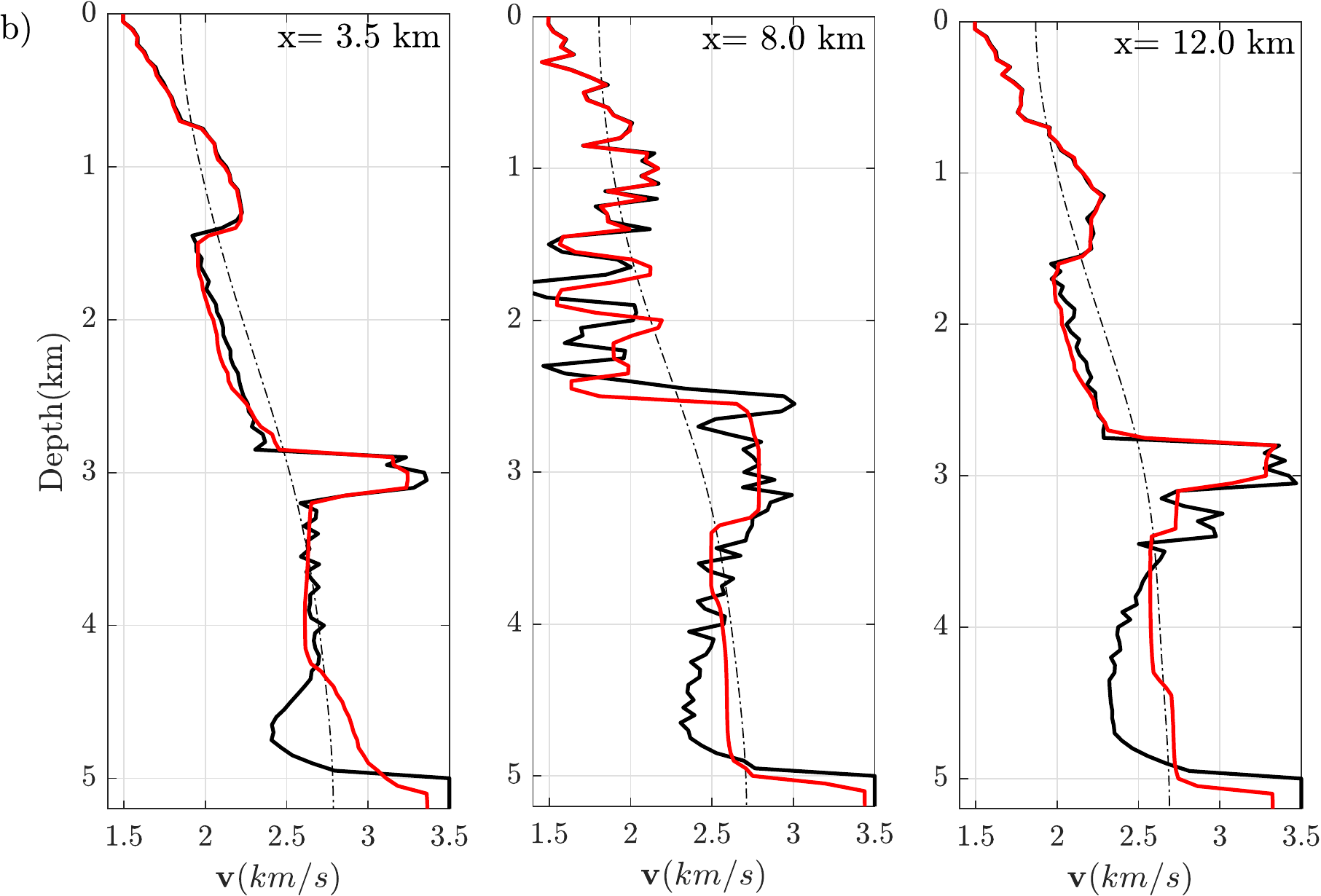}
\includegraphics[width=0.47\textwidth]{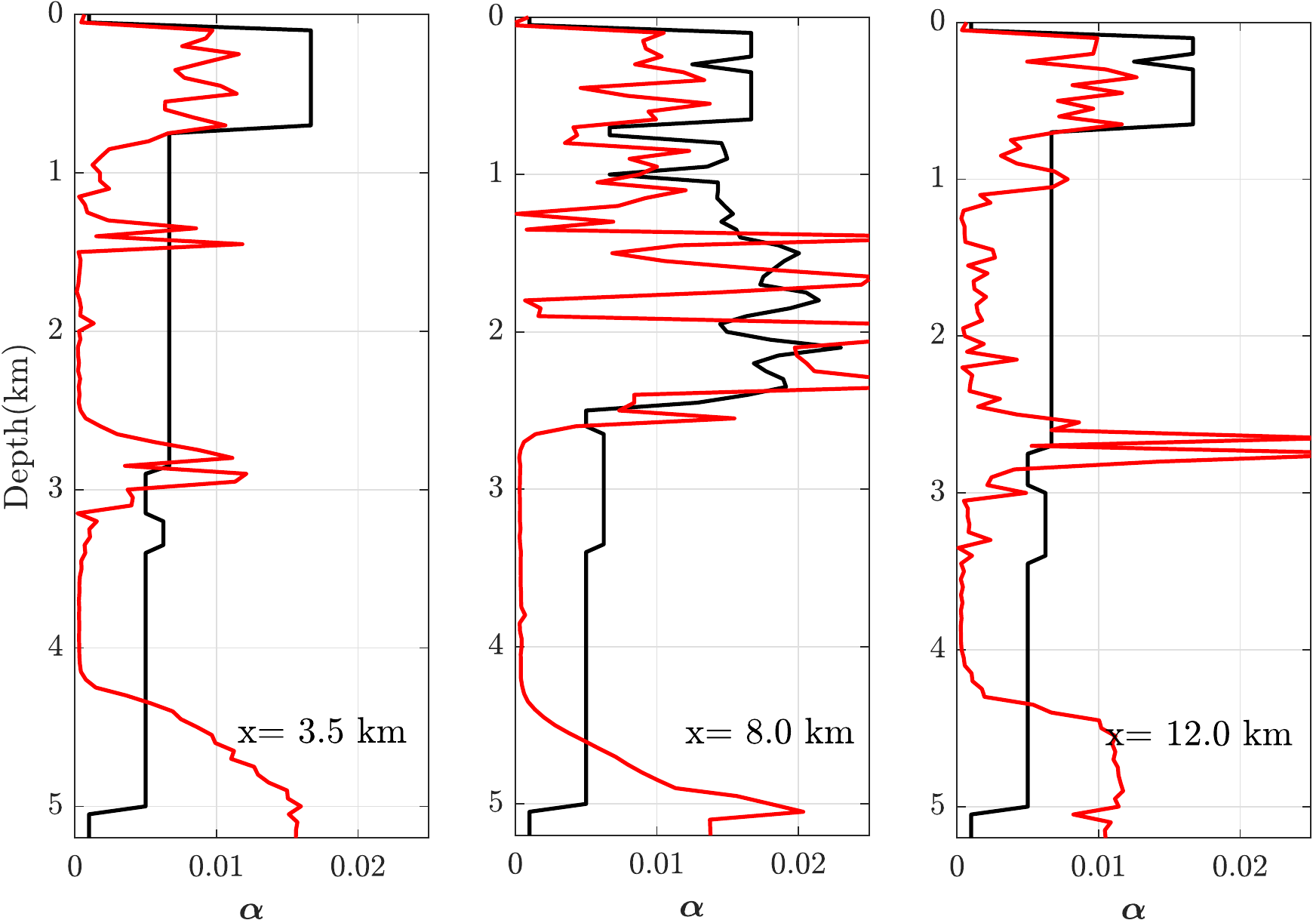}\\
\includegraphics[width=0.49\textwidth]{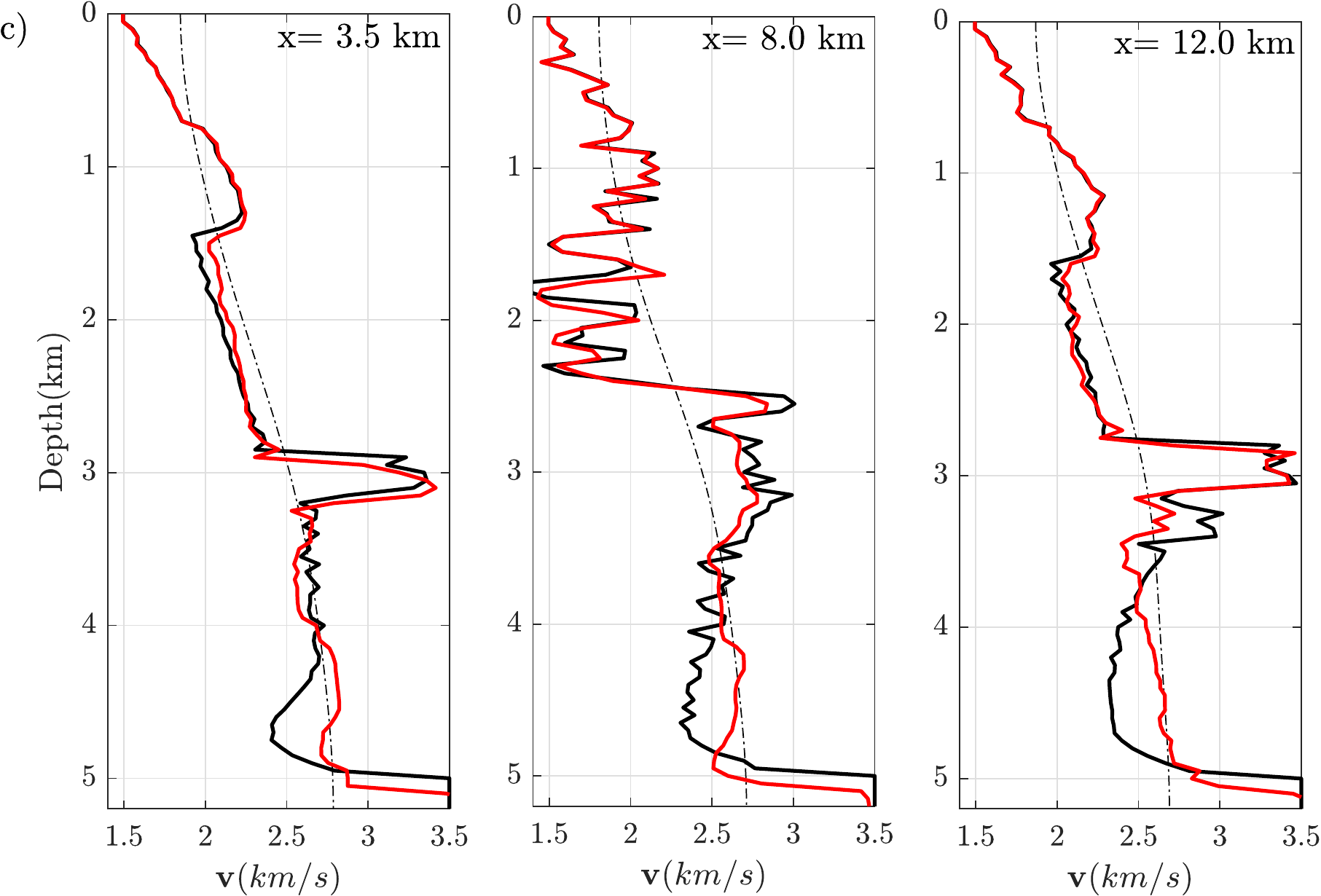}
\includegraphics[width=0.47\textwidth]{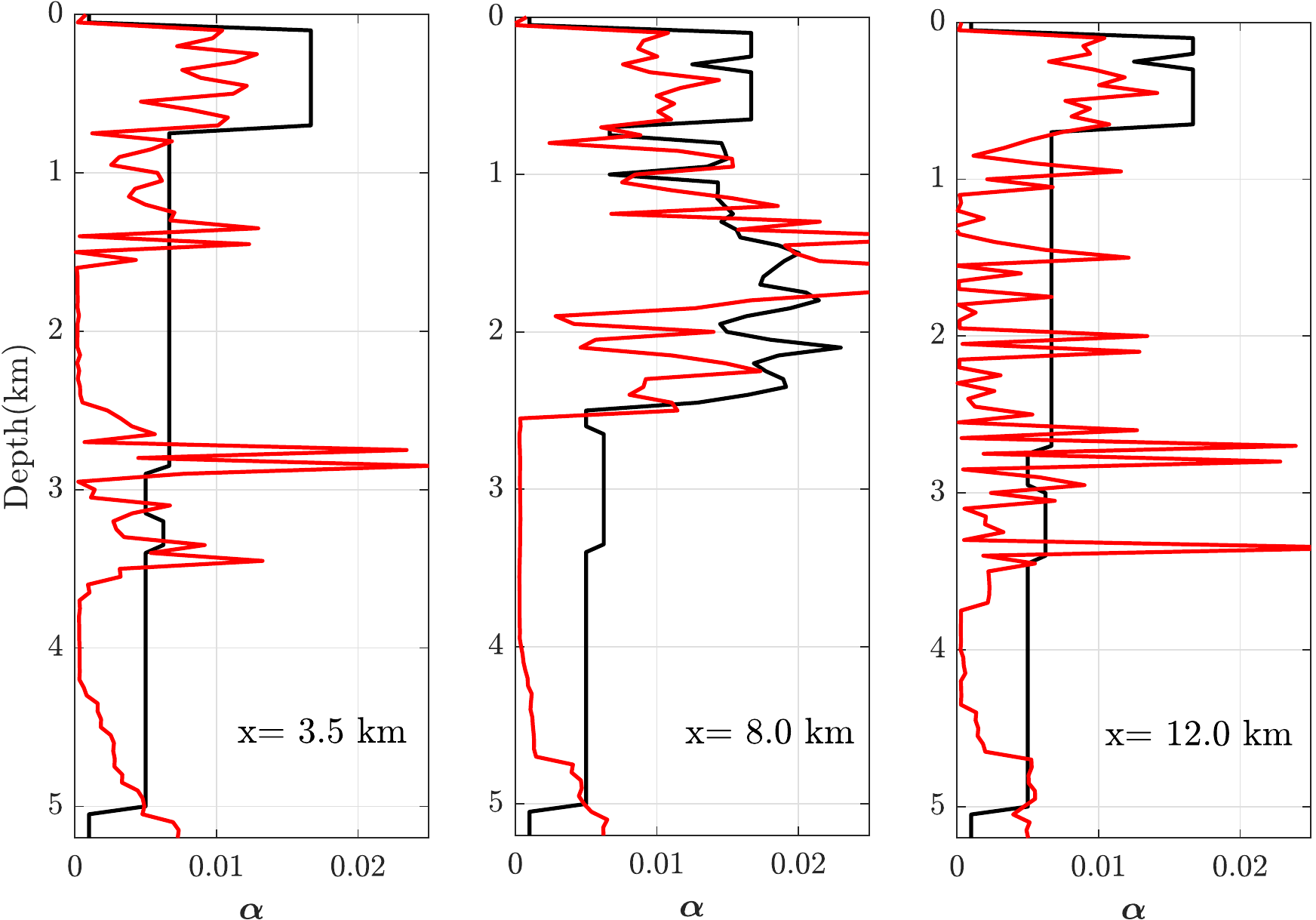}\\
\includegraphics[width=0.49\textwidth]{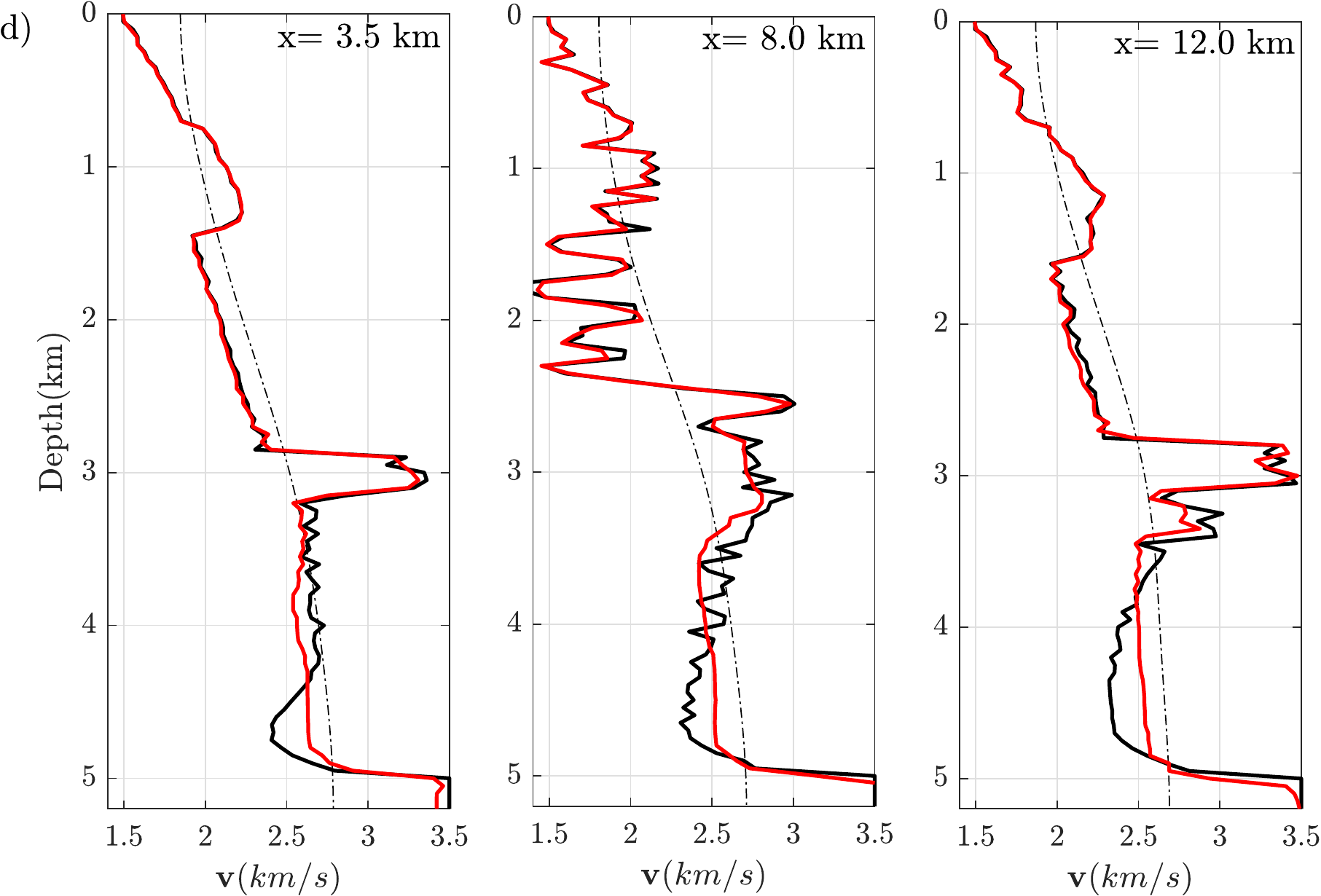}
\includegraphics[width=0.47\textwidth]{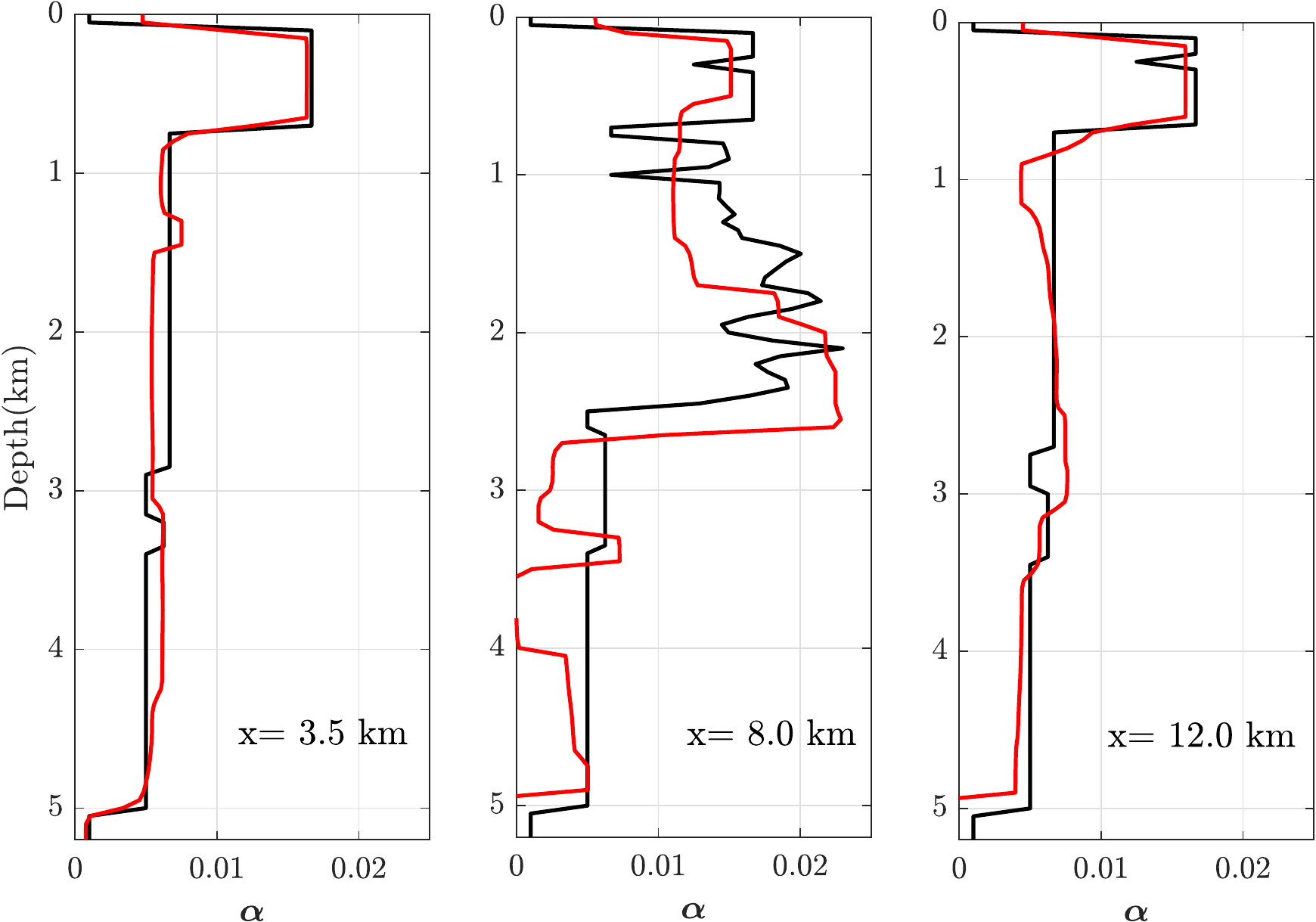}\\
\caption{North Sea case study with noiseless data. Direct comparison 
of extracted physical parameters using KF mechanism from visco-acoustic IR-WRI results. (a) without regularization, (b-d) with TV regularization using (b) Algorithm \ref{Alg1}, (c) Algorithm \ref{Alg2}, and (d)  Algorithm \ref{Alg3}. 
The first three logs are for velocity at $x=3.5, 8.0$ and $12.0 ~km$ respectively, and the second three logs are for attenuation at the same positions.}
\label{fig:north_test_log}
\end{figure}

\begin{figure}
\includegraphics[width=1\textwidth]{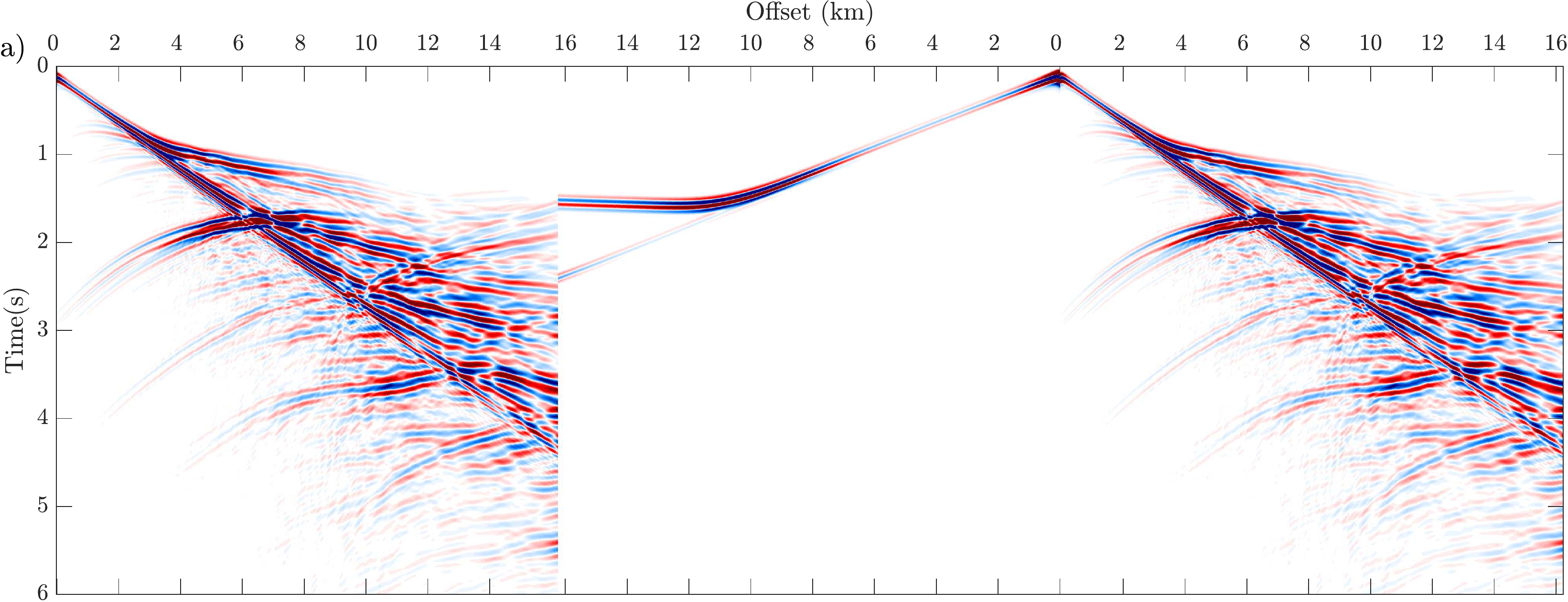}\\
\includegraphics[width=1\textwidth]{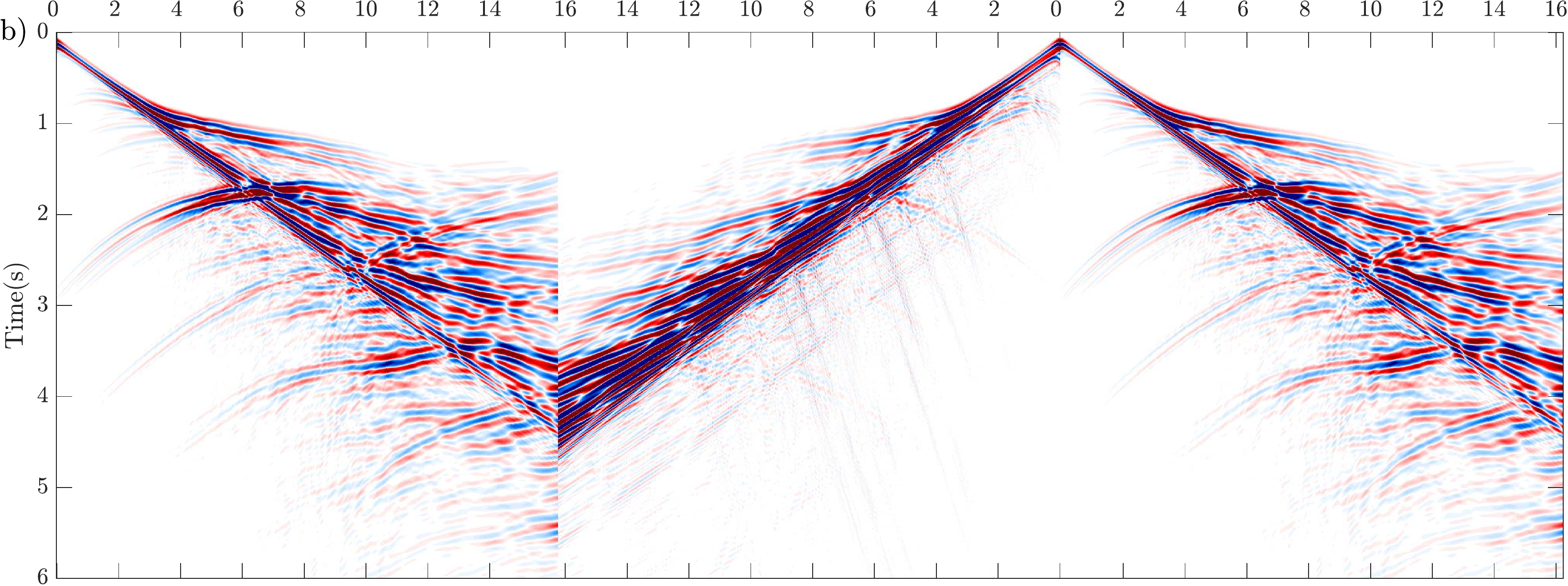}\\
\includegraphics[width=1\textwidth]{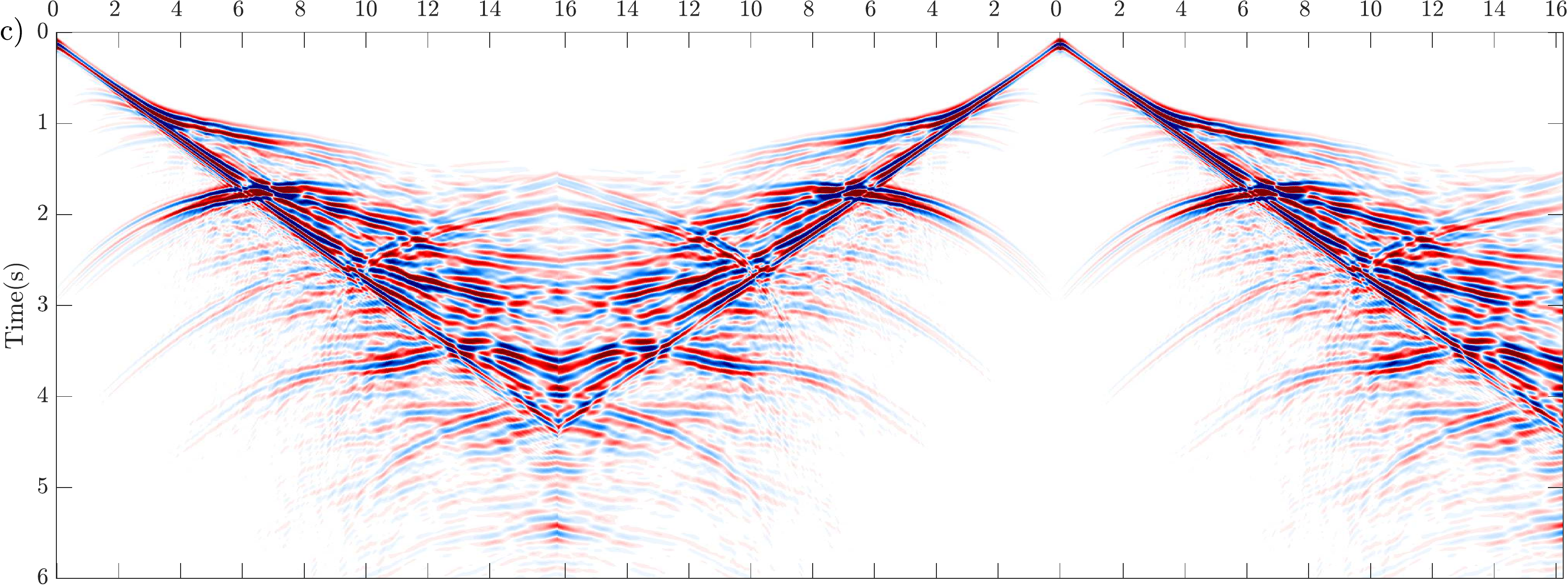}\\
\caption{North Sea case study. Time domain seismograms computed SLS attenuation mechanism in (a) the initial model, (b) IR-WRI without any regularization and (c) IR-WRI model with bound constraints and TV regularization using Algorithm \ref{Alg3}. The true seismograms are shown in the first and the last panel and the above mentioned seismograms are shown in the middle panel with a folded representation to allow for the comparison with the true seismograms at at short and long offsets.  The seismograms are plotted with a reduction velocity of 2.5 km/s for sake of time axis compression. The source is located at 16.0~km of distance on the sea bottom. }
\label{fig:north_test_seis}
\end{figure}

\begin{figure}
\includegraphics[width=1\textwidth]{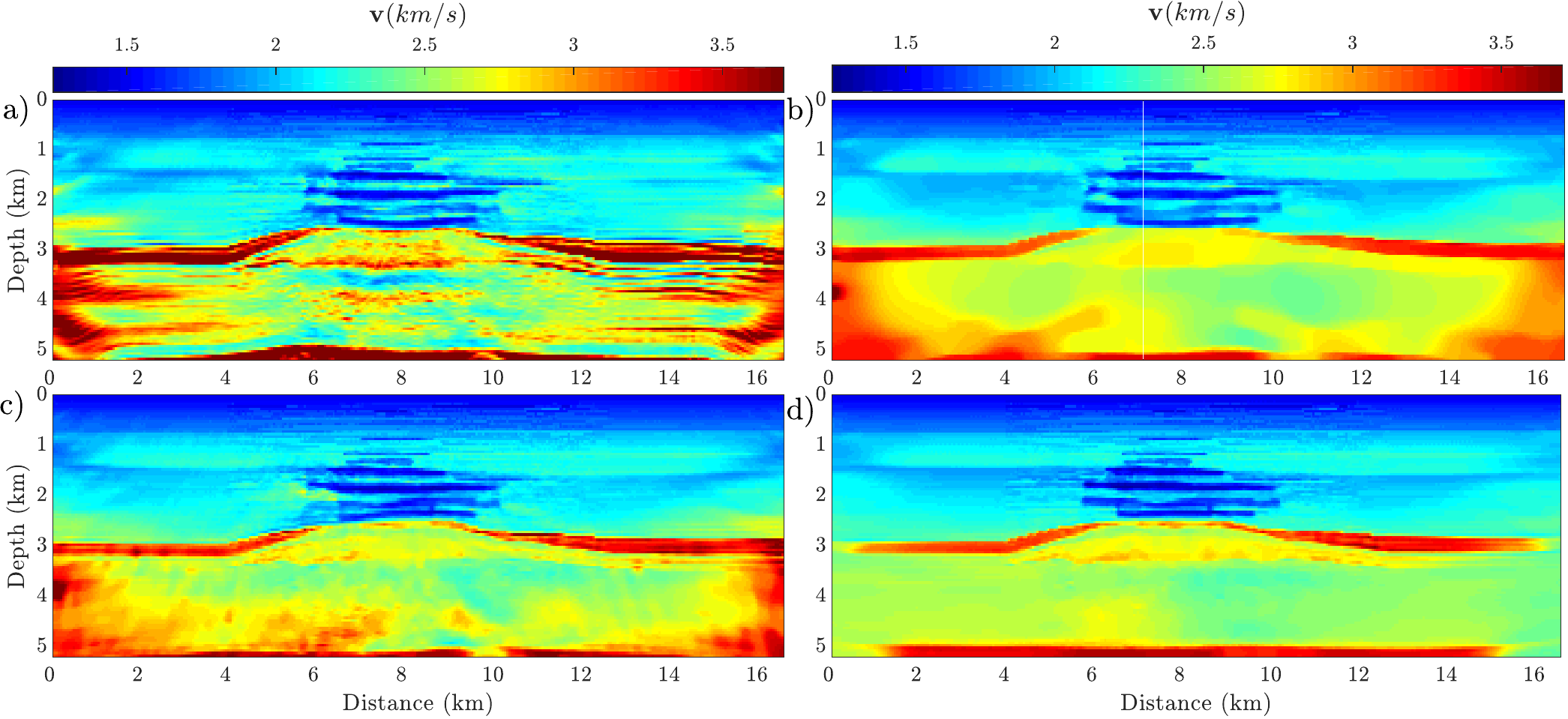}
\caption{North Sea case study. Extracted velocity using SLS mechanism from visco-acoustic IR-WRI results, (a) without regularization, (b-d) with TV regularization using (b) Algorithm \ref{Alg1}, (c) Algorithm \ref{Alg2}, and (d)  Algorithm \ref{Alg3}.}
\label{fig:north_test1}
\end{figure}

\begin{figure}
\includegraphics[width=1\textwidth]{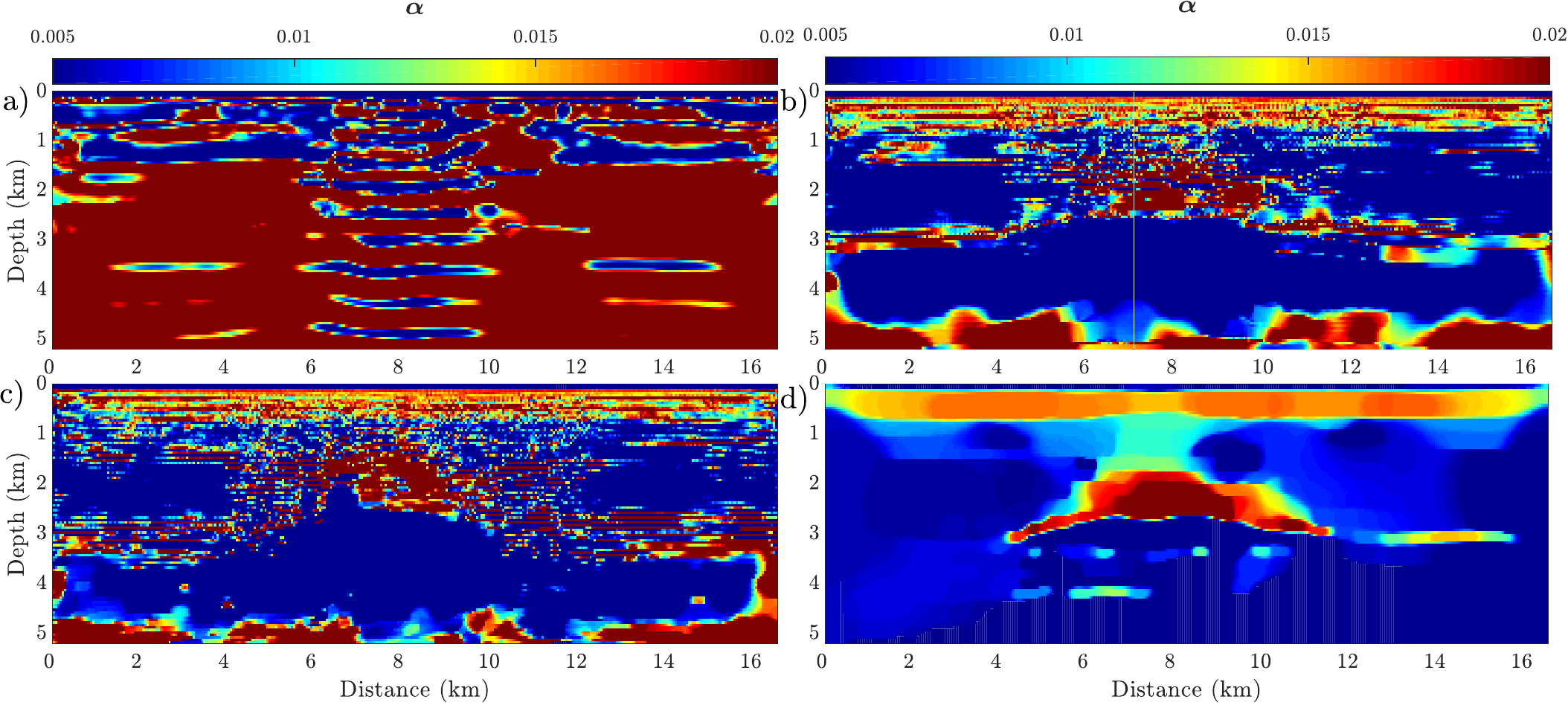}
\caption{Same as Fig. \ref{fig:north_test}, but for extracted $\boldsymbol{\alpha}$.}
\label{fig:north_test_Q1}
\end{figure}

\begin{figure}
\includegraphics[width=0.49\textwidth]{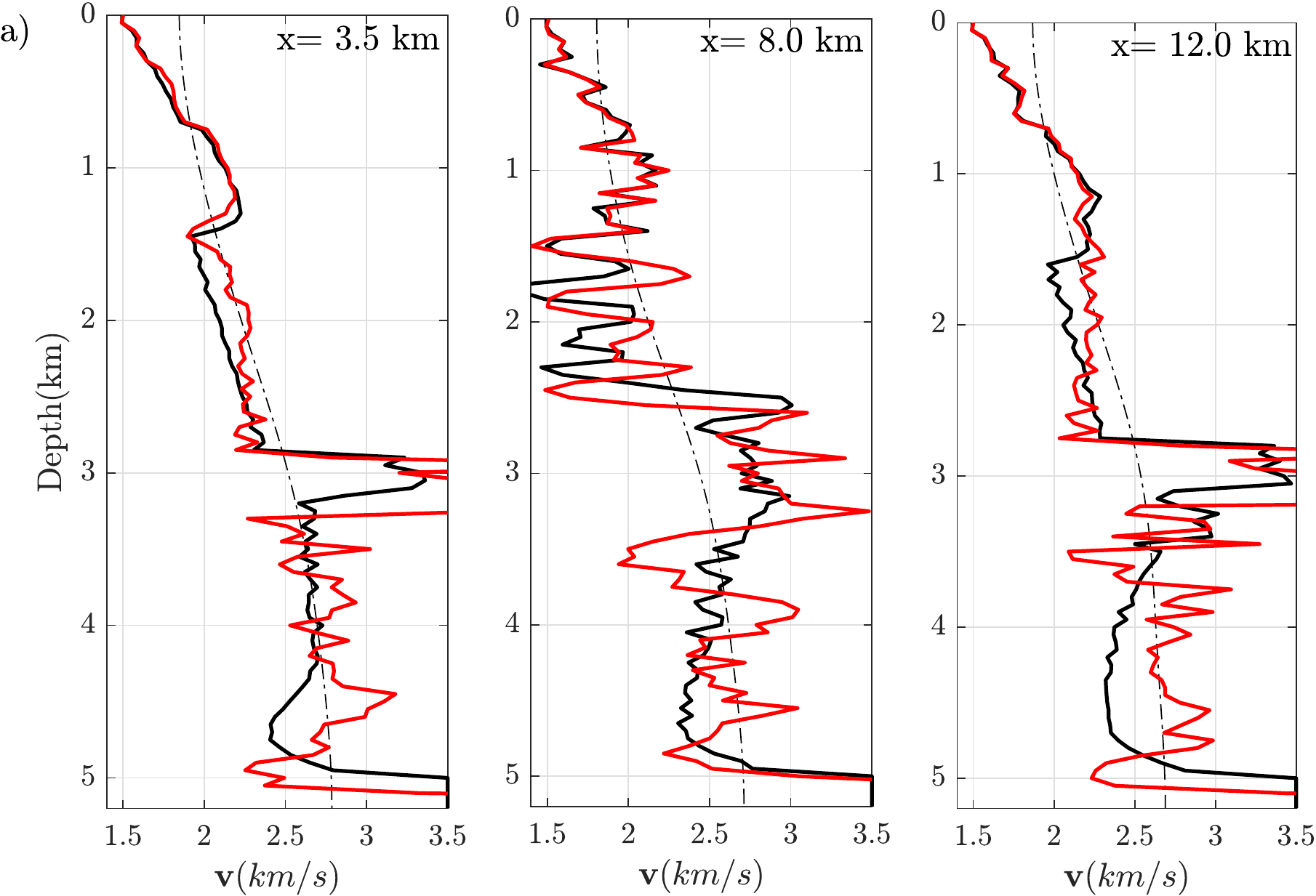}
\includegraphics[width=0.47\textwidth]{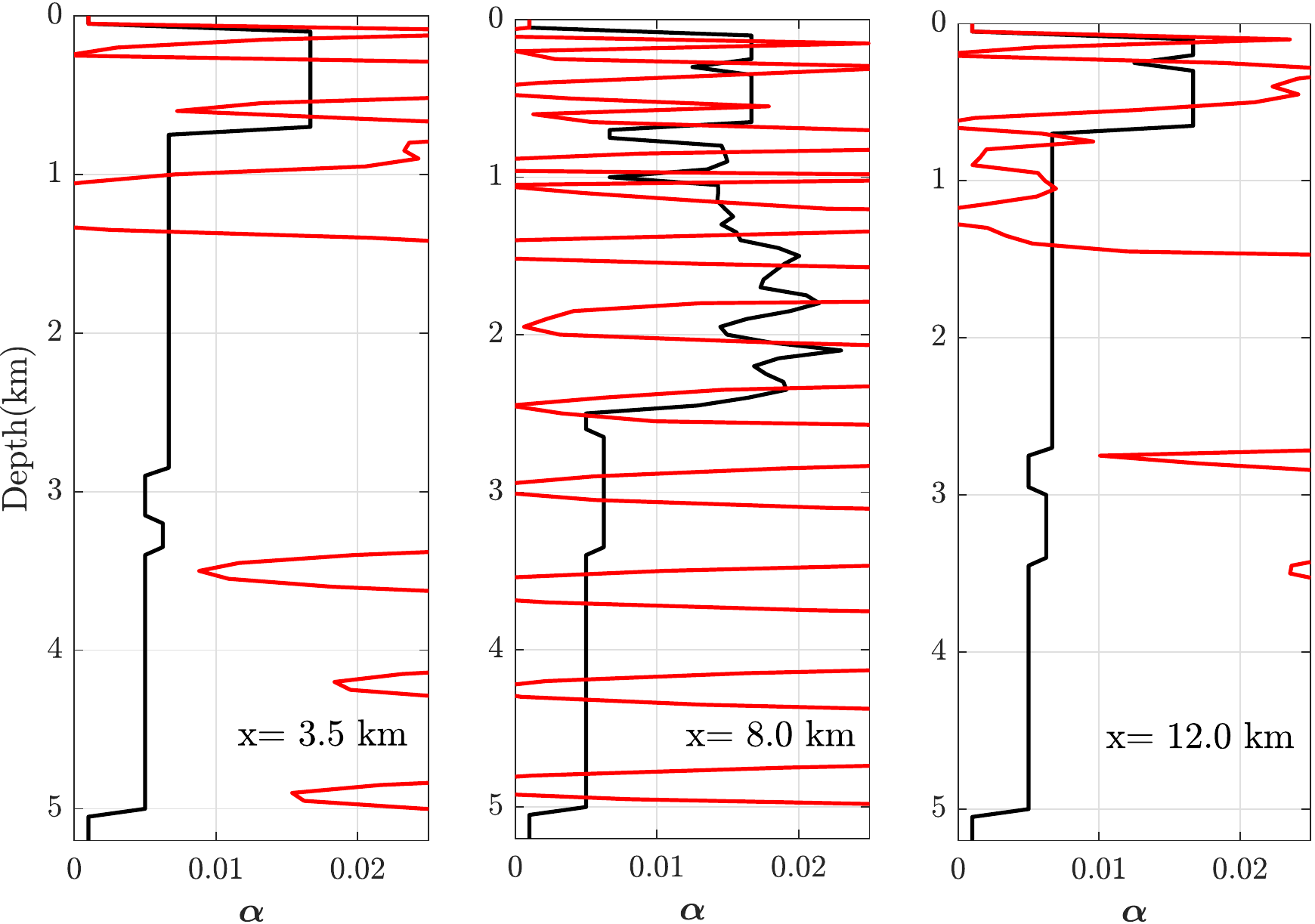}\\
\includegraphics[width=0.49\textwidth]{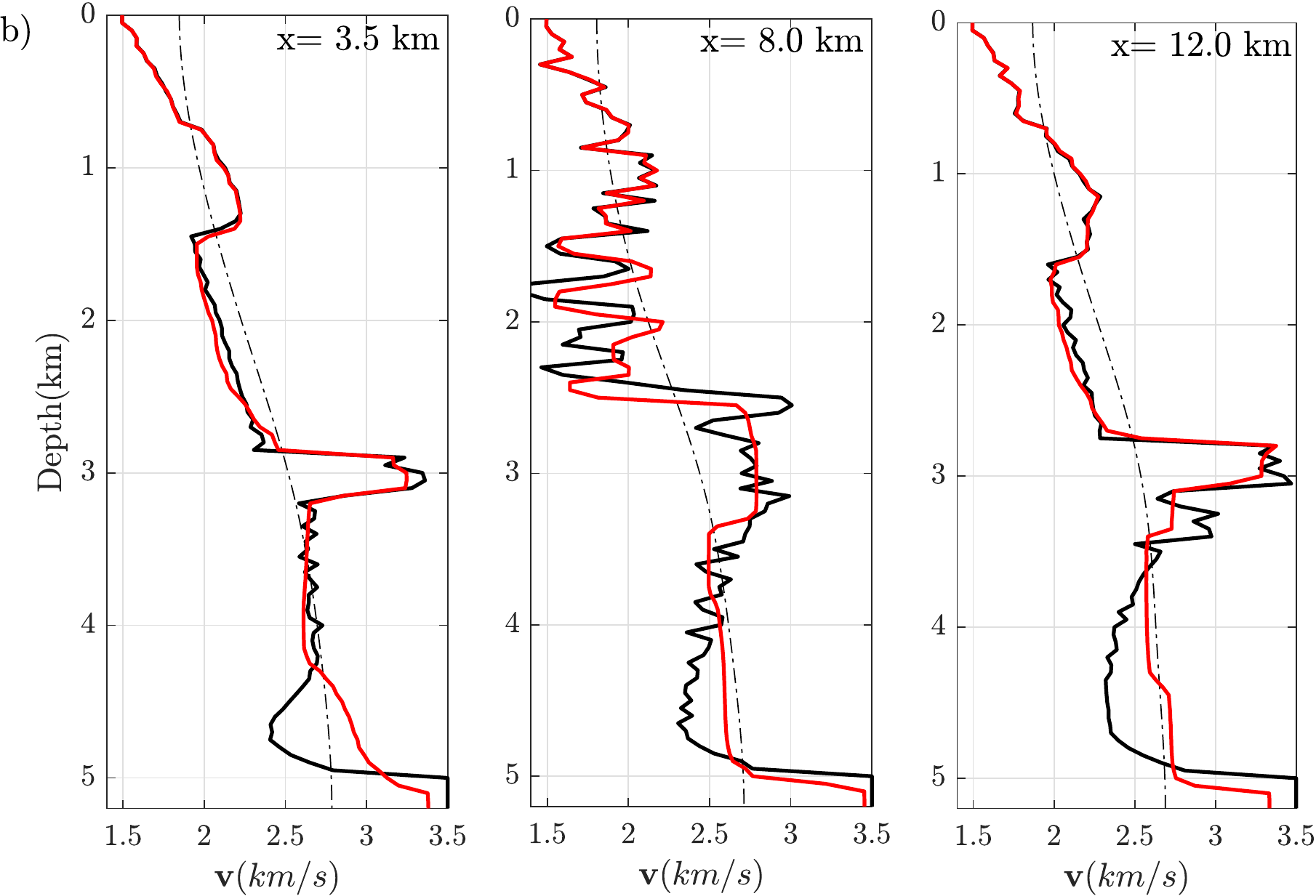}
\includegraphics[width=0.47\textwidth]{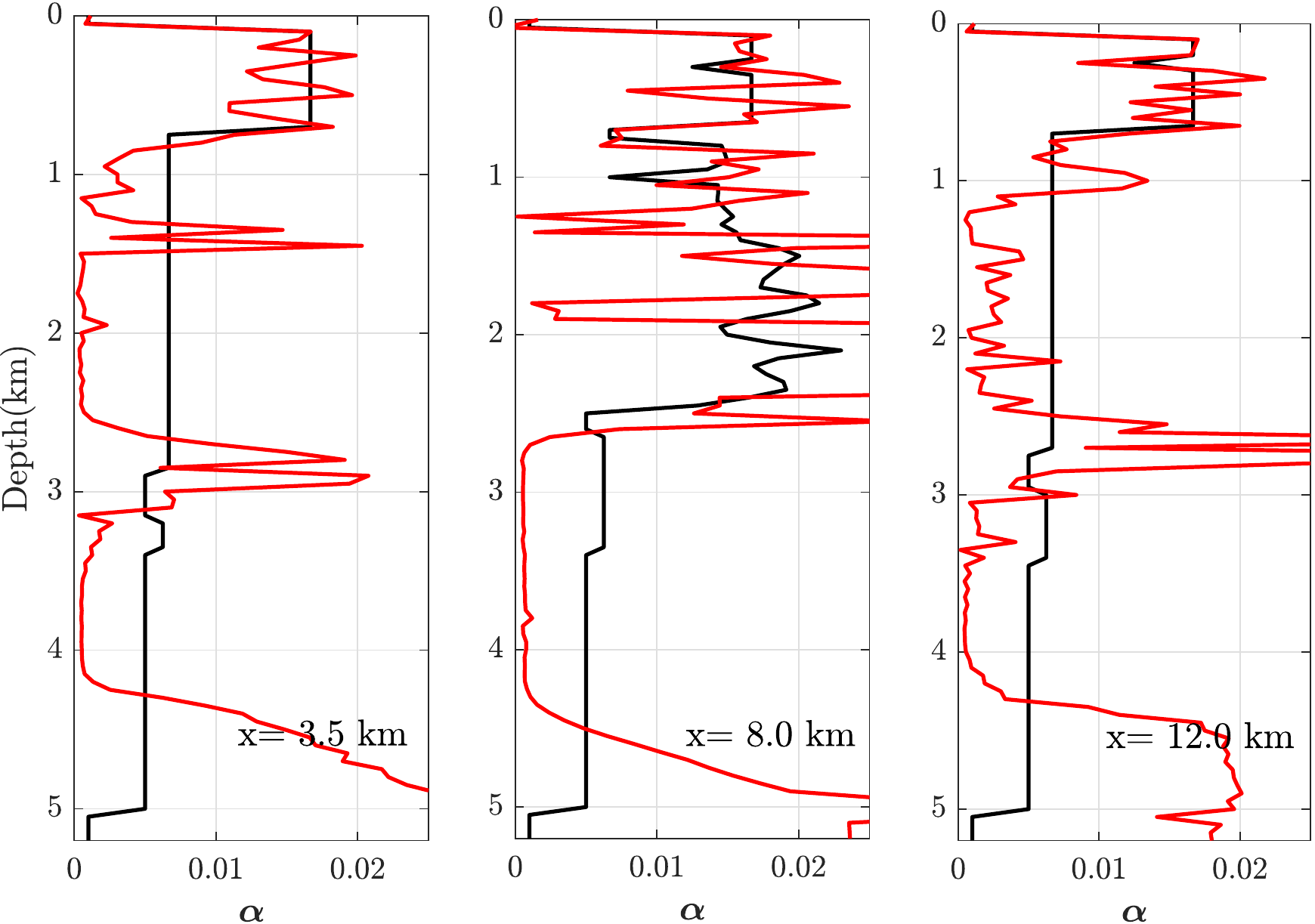}\\
\includegraphics[width=0.49\textwidth]{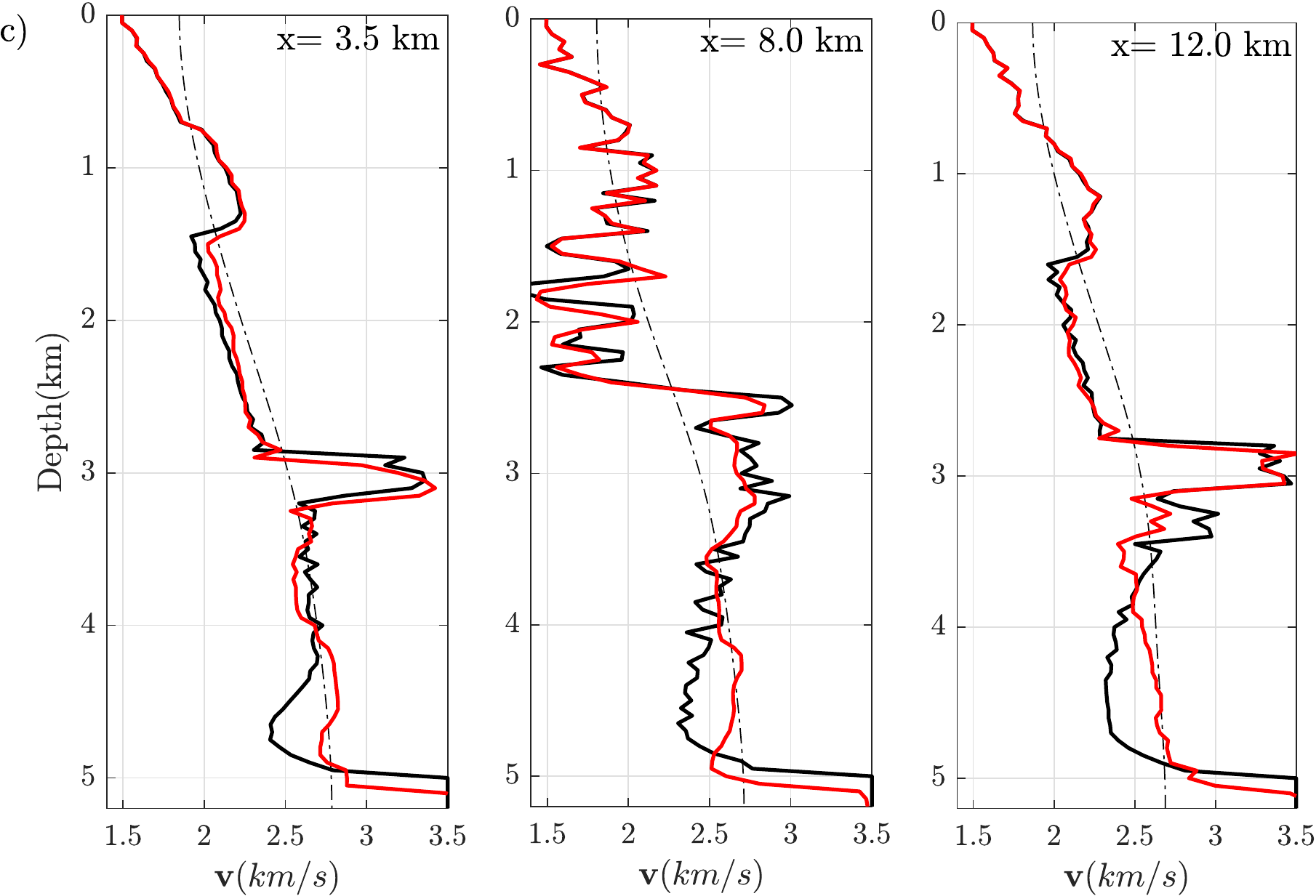}
\includegraphics[width=0.47\textwidth]{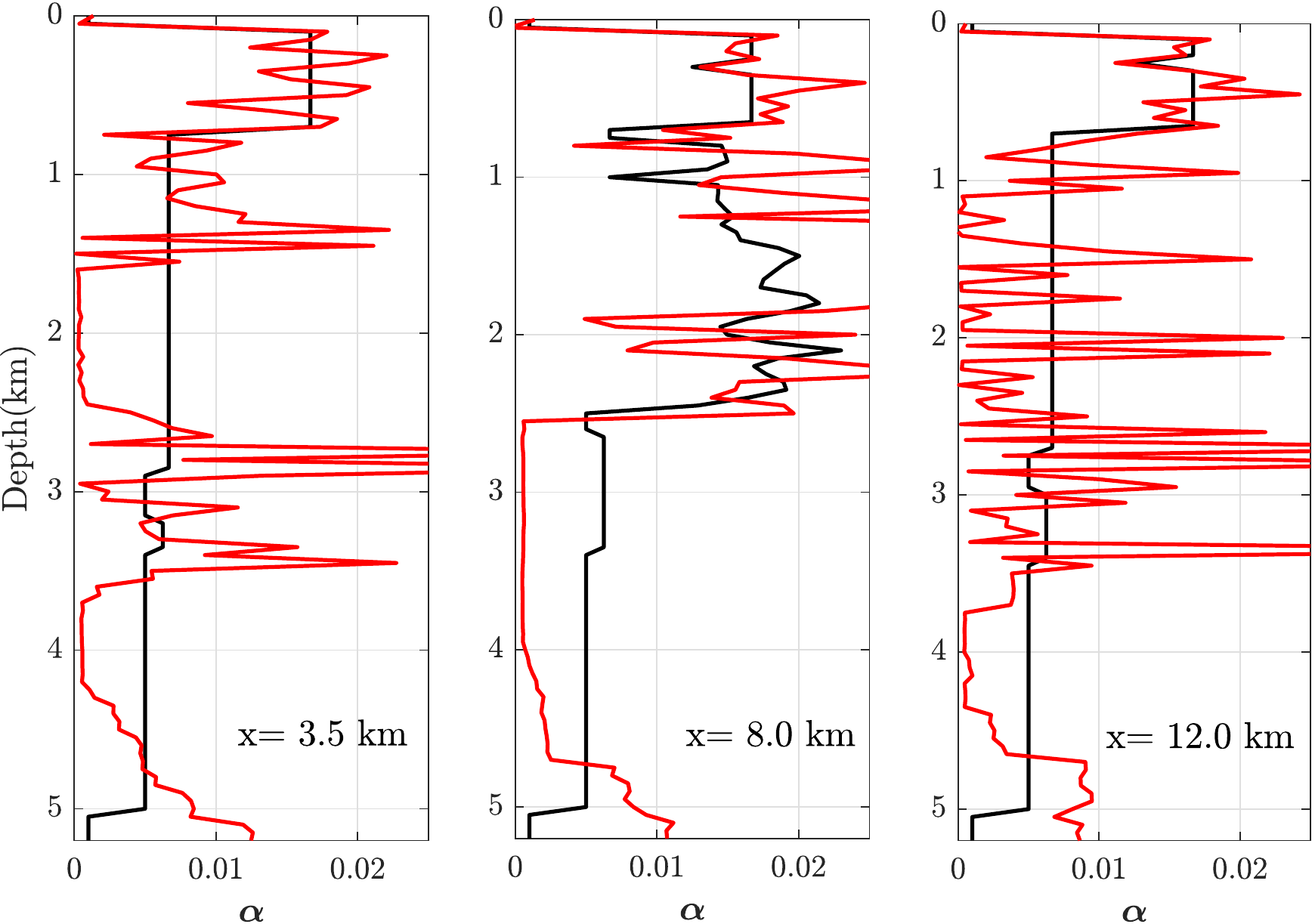}\\
\includegraphics[width=0.49\textwidth]{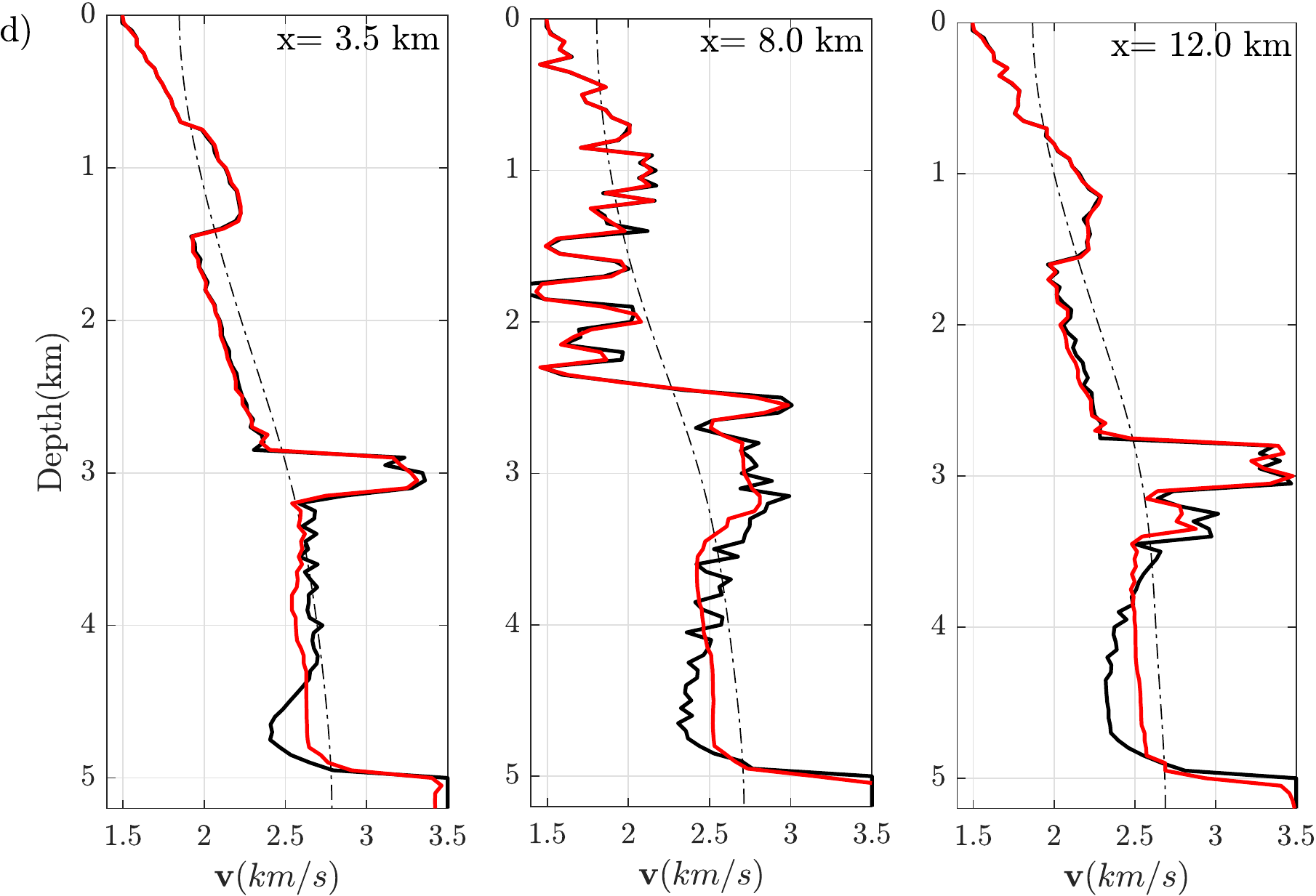}
\includegraphics[width=0.47\textwidth]{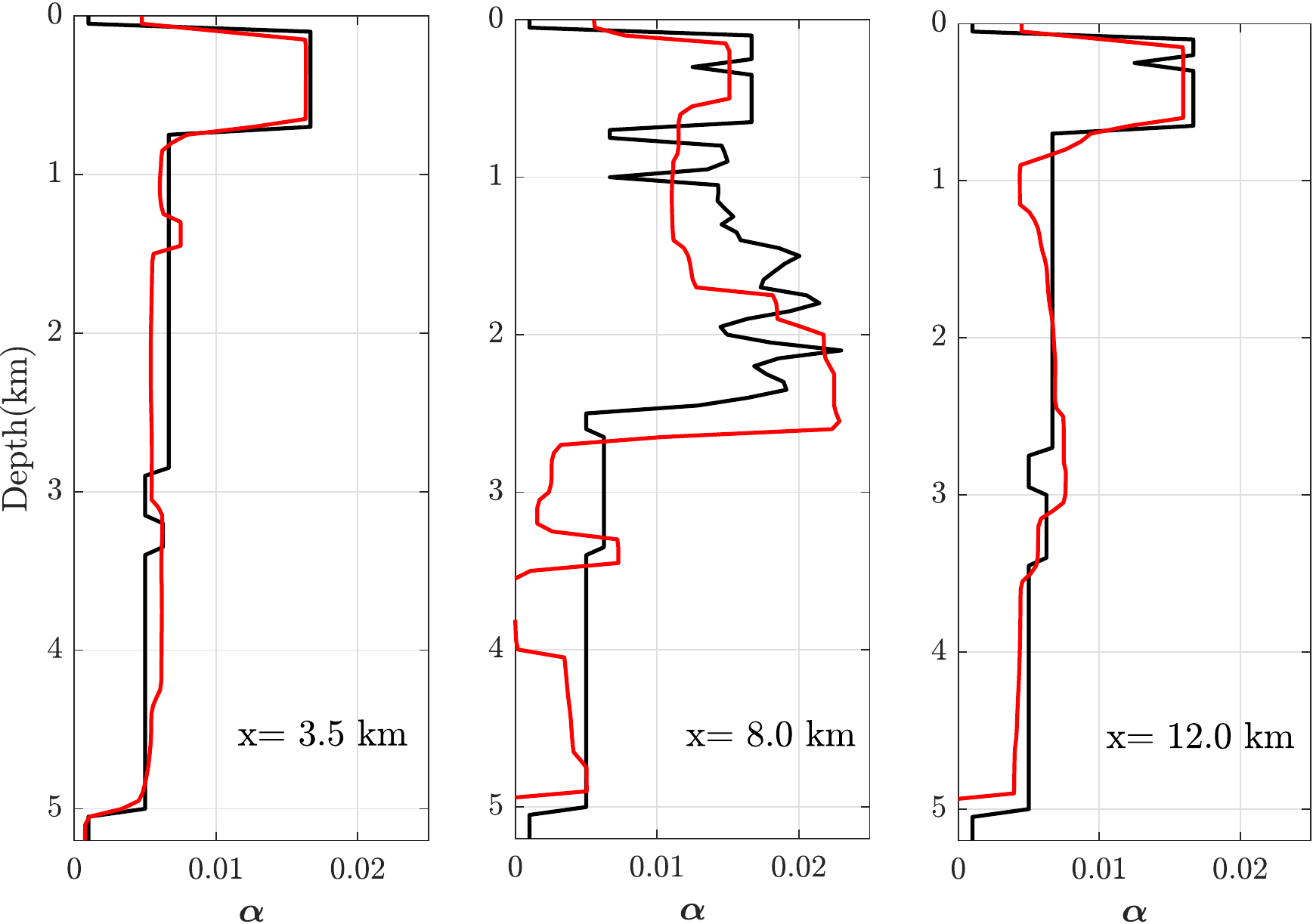}\\
\caption{North Sea case study with noiseless data. Direct comparison 
of extracted physical parameters using SLS mechanism from visco-acoustic IR-WRI results. (a) without regularization, (b-d) with TV regularization using (b) Algorithm \ref{Alg1}, (c) Algorithm \ref{Alg2}, and (d)  Algorithm \ref{Alg3}. 
The first three logs are for velocity at $x=3.5, 8.0$ and $12.0 ~km$ respectively, and the second three logs are for attenuation at the same positions.}
\label{fig:north_test_log1}
\end{figure}

%
%
The fixed-spread surface acquisition consists of 80 (reciprocal) explosive sources spaced 200~m apart at 25~m depth and 320 (reciprocal) hydrophone receivers spaced 50~m apart at 75~m depth on the sea bottom. 
A free-surface boundary condition is used on top of the grid, the source signature is a Ricker wavelet with a 10~Hz dominant frequency, and the SLS attenuation mechanism (\ref{SLSmodel0}) with $\omega_r=100\pi$ Rad is used to generate data.   
%
%
We compare the results obtained from IR-WRI without and with TV regularization using Algorithms \ref{Alg1}-\ref{Alg3} when the physical parameters $\boldsymbol{v}$ and $\boldsymbol{\alpha}$ are extracted from the estimated complex velocity using KF attenuation mechanism in \eqref{IKFv} and \eqref{IKFalpha}. 
We perform the inversion with small batches of three frequencies with one frequency overlap between two consecutive batches, moving from the low frequencies to the higher ones according to the frequency continuation strategy described in section \ref{MFI}. The starting and final frequencies are 3~Hz and 15~Hz and the sampling interval in each batch is 0.5~Hz. Compared to \cite{Keating_2019_PCA}, we use narrower frequency bands (1~Hz instead of 2~Hz) because the monovariate complex domain inversion should be less sensitive to parameter cross-talks. Under this assumption, narrowing the frequency bands allows us to mitigate the footprint of the piecewise frequency-independent approximation of our scheme (Figure~\ref{fig:SLS_KF}). The stopping criterion for each batch is given to be either reaching a maximum iteration 15 or
\begin{eqnarray}
\label{Stop}
\sum_{l=1}^L\| \bold{A(m}_{\omega_l}^{k+1})\bold{u}_{\omega_l}^{k+1}-\bold{b}_{\omega_l}\|_2^2 \leq \epsilon_b ~~ \nonumber \text{and} ~~  \sum_{l=1}^L\|\bold{Pu}_{\omega_l}^{k+1}-\bold{d}_{\omega_l}\|_2^2 \leq \epsilon_d,
\end{eqnarray} 
where $\epsilon_b=10^{-3}$ and $\epsilon_d=10^{-5}$.
We perform visco-acoustic IR-WRI for noiseless data. The extracted $\boldsymbol{v}$ and $\boldsymbol{\alpha}$ models using KF mechanism in \eqref{IKFv} and \eqref{IKFalpha} inferred from IR-WRI without regularization are shown in Figs. \ref{fig:north_test}a and \ref{fig:north_test_Q}a, respectively, while those obtained with TV regularization using Algorithm \ref{Alg1}-\ref{Alg3} are shown in Figs. \ref{fig:north_test}(b-d) and \ref{fig:north_test_Q}(b-d), respectively. The direct comparisons between the logs extracted from the true models, the initial model (just for $\boldsymbol{v}$) and the IR-WRI models at $x=3.5~km$, $x=8.0~km$ and $x=12.0~km$ are shown in Fig. \ref{fig:north_test_log} with same order as Fig \ref{fig:north_test} and \ref{fig:north_test_Q}. \\
Despite using a crude initial models, the shallow sedimentary part and the gas layers are fairly well reconstructed in all $\boldsymbol{v}$ models (without and with regularization). The main differences are shown at the reservoir level and below. Without regularization, the reconstruction at the reservoir level is mispositioned and the inversion fails to reconstruct the smoothly-decreasing velocity below the reservoir due to the lack of diving wave illumination at these depths. This in turn prevents the focusing of the deep reflector at 5~km depth by migration of the associated short-spread reflections. When regularization is used, visco-acoustic IR-WRI provides a more accurate and cleaner images of the reservoir and better reconstructs the sharp contrast on top of it (especially using Algorithm \ref{Alg3}). The extracted $\boldsymbol{v}$ model with Algorithm \ref{Alg3}, Fig. \ref{fig:north_test}d, also reconstructs the deep reflector at the correct depth in the central part of the model. \\
The estimated $\boldsymbol{\alpha}$ models with regularization are more accurate in the  shallow part (less than 750~m) compared to the deep part. The estimated $\boldsymbol{\alpha}$ model with Algorithm \ref{Alg3}, Fig. \ref{fig:north_test_Q}d, is better reconstructed compared to the estimated model without/with regularization using Algorithm \ref{Alg1} and \ref{Alg2}, hence validating the conclusions of the previous tests. The higher sensitivity of $\boldsymbol{\alpha}$ to the regularization scheme relative to $\boldsymbol{v}$ highlights the higher sensitivity of the data to the later, which controls the kinematic aspects of wave propagation. 
We further assess the IR-WRI results in Figs. \ref{fig:north_test_seis} by comparing time-domain seismograms computed in the true $\boldsymbol{v}$ and $\boldsymbol{\alpha}$ models (Figs. \ref{fig:north_test_true}a, \ref{fig:val_initial}a) with those computed in the initial models (Figs. \ref{fig:north_test_true}b, \ref{fig:val_initial}b) and the final IR-WRI models obtained without regularization (Figs. \ref{fig:north_test}a,\ref{fig:north_test_Q}a) and with TV regularization (Algorithm \ref{Alg3}) (Figs. \ref{fig:north_test}d, \ref{fig:north_test_Q}d). In any case, the seismograms are computed with the SLS model (eq. \eqref{SLSmodel0}) with $\omega_r=100 \pi$ Rad. This means that, although we assume that $\boldsymbol{m}$ was independent to frequency during the inversion of one frequency batch, we perform the forward simulation with this frequency dependency to assess whether the reconstructed $\boldsymbol{v}$ and $\boldsymbol{\alpha}$ models allows for data fitting when the true attenuation model is used. 
Accordingly, we stress that our implementation of visco-acoustic IR-WRI does not rely on an inverse crime procedure in the sense that the forward problem used during inversion embeds an implicit approximation associated with the band-wise frequency dependence $\boldsymbol{m}$.
Seismograms computed in the initial model (Fig. \ref{fig:north_test_seis}a) mainly show the direct wave and the diving waves, the latter being highly cycle skipped relative to those computed in the true model. 
Seismograms computed in the unregularized IR-WRI model (Fig. \ref{fig:north_test_seis}b) don't match recorded amplitudes due to irrelevant $\boldsymbol{\alpha}$ (Fig. \ref{fig:north_test_Q}b). Seismograms computed in the IR-WRI model regularized with Algorithm \ref{Alg3} match remarkably well the traveltimes and amplitude of the recorded arrivals at short and long offsets, hence validating the piecewise frequency-independent approximation. \\
For sake of completeness, we show  $\boldsymbol{v}$ and $\boldsymbol{\alpha}$ models when they are extracted with the SLS model (that used to generate the data) in Figs.~\ref{fig:north_test1}-\ref{fig:north_test_log1}. These models can be compared with those extracted with the KF model shown in Figs.~\ref{fig:north_test}-\ref{fig:north_test_log}. Consistently with the results of the inclusion test, we show that the attenuation model used for extraction has a little imprint on the recovered $\boldsymbol{v}$ (Figs.~\ref{fig:north_test1},\ref{fig:north_test_log1}). It has a negligible imprint on the recovered $\boldsymbol{\alpha}$ when the complex velocity has been accurately recovered with the TV regularization on phase and amplitude (Algorithm \ref{Alg3}) (Figs.~\ref{fig:north_test_Q1}d-\ref{fig:north_test_log1}d). The imprint is more visible on the amplitudes of $\boldsymbol{\alpha}$ when the complex velocity is recovered less accurately with Algorithms \ref{Alg1}-\ref{Alg2} (Figs.~\ref{fig:north_test_log1}(a-c)). This imprint manifests by underestimated $\boldsymbol{\alpha}$ in Fig.~\ref{fig:north_test_log} relative to Fig.~\ref{fig:north_test_log1}. Note that these underestimated $\boldsymbol{\alpha}$ have a negligible effects on the recovered $\boldsymbol{v}$, which is another illustration of the higher sensitivity of the data to  $\boldsymbol{v}$ relatively to $\boldsymbol{\alpha}$.

%


\section{Conclusions} \label{conclusions}
We implemented the joint estimation of velocity and attenuation in the frequency-domain IR-WRI using complex velocity as the optimization parameter.  This differs from traditional methods that process separately  real-valued phase velocities and attenuation in the real domain (C2R method). Optimization for complex-valued variables allows us to preserve the bi-convexity of the IR-WRI by considering two variables (wavefield and complex-valued velocity) instead of three variables (wavefield,  real-valued velocity and attenuation). Dispersion effects, which make complex velocities frequency dependent, are approximately account for through a hierarchical multiscale inversion proceeding over small frequency batches. This design amounts to assume a band-wise frequency dependency of complex velocity.
Although this narrow band-by-band design is suboptimal to uncouple real-valued velocity and attenuation when visco-acoustic FWI is implemented in the real domain, it should not be detrimental in the complex domain where FWI is recast as a mono-parameter problem.
Regularizations are implemented with ADMM and proximal methods such that they can be applied separately on different parts of the complex-valued parameters (real, imaginary, magnitude, and phase). This gives the necessary flexibility to tailor this regularization to each of these parts.
Unlike C2R methods, this new inversion algorithm in the complex domain does not rely on an a priori empirical relation between the complex velocity and the real phase velocity and attenuation (i.e., KF or SLS mechanism). Instead, the real phase velocity and attenuation models are extracted a posteriori from the final complex velocity model with arbitrary empirical relation, hence allowing the assessment of several of them. The method was validated against a series of synthetic experiments. All of them supports that regularization of the magnitude and phase of the complex velocity provides the most reliable results. This probably results because the amplitude and phase of the complex velocity better separate the real velocity and attenuation, which are the physical parameters we seek to recover. These parameters have a different signature in the data in terms of strength and trend (kinematic, dynamic, dispersion), and hence required tailored regularization.  A realistic synthetic test inspired from the North sea further validates the workflow. In particular, it shows that time-domain seismograms computed in the real velocity and attenuation models extracted from the final complex-valued velocity model match the data remarkably well when the time-domain simulation is performed with the same empirical relation as that used to generate the recorded data. The numerical examples also support that the attenuation mechanism used to extract the real parameters has a minor impact on the recovered parameters as long as the complex velocities have been reconstructed sufficiently accurately. This results from the fact that the final complex velocity from which the real parameters are extracted has been estimated from a narrow batch of high frequencies.

\section*{Acknowledgments}
This study was partially funded by the WIND consortium (\textit{https://www.geoazur.fr/WIND}), sponsored by Chevron, Shell, and Total. This study was granted access to the HPC resources of SIGAMM infrastructure (http://crimson.oca.eu), hosted by Observatoire de la C\^ote d'Azur and which is supported by the Provence-Alpes C\^ote d'Azur region, and the HPC resources of CINES/IDRIS/TGCC under the allocation A0050410596 made by GENCI."
\appendix 
\section{$\mathcal{M}$ for KF and SLS attenuation mechanism} \label{Appa}
Basically, $\mathcal{M}$ is a frequency-dependent non-linear mapping between velocity and attenuation factor at the reference frequency \cite{Ursin_2002_CDA}. \cite{Ursin_2002_CDA} have compiled eight different $\mathcal{M}$ models
related to some of empirical attenuation mechanisms. Here we review the KF and SLS which are used in this paper.  
The most traditional form of $\mathcal{M}$ is the KF relation:
\begin{equation} \label{KFmodel0}
\boldsymbol{m}=\mathcal{M}(\omega,\boldsymbol{v}, \boldsymbol{\alpha})=\frac{1}{\boldsymbol{v}^2}\left( 1- \frac{\boldsymbol{\alpha}}{\pi } \ln |\frac{\omega}{\omega_r}| + i \frac{\boldsymbol{\alpha}}{2 }\right)^2,
\end{equation} 
where $\omega_r$ is a reference frequency  \cite{Aki_2002_QST}. Note that $\mathcal{M}$ depends on frequency through the logarithmic causality correction term.\\
The velocity and attenuation factor parameters can be obtained form the estimated complex velocity $\boldsymbol{\hat{m}}$ by the inverse mapping of $\mathcal{M}$ as
\begin{eqnarray}
\hat{\boldsymbol{v}}= \frac{1}{\Re(\sqrt{\boldsymbol{\hat{m}}}) + \frac{2}{\pi}\ln |\frac{\omega}{\omega_r}| \Im(\sqrt{\boldsymbol{\hat{m}}})}, \label{IKFv}\\
\hat{\boldsymbol{\alpha}}=\frac{2\Im(\sqrt{\boldsymbol{\hat{m}}})}{\Re(\sqrt{\boldsymbol{\hat{m}}}) + \frac{2}{\pi}\ln |\frac{\omega}{\omega_r}| \Im(\sqrt{\boldsymbol{\hat{m}}})} \label{IKFalpha}
\end{eqnarray}
For SLS mechanism $\mathcal{M}$ is defined as 
\begin{equation} \label{SLSmodel0}
\boldsymbol{m}=\mathcal{M}(\omega,\boldsymbol{v}, \boldsymbol{\alpha})=\frac{1}{\boldsymbol{v}^2}\Re\left(\sqrt{\frac{1+i\omega_r \tau_\sigma}{1+i\omega_r \tau_\epsilon}} \right)^{-2} \frac{1+i\omega \tau_\sigma}{1+i\omega \tau_\epsilon},
\end{equation} 
where $\tau_\epsilon$ and $\tau_\delta$ are relaxation times related to the constants of the effective springs and dash-pot of the model \cite{Zhu_2013_ACS} and defined as 
\begin{eqnarray}
\tau_\epsilon=\frac{1}{\omega_r}\left(\sqrt{1+\boldsymbol{\alpha}^2}+\boldsymbol{\alpha}\right),\label{tau_epsi}\\
\tau_\sigma=\frac{1}{\omega_r}\left(\sqrt{1+\boldsymbol{\alpha}^2}-\boldsymbol{\alpha}\right). \label{tau_sigma}
\end{eqnarray}  
Finally, the attenuation factor can be obtained form the estimated complex velocity $\boldsymbol{\hat{m}}$ by the inverse mapping of $\mathcal{M}$ as
\begin{equation}
\boldsymbol{\hat{\alpha}}=\frac{\omega^2 + \omega_r^2}{2\omega \omega_r \frac{\Re(1/\bold{\hat{m}})}{\Im(1/\bold{\hat{m}})}} \label{ISLSalpha}
\end{equation}
and after extracting $\tau_\epsilon$ and $\tau_\delta$ using \eqref{tau_epsi} and \eqref{tau_sigma}, $\hat{\boldsymbol{v}}$ will be
\begin{equation}
\hat{\boldsymbol{v}}=\sqrt{\Re(\frac{1}{\bold{\hat{m}}})\Re\left(\sqrt{\frac{1+i\omega_r\tau_\sigma}{1+i\omega_r\tau_\epsilon}}\right)^{-2}\frac{1+\omega^2\tau_\sigma^2}{1+\omega^2\tau_\sigma\tau_\epsilon}}. \label{ISLSv}
\end{equation}

\bibliographystyle{siamplain}
\newcommand{\SortNoop}[1]{}

\end{document}